\documentclass[12pt]{article}
\usepackage{latexsym}
\usepackage{graphics}
\usepackage{color}
\begin{document}

\begin{titlepage}
\begin{flushright}
       {\bf UK/98-03}  \\
       June 1998      \\
\end{flushright}
\begin{center}

{\bf {\LARGE  Valence QCD: Connecting QCD to the Quark Model}}

\vspace{1.5cm}

{\bf  K.F. Liu$^a$, S.J. Dong$^a$, T. Draper$^a$, D. Leinweber$^b$,
J. Sloan$^a$, \mbox{W. Wilcox}$^c$, and R. M. Woloshyn$^d$} \\ [0.5cm]
{\it $^a$ Dept. of Physics and Astronomy,  
  Univ. of Kentucky, Lexington, KY 40506\\
  $^b$ Special Research Centre for the Subatomic Structure of Matter
            and Department of Physics and Mathematical Physics,
            University of Adelaide, Australia 5005 \\
  $^c$ Dept. of Physics, Baylor Univ., Waco, TX 76798 \\
  $^d$ TRIUMF, 4004 Wesbrook Mall, Vancouver, B. C., Canada V6T 2A3}
\end{center}

\begin{abstract}
A valence QCD theory is developed to study the valence quark properties of
hadrons. To keep only the valence degrees of freedom, the pair creation 
through the Z graphs is deleted 
in the connected insertions; whereas, the sea quarks are eliminated in the
disconnected insertions. This is achieved with a new ``valence QCD'' lagrangian
where the action in the time direction is modified so that
the particle and antiparticle decouple. 
It is shown in this valence version of QCD that
the ratios of isovector to isoscalar matrix
elements (e.g. $F_A/D_A$ and $F_S/D_S$ ratios) in the nucleon
reproduce the $SU(6)$ quark model
predictions in a lattice QCD calculation. We also consider 
how the hadron masses are affected on the lattice and discover new insights 
into the origin of dynamical mass generation. It is found that, within 
statistical errors, the nucleon and the $\Delta$ become degenerate for the 
quark masses we have studied (ranging from one to four times the strange mass).
 The $\pi$ and $\rho$ become nearly degenerate in this range. 
It is shown that valence QCD has the C, P, T symmetries. The lattice version
is reflection positive. It also has the vector and axial symmetries, 
the latter leads to a modified partially conserved axial Ward identity. As 
a result, the theory has a $U(2N_F)$ symmetry in the particle-antiparticle 
space. Through lattice simulation, it appears that this is dynamically
broken down to $U_q(N_F) \times U_{\bar{q}}(N_F)$. Furthermore, the
lattice simulation reveals spin degeneracy in the hadron masses and
various matrix elements. This leads to an approximate 
$U_q(2N_F) \times U_{\bar{q}}(2N_F)$ symmetry which is the basis for the
valence quark model. 
In addition, we find that the masses of N, $\Delta, \rho, \pi, a_1$, and 
$a_0$ all drop precipitously compared to their counterparts in the quenched
QCD calculation. This is interpreted as due to the disapperance of the
`constituent' quark mass which is dynamically generated through tadpole
diagrams. The origin of the hyper-fine splitting in the baryon is largely 
attibuted to the Goldstone boson exchanges between the quarks. Both of 
these are the consequences of lacking chiral symmetry in valence QCD. 
We discuss its implication on the models of hadrons.

\bigskip
 
 
\end{abstract}

\vfill
\end{titlepage}

\section{Introduction}
In addition to its classification scheme, the quark model is,
by and large, quite successful in delineating the
spectrum, structure, and decays of mesons and baryons. One
often wonders what the nature of the approximation is, especially
in view of the advent of quantum chromodynamics (QCD) which is
believed to be, after all, the fundamental theory of the quarks and gluons.
In order to address this question, we need to understand first where
the quark model is successful and where it fails.
 
To begin with, we need to define what 
we mean by the quark model. We consider the simplest approach which 
includes the following ingredients:
\begin{itemize}
\item
The Fock space is restricted to the valence quarks only, i.e.
3 quarks for the baryon and a quark-antiquark pair for the meson.
Although there
are variations by including quark self-energy and so on which
go beyond the instantaneous interaction and will invoke higher
Fock space (e.g. $q^4\bar{q}$ for the baryon and $q^2\bar{q}^2$ for
the meson), we will not consider them here.
\item
 These valence quarks,
be they the dressed constituent quarks or the bare quarks, are
confined in a potential or a bag. To this zeroth order, the hadron
wave functions involving $u,d,$ and $s$ quarks and anti-quarks
are classified by the flavor-spin and spatial coordinates according
to the $SU_q(6) \times SU_{\bar{q}}(6) \times O(3)$ group
(for brevity, we shall refer it as $SU(6)$).
The wave functions are totally antisymmetric in the color space for the
baryons and symmetric in the color-anticolor combinations for the
meson. For example, the S- and D-wave baryons are described by
the 56-plets and the P-wave baryons by the 70-plets. Similarly, the
S- and P-wave mesons are described by the 36-plets~\cite{clo79}.
\item
The $SU(6)$ symmetry is broken down to $SU(3) \times SU(2)$ by the
residual interaction between the quarks which is
weak compared to the confining potential.
The degeneracies within the multiplets are lifted by these
residual interactions.
 
Of course, additional breakings of flavor $SU(3)$ due to the
quark masses are responsible for the detailed splitting within the
octet/decuplet baryon multiplets and the meson nonets.
\end{itemize}
 
   There are many different versions of the
quark model which share these attributes. They have been called naive
quark model, non-relativistic quark model, constituent quark model,
bag model, etc. in the literature. Here we shall refer them generically
as the {\it\bf Valence Quark Model} with the defining features of the
lowest Fock space (or valence Fock space) and the $SU(6)$ flavor-spin
symmetry, albeit approximate, as their common denominator.
In this work, we shall concentrate our discussion on the light quark
systems where the valence quark picture is less well understood.
For mesons with heavy quarks, such as the charmoniums and upsilons,
the valence picture based on the non-relativistic potential model which
fits experiments reasonably well is confirmed by the
non-relativisitc lattice QCD calculations~\cite{slo95,shi97,ses98,dav98}.
We shall not address them in this study.

Given the definition of the valence quark model, it is easier to
understand where it succeeds as well as where it fails. For example,
with the one-gluon like exchange potential~\cite{dgg75} as the residual
interaction between the non-relativistically confined quarks to describe
the hyper-fine and fine splittings of the hadron masses,
the valence quark model is very successful in fitting meson and baryon 
masses \cite{lw78b,lw83,ik78,ik79} and baryon magnetic 
moments~\cite{dgg75,br88,pon96}.
It is also successful in delineating the pattern of 
electro-magnetic~\cite{clo79,ono76,ki80},
semileptonic and nonleptonic weak decays~\cite{dgh92},
and the Okubo-Zweig-Iizuka (OZI) rule~\cite{ozi63}, etc.
Similarly it is true for the MIT bag model where the relativistic
quarks and anti-quarks are confined in a bag with one-gluon exchange
interaction~\cite{cjj74,dh84}.
 
It is worthwhile
noting that all these are based on the valence quark picture aided with
$SU(6) \times O(3)$ group for its flavor-spin and space group. On the
other hand, there are notable failures. For example, it fails to account
for the $U(1)$ anomaly (the $\eta'$ mass), the proton spin crisis and the
$\pi N \sigma$ term. All these
problems are associated with large contributions from disconnected insertions
involving sea-quarks~\cite{wv79,dll95,dll96}. These are places where
the OZI rule fails badly. Consequently, it is natural not to expect the
valence quark model to work in these cases.
 
There are also other places where the valence quark model does not 
work well.
They include hadron scatterings, couplings, and form factors which are
well described by models utilizing the chiral symmetry inherent in QCD.
Examples of successful approaches based on chiral symmetry
include $\pi\pi$ scattering~\cite{wei66,gl85}, vector dominance~\cite{sak66},
KSRF relation \cite{ksr66}, low-energy $\pi N$ scatterings
~\cite{wei66,tom66}, $\pi N$ scattering
up to about 1 GeV with the skyrmion \cite{mat87}, nucleon static
properties~\cite{anw83}, electromagnetic~\cite{anw83} form factors,
$\pi NN$ form factor \cite{ll87}, NN interaction, and
Goldberger-Treiman relation~\cite{gt58}. All these have been worked out quite
successfully by the parallel developments which explore the chiral
symmetry of QCD. These include the $\sigma$ model, current algebra, PCAC,
chiral perturbation theory, and the more recent developments incorporating
large $N_c$ QCD~\cite{th74,wit79}, such as the skyrmion~\cite{anw83,liu87}
and the contracted current algebra~\cite{djm94}.

The common theme
of these models is chiral symmetry which involves meson cloud in the
baryon and thus, the higher Fock space beyond the valence. This
cloud degree of freedom is essential in the case of vector dominance 
of EM form factors, the
pion cloud for the Goldberger-Treiman relation, and the non-vanishing
neutron electric form factor. Therefore, it is a challenge to
understand why the valence quark model `works' without spontaneously
broken chiral symmetry, and where the hyper-fine splitting
in hadron spectroscopy and the constitute quark mass come from.

   From the above discussion, it is clear that the Fock space beyond the
valence is important and we mentioned two degrees of freedom,
namely cloud and sea. How to relate these degrees of freedom back to QCD
unambiguously, how to find out their roles in
physical quantities, and, more importantly, how to relate them to
chiral symmetry are the main subjects of this paper. It turns out that
chiral symmetry plays essential roles in light hadron spectroscopy as well
as hadron structure. We find that both 
the `constitutent' quark mass and the hyper-fine splitting in light baryons 
are more of a consequence of spontaneous chiral symmetry breaking than
that of gluons and sea quarks.

In Sec. 2, we will define these dynamical degrees of freedom in the Euclidean
path integral formalism for the hadronic tensor in deep inelastic
scattering. In Sec. 3, we will demonstrate the effects of these
degrees of freedom in hadron structure. In Sec. 4, we introduce a 
valence QCD theory which modifies QCD to suppress quark-antiquark 
pair production. We will also explore the discrete and continuous 
symmetries of valence QCD in Sec. 4.
In Sec. 5, we adopt a lattice action for valence QCD and prove its
reflection posivity and hermiticity. The pion mass, the pion decay constant, 
and the current quark mass from the axial Ward identity are
used to define the zero quark mass limit on the lattice. In Sec. 6, we 
calculate various ratios of matrix elements to check the $SU(6)$ relations. 
The nucleon form factors are calculated and presented in Sec. 7. 
We then study hadron spectroscopy in comparison with that of QCD to 
explore the origin of the hyper-fine splitting and the `costitutent' quark mass 
in Sec. 8. Perhaps the most exciting aspect of valence QCD is a new 
understanding of the origin of dynamical mass generation, something missing 
in the valence quark model and put in by hand via a constituent quark mass. 
In Sec. 9, we compare the symmetry breaking patterns in valence
QCD and QCD. Finally, in Sec. 10, we summarize the lessons 
learned from the valence QCD and draw an analogy between the valence
quark model and the nuclear shell model. We will also discuss the
implication on model building for hadrons.

\section{Quark Dynamical Degrees of Freedom}  \label{dof}

We have so far alluded to the meson clouds and sea quarks in 
additional to the valence quarks. They appear in various QCD-inspired 
hadronic models and effective theories. How does one define the valence, 
the cloud, and the sea quarks unambiguously and in a model independent 
way in QCD?  It turns out that the best way of revealing these
dynamical degrees of freedom is in deep inelastic scattering where 
the quarks show up as the parton densities.

The deep inelastic scattering of a muon on a nucleon involves the hadronic
tensor which, being an inclusive reaction, involves all intermediate states
\begin{equation}   \label{w}
W_{\mu\nu}(q^2, \nu) = \frac{1}{2M_N} \sum_n  (2\pi)^3 
\delta^4 (p_n - p - q) \langle N|J_{\mu}(0)|n\rangle
\langle n|J_{\nu}(0) | N\rangle_{spin\,\, ave.}. 
\end{equation}
Since deep inelastic
scattering measures the absorptive part of the  
Compton scattering, it is the imaginary part of the forward amplitude and
can be expressed as the current-current correlation function in the nucleon, 
i.e.
\begin{equation}  \label{wcc}
W_{\mu\nu}(q^2, \nu) = \frac{1}{\pi} Im T_{\mu\nu}(q^2, \nu)
= \frac{1}{2M_N}
\langle N| \int \frac{d^4x}{2\pi}  e^{i q \cdot x} J_{\mu}(x)
J_{\nu}(0) | N\rangle_{spin\,\, ave.}.
\end{equation} 

It has been shown~\cite{ld94} that the hadronic tensor 
$W_{\mu\nu}(q^2, \nu)$ can be obtained from the Euclidean path-integral
formalism where the various quark dynamical degrees of freedom are
readily and explicitly revealed.
In this case, one considers the ratio of the four-point function 
\mbox{$\langle O_N(t) J_{\nu}(\vec{x},t_2) J_{\mu}(0,t_1)
O_N(0)\rangle$} and the two-point function
\mbox{$\langle O_N(t-(t_2-t_1)) O_N(0)\rangle$},
where $O_N(t)$ is an interpolation
field for the nucleon with momentum $p$ at Euclidean time $t$.
For example, $O_N(t)$ can be taken to be the 3 quark fields
with nucleon quantum numbers, 
\begin{equation}
O_N = \int d^3x e^{i \vec{p}
\cdot \vec{x}} \varepsilon^{abc} \Psi^{(u)a} (x) ((\Psi^{(u)b}(x))^TC
\gamma_5\Psi ^{(d)c}(x)), 
\end{equation}
for the proton.

As both $t - t_2 >> 1/\Delta M_N$ and $t_1 >> 1/\Delta M_N$, where
$\Delta M_N$ is the mass gap between the nucleon and the next
excitation (i.e. the threshold of a nucleon and a pion in the $p$-wave),
the intermediate state contributions will be dominated by
the nucleon with the Euclidean propagator $e^{-M_N (t-(t_2 - t_1))}$.
Hence,
\begin{eqnarray}  \label{wmunu}
\widetilde{W}_{\mu\nu}(\vec{q}^{\,2},\tau) &=&
 \frac{\frac{1}{2M_N}< O(t) \int \frac{d^3x}{2\pi} e^{-i \vec{q}\cdot
 \vec{x}}
 J_{\mu}(\vec{x},t_2)J_{\nu}(0,t_1) O(0)>}{<O(t- \tau) O(0)>} \,
 \begin{array}{|l} \\  \\  \footnotesize{t -t_2 >> 1/\Delta M_N} \\
 \footnotesize{t_1 >> 1/\Delta M_N} \end{array} \nonumber \\
 &=& \frac{\frac{f^2}{2M_N} e^{-M_N(t-t_2)}<N| \int \frac{d^3x}{2\pi}
 e^{-i\vec{q}\cdot \vec{x}} J_{\mu}(\vec{x},t_2) J_{\nu}(0,t_1)|N>
e^{-M_Nt_1}}{f^2 e^{-M_N(t-\tau)}} \nonumber \\
&=&\frac{1}{2M_N V} <N|\int \frac{d^3x}{2\pi} e^{-i\vec{q}\cdot \vec{x}}
J_{\mu}(\vec{x},t_2) J_{\nu}(0,t_1)|N>,
\end{eqnarray}
where $\tau = t_2 - t_1$, f is the transition matrix element
$\langle 0|O_N|N\rangle$, and V is the 3-volume.
Inserting intermediate states, $\widetilde{W}_{\mu\nu}(\vec{q}^{\,2},\tau)$
becomes
\begin{equation}   \label{wtilde}
\widetilde{W}_{\mu\nu}(\vec{q}^{\,2},\tau)
= \frac{1}{2M_N V} \sum_n (2\pi)^2 
\delta^3 (p_n - p + q)  \langle N|J_{\mu}(0)|n\rangle
\langle n|J_{\nu}(0) | N\rangle_{spin\,\, ave.} e^{- (E_n - E_N) \tau}.
\end{equation}

We see from Eq. (\ref{wtilde}) that the time dependence is in the
exponential factor $e^{- (E_n - E_N) \tau}$. To go back to the 
delta function $\delta(E_n - E_N + \nu)$ in Eq. (\ref{w}), one needs
to carry out the inverse Laplace transform~\cite{wil93,ld94}
\begin{equation} 
W_{\mu\nu}(q^2,\nu) = \frac{V}{i} \int_{c-i \infty}^{c+i \infty} d\tau
e^{\nu\tau} \widetilde{W}_{\mu\nu}(\vec{q}^{\,2}, \tau).
\end{equation} 
This is basically doing the anti-Wick rotation back to the 
Minkowski space. 
 
In the Euclidean path-integral formulation of 
$\widetilde{W}_{\mu\nu}(\vec{q}^{\,2}, \tau)$,
contributions to the four-point function can be classified
according to different topologies of the quark paths between
the source and the sink of the proton. They represent different
ways the fields in the currents $J_{\mu}$ and $J_{\nu}$ contract with
those in the nucleon interpolation operator $O_N$.
This is so because the quark action and the electromagnetic currents
are both bilinear in quark fields, i.e. of the form 
$\overline{\Psi}M \Psi$, so that the quark number is conserved and
as a result the quark line does not branch the way a gluon line does.
Fig. 1(a) and 1(b) represent connected insertions (C.I.) of the
currents.  Here the quark fields from the interpolators $O_N$ contract
with the currents such the the quark lines flow continuously from $t =
0$ to $t =t$. Fig. 1(c), on the other hand, represents a disconnected
insertion (D.I.) where the quark fields from $J_{\mu}$ and $J_{\nu}$
self-contract and are hence disconnected from the quark paths between
the proton source and sink. Here, ``disconnected'' refers only to the 
quark lines. Of course, quarks surf in the
background of the gauge field and all quark paths are ultimately
connected through the gluon lines.

\begin{figure}[h]
\[
\hspace*{1.5in}\setlength{\unitlength}{0.01pt}
\begin{picture}(45000,20000)
\put(-10000, 8800){\circle*{700}}
\put(-10000, 8800){\line(1,1){3182}}
\put(-8410, 10391){\vector(1,1){200}}
\put(-7258,11700){{\bf $\times$}}
\put(-6818,11982){\line(1,0){3600}}
\put(-5018,11982){\vector(1,0){200}}
\put(-3658,11700){{\bf $\times$}}
\put(0000, 8800){\line(-1,1){3182}}
\put(-1591,10391){\vector(1,-1){200}}
\put(-10000, 8800){\line(1,0){10000}}
\put(-5000,  8800){\vector(1,0){300}}
\put(0000, 8800){\circle*{700}}
\qbezier(-10000, 8800)(-5000, 5000)(00, 8800)
\put(-5000, 6900){\vector(1,0){300}}
\put(-6000, 5000){\vector(1,0){2000}}
\put(-3500, 5000){$t$}
\put(-10000,10000){$0$}
\put(0,10000){$t$}
\put(-7000,11000){$t_1$}
\put(-3600,11000){$t_2$}
\put(-7200,13000){$J_\nu$}
\put(-3400,13000){$J_\mu$}
\put(-5600, 3000){(a)}
\put(6000, 8800){\circle*{700}}
\put(18000, 8800){\circle*{700}}
\qbezier(9000, 12850)(9000 ,14850)(12000,14850)
\qbezier(12000,14850)(15000,14850)(15000,12850)
\qbezier(6000 , 8800)(15000,10800)(15000,12850)
\qbezier(9000 ,12850)(9000 ,10850)(18000,08800)
\put(12000,14850){\vector(-1,0){200}}
\put(9000,  9550){\vector(3,1){200}}
\put(15000, 9550){\vector(3,-1){200}}
\put(9700, 14250){{\bf $\times$}}
\put(13450,14250){{\bf $\times$}}
\put(6000, 8800){\line(1,0){12000}}
\put(12000, 8800){\vector(1,0){200}}
\qbezier(6000,08800)(12000, 5000)(18000, 8800)
\put(12000, 6900){\vector(1,0){200}}
\put(13500, 5000){$t$}
\put(6000,10000){$0$}
\put(18000,10000){$t$}
\put(11000, 5000){\vector(1,0){2000}}
\put(9800 ,13200){$t_1$}
\put(13600,13200){$t_2$}
\put(9700 ,15300){$J_\nu$}
\put(13400,15300){$J_\mu$}
\put(10600, 3000){(b)}
\put(24000, 8000){\circle*{700}}
\put(24000, 8000){\line(1,0){12000}}
\put(30000, 8000){\vector(1,0){400}}
\put(36000, 8000){\circle*{700}}
\qbezier(24000, 8000)(30000, 5000)(36000, 8000)
\put(30000, 6500){\vector(1,0){400}}
\qbezier(24000, 8000)(30000,11000)(36000, 8000)
\put(30000, 9500){\vector(1,0){400}}
\put(30000,13000){\circle{4000}}
\put(27600,13300){{\bf $\times$}}
\put(31400,13300){{\bf $\times$}}
\put(30000,15000){\vector(1,0){200}}
\put(30000,11000){\vector(-1,0){200}}
\put(29000, 5000){\vector(1,0){2000}}
\put(31500, 5000){$t$}
\put(24000,10000){$0$}
\put(36000,10000){$t$}
\put(27000,12000){$t_1$}
\put(32000,12000){$t_2$}
\put(27000,14700){$J_\nu$}
\put(32000,14700){$J_\mu$}
\put(29500,3000){(c)}
%
\end{picture}
\]
\caption{Quark skeleton diagrams in Euclidean path integral formalism
for evaluating $W_{\mu\nu}$ from the four-point function defined
in Eq. (\ref{wmunu}). These represent the lowest twist contributions to
$W_{\mu\nu}$. (a) and (b) are the connected insertions and (c) is the
disconnected insertion.}
\end{figure}
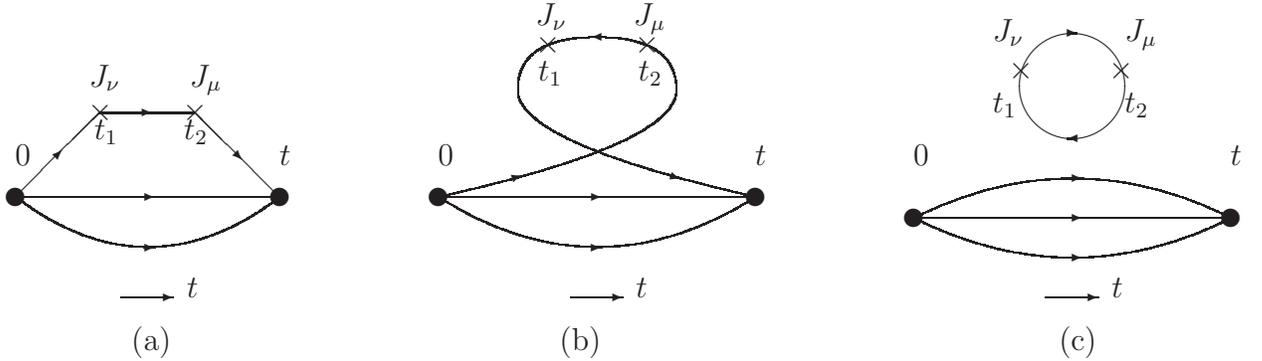

\newpage
\clearpage

 The infinitely many possible
gluon lines and additional quark loops are implicitly there in Fig. 1
but are not explicitly drawn. Fig. 1 represents the contributions of
the class of ``handbag" diagrams where the two currents are hooked
on the same quark line. These are leading twist contributions in deep
inelastic scattering.

The other contractions involving the
two currents hooking onto different quark lines are represented in
Fig. 2. Given a renormalization scale, these are higher twist
contributions in the Bjorken limit. Although details of the
operator product expansion will be given elsewhere~\cite{liu98}, it
is easy to see why these contributions in Fig. 2 are suppressed in deep 
inelastic scattering. Since the hadronic tensor $W_{\mu\nu}(q^2, \nu)$ in
Eq. (\ref{wcc}) is the current-current correlation for the forward matrix
element, the large momentum that is injected through the current at
$t_1$ must be carried out by the current at $t_2$ to avoid flowing to
the nucleon sink at $t$. In the case of Fig. 1, the momentum will simply
be conducted through the quark propagators between the currents. However, 
in Fig. 2, no connected quark propagators exist between the current. Thus,
the momentum has to be exchanged through the gluon propagator, leading
to the $O(1/Q^2)$ and $O(\alpha_s)$ suppression. This is the reason
the parton density interpretation is valid for DIS which involves 
diagonal quark propagators, since the cross terms from the quark Fock space
wave functions are suppressed. From now on, we will neglect
these ``cat's ears" diagrams in Fig. 2. 
 
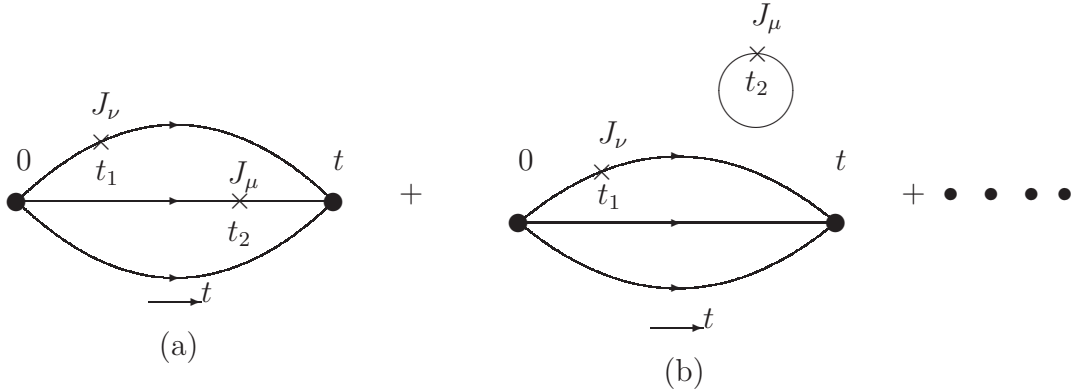
\begin{figure}[h]
\[
\hspace*{1.5in}
\setlength{\unitlength}{0.01pt}
\begin{picture}(45000,20000)
\put(-10000,8800){\circle*{700}}
\put(-7258,10750){{\bf $\times$}}
\put(-2000, 8500){{\bf $\times$}}
\put(-10000,8800){\line(1,0){12000}}
\put(-4000, 8800){\vector(1,0){200}}
\put(-4000,  5900){\vector(1,0){200}}
\put(-4000, 11700){\vector(1,0){200}}
\put(2000,8800){\circle*{700}}
\qbezier(-10000,8800)(-4000, 3000)(2000, 8800)
\qbezier(-10000,8800)(-4000,14600)(2000, 8800)
\put(-5000, 5000){\vector(1,0){2000}}
\put(-3000, 5000){$t$}
\put(-10000,10000){$0$}
\put(2000,10000){$t$}
\put(-7000, 9600){$t_1$}
\put(-2000, 7300){$t_2$}
\put(-7200,12100){$J_\nu$}
\put(-2000, 9500){$J_\mu$}
\put(-4600, 3000){(a)}
\put(4500,8800){$+$}
\put(9000,8000){\circle*{700}}
\put(9000,8000){\line(1,0){12000}}
\put(15000,8000){\vector(1,0){200}}
\put(21000,8000){\circle*{700}}
\qbezier(9000,8000)(15000, 3000)(21000, 8000)
\put(15000, 5500){\vector(1,0){200}}
\qbezier(9000,8000)(15000,13000)(21000, 8000)
\put(15000,10500){\vector(1,0){200}}
\put(18000,13000){\circle{3000}}
\put(17600,14100){{\bf $\times$}}
\put(11700, 9600){{\bf $\times$}}
\put(14000,4000){\vector(1,0){2000}}
\put(16000,4000){$t$}
\put(9000,10000){$0$}
\put(21000,10000){$t$}
\put(12000, 8700){$t_1$}
\put(17600,13000){$t_2$}
\put(12000,11000){$J_\nu$}
\put(17800,15500){$J_\mu$}
\put(14500, 2000){(b)}
\put(23500,8800){$+~\bullet~\bullet~\bullet~\bullet$}
%
\end{picture}
\]
\caption{Quark skeleton diagrams similar to those in Fig. 1, except that
the two current insertions are on different quark lines. These correspond 
to higher twist contributions to $W_{\mu\nu}$ and are suppressed by
$1/Q^2$.}  
\end{figure}

In the deep inelastic limit where $x^2 \leq O(1/Q^2)$ (we are using
the Minkowski notation here), the leading light-cone singularity
of the current product (or commutator) gives rise to a free quark
propagator between the currents. In the time-ordered diagrams in
Fig.1, Fig. 1(a)/1(b) involves only a quark/antiquark propagator
between the currents; whereas, Fig. 1(c) has both quark and antiquark
propagators. Hence, there are two distinct classes of diagrams
where the antiquarks contribute. One comes from the D.I.
; the other comes from the C.I.. It is
frequently assumed that connected insertions involve only ``valence"
quarks which are responsible for the baryon number. 
This is not true. To define the quark distribution functions more
precisely, we shall call the antiquark distribution from the
D.I. (which are connected to the ``valence''  quark propagators and
other quark loops through gluons) the ``sea'' quark. We shall refer to
the antiquark in the backward time going quark propagator between
$t_1$ and $t_2$ in Fig. 1(b) as the ``cloud'' antiquark. On the
other hand, the quark in the time forward propagator between
$t_2$ and $t_1$ in Fig. 1(a) includes both the valence and the
cloud quarks. This is because a quark propagator from $t = 0$ to
$t = t (t > 0)$ involves both the time forward and backward zigzag motions so 
that one cannot tell if the quark propagator between $t_2$ and $t_1$ 
is due to the valence or the cloud. All one knows is that it is a quark 
propagator. 
In other words, one needs to consider cloud quarks in addition to the valence
to account for the production of cloud quark-antiquark pairs in 
a connected fashion (Fig. 1(a)); whereas, the
pair production in a disconnected fashion is in Fig. 1(c).

One important point to raise at this stage is that this separation
into valence, anti-cloud and sea is gauge invariant (i.e. in the
path-integral formalism of Eq. (\ref{wtilde}), no gauge fixing is
required) and topologically distinct as far as the quark skeleton diagrams
in Fig. 1 are concerned. However, the separation depends on the
frame of the nucleon. It is expected that the parton model acquires
its natural interpretation in the large momentum frame of the
nucleon, i.e. $p \ge q$. 
Consequently, in the large momentum frame, the parton density for the 
$u$ and $d$ antiquarks come from two sources. 
\begin{equation}   \label{antiparton}
\overline{q}(x) = \overline{q}_c(x) + \overline{q}_s(x),
\end{equation}
where $\overline{q}_c(x)$ is the cloud antiparton distribution from the 
C. I. in Fig. 1(b) and $\overline{q}_s(x)$ denotes the sea antiparton
distribution from the D. I. in Fig. 1(c). The strange and charm quarks 
would only contribute in the D. I. in Fig. 1(c). Similarly, the $u$ and $d$
quarks have 2 sources, i.e.
\begin{equation}   \label{quark}
q(x) = q_{V+c}(x) + q_s(x),
\end{equation}
where $q_{V+c}(x)$ denoting the valence and cloud quarks and $q_s(x)$ denoting
the sea quark are from Fig. 1(a) and Fig. 1(c)
respectively. Upon defining $q_c(x) = \overline{q}_c(x)$ (note the subscript
``c'' denotes the cloud not charm), the 
valence parton distribution is obtained by 
\begin{equation}
q_V(x) = q_{V+c}(x) - \overline{q}_c(x),
\end{equation}
and is responsible for the baryon number,
i.e. $\int u_V(x) dx = \int [u(x) - \overline{u}(x)] dx = 2$ and
$\int d_V(x) dx = \int [d(x) - \overline{d}(x)] = 1$ for the proton.

It has been shown~\cite{ld94} that the sea partons in Fig. 1(c) cannot 
give rise to a large Gottfried sum rule violation, i.e. 
$\bar{u}_s(x) = \bar{d}_s(x)$. Whereas, the origin of $\bar{u}(x) \neq 
\bar{d}(x)$ can come primarily from the cloud antipartons in Fig. 1(b).

After the dynamical degrees of freedom are established in 
deep inelastic scattering, we need to address their relevance to
the quark model. The quark model is designed to delineate hadron properites
in the rest frame or at low energies, such as hadron masses, decay constants,
form factors, electroweak transitions, etc. Unlike the hadronic tensor
which entails the calculation of 4-point functions as illustrated in
Eq. (\ref{wtilde}), these quantities involve 2-point and 3-point
functions. The question is where do the dynamical degrees of freedom
reside in the 3-point functions which describe the matrix elements of
hadrons? To track the degrees of freedom, we can consider
the operator product expansion as an illustration.

Since the momentum transfer $|\vec{q}|$ and energy transfer $\nu$ are
large in DIS, the product of currents in the forward Compton amplitude 
$T_{\mu\nu}(q^2, \nu)$ can be expanded as a series of local operators.
The matrix elements of these local quark bilinear operators are then
related to the moments of the parton distribution. This is relatively
easy to carry out in the path integral formalism since it treats the fermions
as Grassmann numbers instead of operators. While the details are not
important here and will be given elsewhere~\cite{liu98}, all we need
to know is that the effect of expanding in terms of $1/Q^2$ pinches 
the separation of the two currents (i.e. $t_1$ and $t_2$ in Fig. 1) into one
space-time point. Thus the topologically distinct contributions
to the four-point functions extracted from Fig. 1(a), 1(b), and 1(c) are 
related to the matrix elements obtainable from the three-point
functions in Fig. 3(a), 3(b), and 3(c) respectively. The latter represents
matrix elements of the series of the local operators. Notice
that for any single matrix element related to the quark bilinear operator
$\bar{\Psi} \Gamma \Psi$, i.e. $ \langle N| \Gamma | N \rangle$, 
Fig. 3(c) which inherits the sea degree of freedom from Fig. 1(c) is
still distinct from Fig. 1(a) and 1(b) and continues to be a D. I..
On the other hand, Fig. 1(a) and 1(b) are no longer topologically distinct.
In fact, they represent the {\it same} C. I. for the local operator.
Therefore, the valence and cloud degrees of freedom from Figs. 1(a) and 1(b)
are lumped together in the C. I. of 3-point functions. They cannot be
separated in a single matrix element in contrast to the case for 
four-point functions.

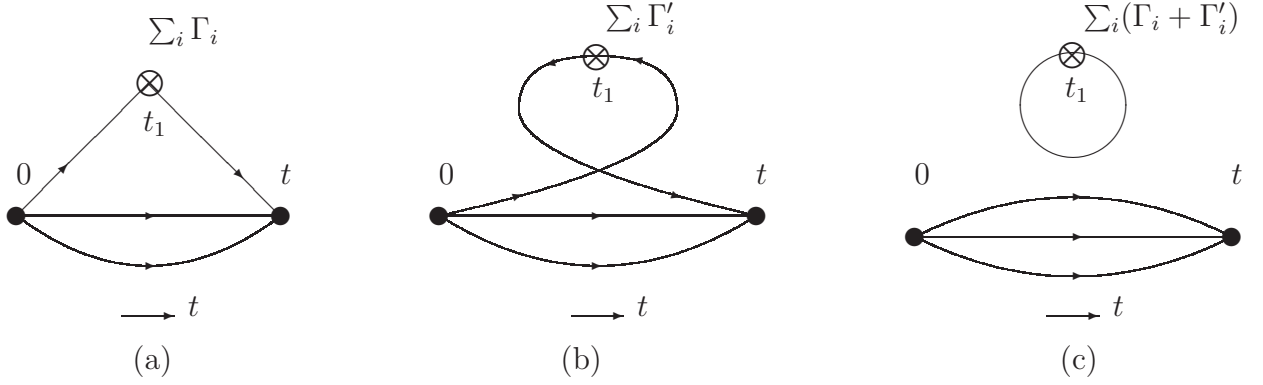
\begin{figure}[h]
\[
\hspace*{1.5in}
\setlength{\unitlength}{0.01pt}
\begin{picture}(45000,20000)
\put(-10000,8800){\circle*{700}}
\put(-10000,8800){\line(1,1){5000}}
\put(-8232, 10568){\vector(1,1){200}}
\put(0000,8800){\line(-1,1){5000}}
\put(-1591,10391){\vector(1,-1){200}}
\put(-5500, 13500){{\bf $\bigotimes$}}
\put(-10000,8800){\line(1,0){10000}}
\put(-5000, 8800){\vector(1,0){300}}
\put(0000,8800){\circle*{700}}
\qbezier(-10000,8800)(-5000, 5000)(00,8800)
\put(-5000, 6900){\vector(1,0){300}}
\put(-6000, 5000){\vector(1,0){2000}}
\put(-5000,15500){{\bf $\sum_i \Gamma_i $}}
\put(-5200,12000){$t_1$}
\put(-3500, 5000){$t$}
\put(-10000,10000){$0$}
\put(0,10000){$t$}
\put(-5600, 3000){(a)}
\put(6000,8800){\circle*{700}}
\put(18000,8800){\circle*{700}}
\qbezier(9000, 12850)(9000 ,14850)(12000,14850)
\qbezier(12000,14850)(15000,14850)(15000,12850)
\qbezier(6000 ,8800)(15000,10800)(15000,12850)
\qbezier(9000 ,12850)(9000 ,10850)(18000,8800)
\put(11400,14500){{\bf $\bigotimes$}}
\put(13500,14700){\vector(-4,1){200}}
\put(10200,14560){\vector(-4,-1){200}}
\put(9000,  9550){\vector(3,1){200}}
\put(15000, 9550){\vector(3,-1){200}}
\put(12300,16000){{\bf $\sum_i \Gamma_i^\prime$}}
\put(6000,8800){\line(1,0){12000}}
\put(12000,8800){\vector(1,0){200}}
\qbezier(6000,8800)(12000, 5000)(18000,8800)
\put(12000, 6900){\vector(1,0){200}}
\put(11800,13300){$t_1$}
\put(13500, 5000){$t$}
\put(6000,10000){$0$}
\put(18000,10000){$t$}
\put(11000, 5000){\vector(1,0){2000}}
\put(10600, 3000){(b)}
\put(24000, 8000){\circle*{700}}
\put(24000, 8000){\line(1,0){12000}}
\put(30000, 8000){\vector(1,0){400}}
\put(36000, 8000){\circle*{700}}
\qbezier(24000, 8000)(30000, 5000)(36000, 8000)
\put(30000, 6500){\vector(1,0){400}}
\qbezier(24000, 8000)(30000,11000)(36000, 8000)
\put(30000, 9500){\vector(1,0){400}}
\put(30000,13000){\circle{4000}}
\put(29400,14600){{\bf $\bigotimes$}}
\put(30400,16000){{\bf $\sum_i (\Gamma_i+\Gamma_i^\prime)$}}
\put(29000, 5000){\vector(1,0){2000}}
\put(29700,13300){$t_1$}
\put(31500, 5000){$t$}
\put(24000,10000){$0$}
\put(36000,10000){$t$}
\put(29500, 3000){(c)}
%
\end{picture}
\]
\caption{Quark skeleton diagrams in the Euclidean path integral formalism 
considered in the evaluation of matrix elements for the sum of local 
operators from the 
operator product expansion of $J_{\mu}(x) J_{\nu}(0)$. (a), (b) and (c)
corresponds to the operator product expansion from Fig. 1(a), 1(b) and
1(c) respectively.}
\end{figure}

What we have shown in this section is that for a flavor-singlet current
$\bar{\Psi} \Gamma \Psi$, the matrix element $ \langle N| \Gamma | N' \rangle$
has both the C. I. (Figs. 3(a) and 3(b) represent the same C. I. in this
case) and the D. I. (Fig. 3(c)). While
one can study the sea effect directly and separately from the D. I.,
one cannot separately study the valence and the cloud in the Z-graphs	
since both are included in the C. I. and Fig. 3(a) and 3(b) are
topologically indistinguishable. Similarly one can trace the
quark degrees of freedom in decay
constants and hadron masses which are obtainable from the two-point functions.
The flavor non-singlet hadrons are obtainable from the C. I. depicted
in Fig. 4(a).

\begin{figure}[h]
\[
\hspace*{1.5in}
\setlength{\unitlength}{0.01pt}
\begin{picture}(45000,20000)
\put(-10000, 8800){\circle*{700}}
\put( 2000,  8800){\circle*{700}}
\qbezier(-10000, 8800)(-4000, 5000)(2000, 8800)
\qbezier(-10000, 8800)(-4000,12600)(2000, 8800)
\put(-4000, 6900){\vector(1,0){300}}
\put(-4000,10700){\vector(-1,0){200}}
\put(-5000, 5000){\vector(1,0){2000}}
\put(-3000, 5000){$t$}
\put(-10000,10000){$0$}
\put(2000,10000){$t$}
\put(-4800, 3000){(a)}
\put(8000,  8800){\circle*{700}}
\put(20000, 8800){\circle*{700}}
\qbezier(8000, 8800)(13500,11800)(13500, 8800)
\qbezier(8000, 8800)(13500, 5800)(13500, 8800)
\put(15900, 7300){\vector(-1,0){200}}
\put(15900,10300){\vector(1,0){200}}
\qbezier(14500, 8800)(14500,11800)(20000, 8800)
\qbezier(14500, 8800)(14500, 5800)(20000, 8800)
\put(11900, 10300){\vector(1,0){200}}
\put(11900,  7300){\vector(-1,0){200}}
\put(13000, 5000){\vector(1,0){2000}}
\put(15500, 5000){$t$}
\put(8000, 9500){$0$}
\put(20000, 9500){$t$}
\put(13500, 3000){(b)}
%
\end{picture}
\]
\caption{(a) The connected insertion for a meson propagator. (b) The
disconnected insertion part for the flavor-singlet meson propagator which
involves the annihilation channel.} 
\end{figure}
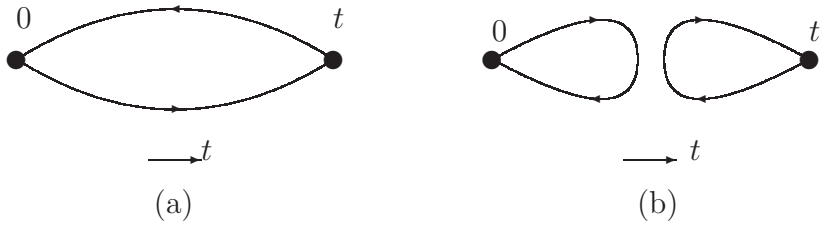

\newpage
\clearpage
The flavor-singlet meson, e.~g. $\eta'$, also involves
the D. I. in 4(b). The quark propagators in the two-point functions in
Fig. 4(a) and 4(b) include the valence and cloud (Z-graphs) only; they do
not involve the sea contribution. The sea effects come only through the
fermion determinant in this case.

\section{Sea and Cloud Quarks in Hadron Structure}

   Now that we have identified the whereabouts of the quark degrees of
freedom, we shall first study the sea effects in hadron structure in order
to examine why the valence quark model succeeds and fails in the places
we alluded to in the introduction. Since the valence and cloud are
tangled together in the C. I., we can only infer the effects of the cloud
indirectly.

\subsection{Matrix Elements and $SU(6)$ Breaking}

\subsubsection{Flavor-singlet $g_A^0$ and $F_A$ and $D_A$} \label{g_A}

Polarized DIS experiments at CERN~\cite{emc88,smc97}
and SLAC~\cite{e14395} have extracted the flavor-singlet axial coupling
constant $g_A^0$ from the $g_1$ sum rule, based on the fact that the latter 
is related to the forward matrix element of the axial current via the operator 
product expansion~\cite{kod80}. Since the
axial current is the canonical spin operator, $g_A^0$ is thus
the quark spin content of the nucleon; i.e.
$g_A^0 = \Delta u + \Delta d + \Delta s$,
where the spin content $\Delta q (q = u,d,s)$ is
defined in the forward matrix element of the axial current,
 $\langle ps|\bar{q}i\gamma_{\mu}\gamma_5 q | ps \rangle
= 2 M_N s_{\mu} \Delta q$.

 
The surprising result the experiments found is a small $g_A^0$
(0.27(10)~\cite{e14395} and 0.28(16)~\cite{smc97}), much smaller than 
the expected value of unity from the
non-relativistic quark model or 0.75 from the $SU(6)$ relation (i.e.
3/5 of the isovector coupling $g_A^3 = 1.2574$). This has attracted
a lot of theoretical attention~\cite{cheng96} and the ensuing confusion 
was dubbed the ``proton spin crisis''.

Direct lattice calculations of $g_A^0$ from the forward matrix element of
the flavor-singlet axial current has been carried out~\cite{dll95,fko95}. 
The details of these calculations have been given elsewhere~\cite{dll95,fko95}.
We shall concentrate on the results here. One word of caution though is that
the present calculations are still in the framework of the quenched 
approximation and the systematic errors due to finite volume and finite lattice size
have not been properly assessed. Nevertheless, the smallness of $g_A^0$
is understood from these calculations~\cite{dll95,fko95}.  
As explained in the preceding section, $g_A^0$ is
composed of two components, i.e. $g_A^0 = g_A^0 (C. I.) + g_A^0 (D. I.)$
where $g_A^0 (C. I.)$ is obtained from the connected insertion in Fig.
5(a) and $g_A^0 (D. I.)$ is obtained from the disconnected insertion with
the axial current in Fig. 5(b).

\newpage
\clearpage
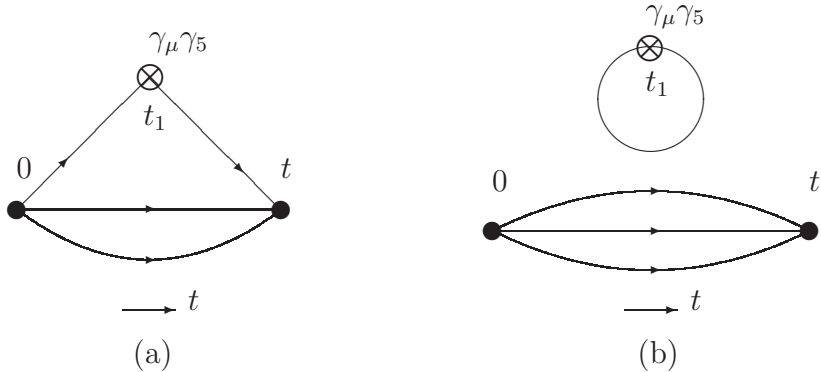
\begin{figure}[ht]
\hspace*{2.0in}
\setlength{\unitlength}{0.01pt}
\begin{picture}(45000,20000)
\put(-10000, 8800){\circle*{700}}
\put(-10000, 8800){\line(1,1){5000}}
\put(-8232, 10568){\vector(1,1){200}}
\put(0000, 8800){\line(-1,1){5000}}
\put(-1591,10391){\vector(1,-1){200}}
\put(-5500, 13500){{\bf $\bigotimes$}}
\put(-10000, 8800){\line(1,0){10000}}
\put(-5000,  8800){\vector(1,0){300}}
\put(0000, 8800){\circle*{700}}
\qbezier(-10000, 8800)(-5000, 5000)(00, 8800)
\put(-5000, 6900){\vector(1,0){300}}
\put(-6000, 5000){\vector(1,0){2000}}
\put(-5000,15000){{\bf $\gamma_\mu \gamma_5$}}
\put(-5200,12000){$t_1$}
\put(-3500, 5000){$t$}
\put(-10000,10000){$0$}
\put(0,10000){$t$}
\put(-5600, 3000){(a)}
\put(8000, 8000){\circle*{700}}
\put(8000, 8000){\line(1,0){12000}}
\put(14000, 8000){\vector(1,0){400}}
\put(20000, 8000){\circle*{700}}
\qbezier(8000, 8000)(14000, 5000)(20000, 8000)
\put(14000, 6500){\vector(1,0){400}}
\qbezier(8000, 8000)(14000,11000)(20000, 8000)
\put(14000, 9500){\vector(1,0){400}}
\put(14000,13000){\circle{4000}}
\put(13400,14600){{\bf $\bigotimes$}}
\put(13800,16000){{\bf $\gamma_\mu \gamma_5$}}
\put(13000, 5000){\vector(1,0){2000}}
\put(13800,13300){$t_1$}
\put(15500, 5000){$t$}
\put(8000, 9500){$0$}
\put(20000, 9500){$t$}
\put(13500, 3000){(b)}
%
\end{picture}
\caption{Quark line diagrams of the three-point function in the Euclidean path 
integral formalism for evaluating $g_A^0$ from the flavor-singlet
axial-vector current. (a) is the connected insertion which contains the
valence and cloud degrees of freedom and (b) is the disconnected insertion 
which contains the quark sea.} 
\end{figure}

Lattice calculation~\cite{dll95} indicates that
each of the $u, d,$ and $s$ flavors contributes $ - 0.12 \pm 0.01$ to the
D.I. (Fig. 5(b)). This vacuum polarization from the
sea quarks is largely responsible for bringing the value of $g_A^0$ from 
$ g_A^0 (C. I.) = 0.62(9)$ to $0.25 \pm 0.12$, in agreement with the
experimental value. This is an example where the sea contributes 
substantially and leads to a large breaking in the $SU(6)$ relation. Thus, it
is understandable that it should come as a surprise to the valence quark model
--- the latter does not have the sea degree of freedom and has simply
ignored it by assuming the OZI rule. It is interesting to
note that although the sea effect is large, $\Delta u_s, \Delta d_s,$
 and $\Delta s$ from the D. I. are almost identical in the lattice 
 calculation. As a result, the sea effects cancel among different flavors and
 are not reflected in $F_A = (\Delta u - \Delta s)/2$, 
 $D_A = (\Delta u - 2 \Delta d + \Delta s)/2$, and
 the $F_A/D_A$ ratio. This can be seen from Table 1 where results from 
the lattice calculations are tabulated and compared with the experiments and
the quark model predictions. Since $\Delta u_s = \Delta d_s = \Delta s$, the
$F_A$, $D_A$, and the $F_A/D_A$ ratio are identical with or without the
sea quarks of the D. I.. Partly for this reason, the relativistic quark model 
prediction based on the input of $g_A^3 = 1.257$ and the 
$SU(6)$, indicated in the last column of Table 1, are quite reasonable for 
$F_A$, $D_A$, and the $F_A/D_A$ ratio.
It was not until experiments revealed a large negative $\Delta s$
which leads to a very small $g_A^0$ that the true shortcomings of the
quark model were disclosed.

\begin{table}[ht]
\caption{Axial coupling constants and quark spin contents of proton in
comparison with experiments, the non-relativistic quark model (NRQM), and
the relativistic quark model (RQM).}
\begin{tabular}{llllll}
 \multicolumn{1}{c}{} &\multicolumn{1}{c}{C. I.} &\multicolumn{1}{c}{C. I.
 + D. I.} &  \multicolumn{1}{c} {Experiments} &  \multicolumn{1}{c} 
 {NRQM} &\multicolumn{1}{c} {RQM}\\
 \hline
 $g_A^0 {\scriptstyle =\Delta u + \Delta d + \Delta s}$ 
 &  0.62(9) & 0.25(12)& 0.28(16) \cite{smc97}, 0.27(10)\cite{e14395}
 & 1 & 0.754  \\
 $g_A^3 {\scriptstyle =\Delta u - \Delta d}$ & 1.20(10) \cite{ldd94a} 
 &1.20(10)& 1.2573(28) & 5/3 & 1.257 \\
 $g_A^8 {\scriptstyle =\Delta u + \Delta d - 2\Delta s}$ & 0.62(9) & 
 0.61(13) &  0.579(25) \cite{cr93} & 1 & 0.754 \\
 $\Delta u $ & 0.91(10) & 0.79(11) & 0.82(5)\cite{smc97}, 0.82(6)\cite{e14395}
  & 4/3 & 1.01 \\
 $\Delta d $ & -.29(10) &- 0.42(11) &  - 0.44(5)\cite{smc97}, - 0.44(6)
 \cite{e14395} & -1/3 & -0.251  \\
 $\Delta s $    &    & - 0.12(1) & - 0.10(5)\cite{smc97}, - 0.10(4)
 \cite{e14395} & 0 & 0 \\
 $F_A {\scriptstyle = (\Delta u - \Delta s)/2}$ & 0.45(6)& 0.45(6) & 0.459(8) 
 \cite{cr93} & 2/3 & 0.503 \\
 $D_A {\scriptstyle = (\Delta u - 2 \Delta d + \Delta s)/2}$ & 0.75(11)
 & 0.75(11) & 0.798(8) \cite{cr93} & 1 & 0.754\\
 $F_A/D_A$ & 0.60(2) & 0.60(2) & 0.575(16) \cite{cr93} & 2/3 & 2/3  \\
 \hline
\end{tabular}
\end{table}

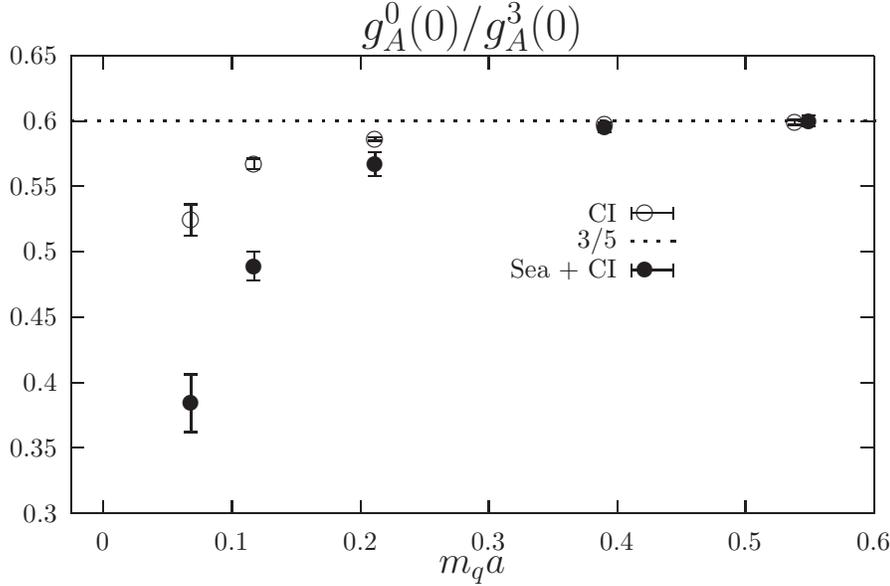
\begin{figure}[h]
\setlength{\unitlength}{0.240900pt}
\ifx\plotpoint\undefined\newsavebox{\plotpoint}\fi
\sbox{\plotpoint}{\rule[-0.200pt]{0.400pt}{0.400pt}}%
\begin{picture}(1500,900)(0,0)
\font\gnuplot=cmr10 at 10pt
\gnuplot
\sbox{\plotpoint}{\rule[-0.200pt]{0.400pt}{0.400pt}}%
\put(176.0,113.0){\rule[-0.200pt]{4.818pt}{0.400pt}}
\put(154,113){\makebox(0,0)[r]{0.3}}
\put(1416.0,113.0){\rule[-0.200pt]{4.818pt}{0.400pt}}
\put(176.0,216.0){\rule[-0.200pt]{4.818pt}{0.400pt}}
\put(154,216){\makebox(0,0)[r]{0.35}}
\put(1416.0,216.0){\rule[-0.200pt]{4.818pt}{0.400pt}}
\put(176.0,318.0){\rule[-0.200pt]{4.818pt}{0.400pt}}
\put(154,318){\makebox(0,0)[r]{0.4}}
\put(1416.0,318.0){\rule[-0.200pt]{4.818pt}{0.400pt}}
\put(176.0,421.0){\rule[-0.200pt]{4.818pt}{0.400pt}}
\put(154,421){\makebox(0,0)[r]{0.45}}
\put(1416.0,421.0){\rule[-0.200pt]{4.818pt}{0.400pt}}
\put(176.0,524.0){\rule[-0.200pt]{4.818pt}{0.400pt}}
\put(154,524){\makebox(0,0)[r]{0.5}}
\put(1416.0,524.0){\rule[-0.200pt]{4.818pt}{0.400pt}}
\put(176.0,627.0){\rule[-0.200pt]{4.818pt}{0.400pt}}
\put(154,627){\makebox(0,0)[r]{0.55}}
\put(1416.0,627.0){\rule[-0.200pt]{4.818pt}{0.400pt}}
\put(176.0,729.0){\rule[-0.200pt]{4.818pt}{0.400pt}}
\put(154,729){\makebox(0,0)[r]{0.6}}
\put(1416.0,729.0){\rule[-0.200pt]{4.818pt}{0.400pt}}
\put(176.0,832.0){\rule[-0.200pt]{4.818pt}{0.400pt}}
\put(154,832){\makebox(0,0)[r]{0.65}}
\put(1416.0,832.0){\rule[-0.200pt]{4.818pt}{0.400pt}}
\put(226.0,113.0){\rule[-0.200pt]{0.400pt}{4.818pt}}
\put(226,68){\makebox(0,0){0}}
\put(226.0,812.0){\rule[-0.200pt]{0.400pt}{4.818pt}}
\put(428.0,113.0){\rule[-0.200pt]{0.400pt}{4.818pt}}
\put(428,68){\makebox(0,0){0.1}}
\put(428.0,812.0){\rule[-0.200pt]{0.400pt}{4.818pt}}
\put(630.0,113.0){\rule[-0.200pt]{0.400pt}{4.818pt}}
\put(630,68){\makebox(0,0){0.2}}
\put(630.0,812.0){\rule[-0.200pt]{0.400pt}{4.818pt}}
\put(831.0,113.0){\rule[-0.200pt]{0.400pt}{4.818pt}}
\put(831,68){\makebox(0,0){0.3}}
\put(831.0,812.0){\rule[-0.200pt]{0.400pt}{4.818pt}}
\put(1033.0,113.0){\rule[-0.200pt]{0.400pt}{4.818pt}}
\put(1033,68){\makebox(0,0){0.4}}
\put(1033.0,812.0){\rule[-0.200pt]{0.400pt}{4.818pt}}
\put(1234.0,113.0){\rule[-0.200pt]{0.400pt}{4.818pt}}
\put(1234,68){\makebox(0,0){0.5}}
\put(1234.0,812.0){\rule[-0.200pt]{0.400pt}{4.818pt}}
\put(1436.0,113.0){\rule[-0.200pt]{0.400pt}{4.818pt}}
\put(1436,68){\makebox(0,0){0.6}}
\put(1436.0,812.0){\rule[-0.200pt]{0.400pt}{4.818pt}}
\put(176.0,113.0){\rule[-0.200pt]{303.534pt}{0.400pt}}
\put(1436.0,113.0){\rule[-0.200pt]{0.400pt}{173.207pt}}
\put(176.0,832.0){\rule[-0.200pt]{303.534pt}{0.400pt}}
\put(806,23){\makebox(0,0){{\large\bf $m_q a$}}}
\put(806,877){\makebox(0,0){{\Large\bf $g_A^0(0)/g_A^3(0)$}}}
\put(176.0,113.0){\rule[-0.200pt]{0.400pt}{173.207pt}}
\put(1033,585){\makebox(0,0)[r]{CI}}
\put(1077,585){\circle{24}}
\put(364,573){\circle{24}}
\put(463,661){\circle{24}}
\put(653,701){\circle{24}}
\put(1014,723){\circle{24}}
\put(1312,727){\circle{24}}
\put(1055.0,585.0){\rule[-0.200pt]{15.899pt}{0.400pt}}
\put(1055.0,575.0){\rule[-0.200pt]{0.400pt}{4.818pt}}
\put(1121.0,575.0){\rule[-0.200pt]{0.400pt}{4.818pt}}
\put(364.0,549.0){\rule[-0.200pt]{0.400pt}{11.804pt}}
\put(354.0,549.0){\rule[-0.200pt]{4.818pt}{0.400pt}}
\put(354.0,598.0){\rule[-0.200pt]{4.818pt}{0.400pt}}
\put(463.0,653.0){\rule[-0.200pt]{0.400pt}{4.095pt}}
\put(453.0,653.0){\rule[-0.200pt]{4.818pt}{0.400pt}}
\put(453.0,670.0){\rule[-0.200pt]{4.818pt}{0.400pt}}
\put(653.0,698.0){\rule[-0.200pt]{0.400pt}{1.204pt}}
\put(643.0,698.0){\rule[-0.200pt]{4.818pt}{0.400pt}}
\put(643.0,703.0){\rule[-0.200pt]{4.818pt}{0.400pt}}
\put(1014.0,719.0){\rule[-0.200pt]{0.400pt}{1.927pt}}
\put(1004.0,719.0){\rule[-0.200pt]{4.818pt}{0.400pt}}
\put(1004.0,727.0){\rule[-0.200pt]{4.818pt}{0.400pt}}
\put(1312.0,723.0){\rule[-0.200pt]{0.400pt}{1.927pt}}
\put(1302.0,723.0){\rule[-0.200pt]{4.818pt}{0.400pt}}
\put(1302.0,731.0){\rule[-0.200pt]{4.818pt}{0.400pt}}
\sbox{\plotpoint}{\rule[-0.500pt]{1.000pt}{1.000pt}}%
\put(1033,540){\makebox(0,0)[r]{3/5}}
\multiput(1055,540)(20.756,0.000){4}{\usebox{\plotpoint}}
\put(1121,540){\usebox{\plotpoint}}
\put(176,729){\usebox{\plotpoint}}
\multiput(176,729)(20.756,0.000){61}{\usebox{\plotpoint}}
\put(1436,729){\usebox{\plotpoint}}
\sbox{\plotpoint}{\rule[-0.200pt]{0.400pt}{0.400pt}}%
\put(1033,495){\makebox(0,0)[r]{Sea + CI}}
\put(1077,495){\circle*{24}}
\put(364,286){\circle*{24}}
\put(463,501){\circle*{24}}
\put(653,661){\circle*{24}}
\put(1014,719){\circle*{24}}
\put(1334,729){\circle*{24}}
\put(1055.0,495.0){\rule[-0.200pt]{15.899pt}{0.400pt}}
\put(1055.0,485.0){\rule[-0.200pt]{0.400pt}{4.818pt}}
\put(1121.0,485.0){\rule[-0.200pt]{0.400pt}{4.818pt}}
\put(364.0,240.0){\rule[-0.200pt]{0.400pt}{21.922pt}}
\put(354.0,240.0){\rule[-0.200pt]{4.818pt}{0.400pt}}
\put(354.0,331.0){\rule[-0.200pt]{4.818pt}{0.400pt}}
\put(463.0,479.0){\rule[-0.200pt]{0.400pt}{10.840pt}}
\put(453.0,479.0){\rule[-0.200pt]{4.818pt}{0.400pt}}
\put(453.0,524.0){\rule[-0.200pt]{4.818pt}{0.400pt}}
\put(653.0,643.0){\rule[-0.200pt]{0.400pt}{8.913pt}}
\put(643.0,643.0){\rule[-0.200pt]{4.818pt}{0.400pt}}
\put(643.0,680.0){\rule[-0.200pt]{4.818pt}{0.400pt}}
\put(1014.0,711.0){\rule[-0.200pt]{0.400pt}{3.854pt}}
\put(1004.0,711.0){\rule[-0.200pt]{4.818pt}{0.400pt}}
\put(1004.0,727.0){\rule[-0.200pt]{4.818pt}{0.400pt}}
\put(1334.0,721.0){\rule[-0.200pt]{0.400pt}{4.095pt}}
\put(1324.0,721.0){\rule[-0.200pt]{4.818pt}{0.400pt}}
\put(1324.0,738.0){\rule[-0.200pt]{4.818pt}{0.400pt}}
\end{picture}
\caption{The ratio $R_A$ in Eq. (\ref{R_A}) is plotted against the
quark mass. The $\bullet$ shows the full result with both the C. I. and
the D. I. The $\circ$ indicates the C. I. result only. The solid line
is valence quark model prediction of 3/5.}
\end{figure}

\newpage
\clearpage
The role of the sea is clear. How about the role of the cloud then?
Since its contribution to the C. I. of three-point funtions is entangled
with the valence, we can
not separate it out as is done for the sea. To see its effect indirectly,
we consider the ratio
\begin{equation}   \label{R_A}
R_A =\frac{g_A^0}{g_A^3} = \frac{\Delta u + \Delta d + \Delta s}
{\Delta u - \Delta d}
= \frac{(\Delta u + \Delta d)(C. I.) + (\Delta u + \Delta d + \Delta s)(D. I.)}
{\Delta u - \Delta d}
\end{equation}
as a function of the quark mass.
Our results  
based on quenched $16^3 \times 24$ lattices with $\beta = 6$ for
the Wilson $\kappa$ ranging between 0.154 to 0.105 which correspond
to strange and twice the charm masses are plotted in Fig. 6 as a
function of the quark mass $ma = ln (4\kappa_c/\kappa -3)$.
The dotted line is the valence quark model prediction of 3/5 for both
the non-relativistic and relativistic cases. For heavy quarks
(i.e. $\kappa \geq 0.133$ or $ma \geq 0.4$ in Fig. 6), we see that 
the ratio $R_A$ is 3/5 irrespective of whether the D. I. is included	
(shown as $\bullet$ in Fig. 6) or not (C. I. alone is indicated as
$\circ$). This is to
be expected because the cloud/sea quarks which are pair produced
via the Z-graphs/loops are suppressed for non-relativistic quarks 
by $O(p/m_q)$ or $ O(v/c)$. As for light quarks, the full result
(C. I. + D. I.) is much smaller than 3/5 largely due to the negatively
polarized sea contribution in the D. I. (Table 1 lists the results at
the chiral limit).  Even for the C. I. alone, 
$R_A$ still dips under 3/5. As we shall see later this is caused
by the cloud quarks. As we remove these cloud quarks in the Z-graphs by
an approximation, $R_A$ will become much closer to 3/5, the $SU(6)$ limit.

   As a cross check, we note that $g_A^8$ from the hyperon semi-leptonic
decay analysis~\cite{cr93} is 0.579(25) which is many sigmas smaller than
the relativistic quark model prediction of 0.754. Since the D. I. of $g_A^8$
is independent of the sea flavors $u, d,$ and $s$, its sea contribution cancels
out. Thus
\begin{equation}
g_A^8 = \Delta u + \Delta d - 2\Delta s = (\Delta u + \Delta d)(C. I.)
+ (\Delta u + \Delta d - 2\Delta s) (D. I.) \sim g_A^0 (C. I.)
\end{equation}
Indeed, the lattice result shows $g_A^0 (C. I.) = 0.62(9)$ which is
consistent with $g_A^8$. Therefore, one concludes that the smallness of
$g_A^8$ is related to the cloud quark effect to be unraveled later.

\subsubsection{$\pi N \sigma$ Term, $\bar{s}s$ in Nucleon and $F_S$ and $D_S$}

   Another place where the sea plays an even larger role is the
$\pi N \sigma$ term.  Like the pion mass in the meson sector, the 
$\pi N \sigma$ term is a measure of the explicit chiral symmetry 
breaking in the baryon sector. There is a long-standing and well-known 
puzzle surrounding the $\pi N \sigma$ term
\begin{equation}
\sigma_{\pi N} =\hat{m}\langle N|\bar{u}u + \bar{d}d|N\rangle ,
\end{equation}
where $\hat{m} = (m_u + m_d)/2$. The isospin even $\pi N$ scattering
amplitude at the Cheng-Dashen point~\cite{cd71} which equals to
$ \sigma_{\pi N}(q^2 = 2 m_{\pi}^2)$ to $O(m_{\pi}^4/m_N^4)$ is determined
to be around 60 MeV. On the other hand, if one assumes that 
$\langle N|\bar{s}s|N\rangle = 0$, a reasonable assumption from the OZI rule, 
the $\sigma_{\pi N}^{(0)}$ obtained from the octet baryon masses gives 
only 32 MeV~\cite{che76}, almost a factor two smaller than 
$ \sigma_{\pi N}(q^2 = 2 m_{\pi}^2)$ extracted from the $\pi N$ scattering.
It has been suggested that the resolution would involve a soft scalar form
factor~\cite{gls91} and a large $\bar{s}s$ content in the nucleon~\cite
{che76,gl82} with $y= 2 \langle N|\bar{s}s|N\rangle/\langle N|\bar{u}u + 
\bar{d}d |N\rangle \sim$ 0.2 -- 0.3. Both aspects of the resolution are
checked quantitatively in the lattice calculations~\cite{dll96,fko95}. 
It is found that the scalar form factor is indeed very soft in the
D. I.~\cite{dll96} which is consistent with a two $\pi$ intermediate state
in the dispersion approach of the chiral perturbation theory~\cite{gls91}.
A large $\bar{s}s$ content in the nucleon is also found~\cite{dll96,fko95}
so that the conjecture of large ratio $y$ to resolve the $\pi N \sigma$ term
discrepancy is verified.

A large $\bar{s}s$ content in the nucleon implies an even larger D. I. for the
$\langle p|\bar{u}u + \bar{d}d|p\rangle$. We see from Table 2 that the 
D. I. of $\langle p|\bar{u}u + \bar{d}d|p\rangle$ is in fact $\sim$ 2 times
larger than the C. I. counterpart. Thus, 2/3 of the $\pi N \sigma$ term comes
from the sea (Table 2). This is the largest sea contribution to any physical
quantity that we are aware of. As for the quark model,  
$\langle p|\bar{u}u + \bar{d}d|p\rangle$ is predicted to be $\leq 3$. 
The reason is the following: the scalar current expanded in the 
plane-wave basis,
\begin{equation}   \label{scurrent}
\int d^3x \overline{\Psi}\Psi(x) = \int d^3k \frac{m}{E}
\sum_s [b^{\dagger}(\vec{k},s)b(\vec{k},s) + d^{\dagger}(\vec{k},s)
d(\vec{k},s)],
\end{equation}
is the sum of the quark and antiquark numbers weighted by the factor
m/E. Since the valence quark model does not contain antiquarks and
${\rm{m}} < {\rm{E}}$, except for the zero modes, the scalar matrix element 
will be less than the valence quark number. Thus, we see that the quark model
underpredicts the $\pi N \sigma$ term by a factor of 3 or more.

Unlike the case of the axial current matrix element, different flavor
contributes differently to the D. I. of the scalar matrix element --
$s$ contributes less than $u$ and $d$. As a result, the $SU(3)$ antisymmetric
and symmetric parameters,
\begin{eqnarray}
F_S \!\!\! &=\!\!\!&
(\langle p|\bar{u}u |p\rangle -\langle N|\bar{s}s|N\rangle)/2
= (\langle p|\bar{u}u |p\rangle (C. I.) +\langle p|\bar{u}u |p\rangle (D. I.)
 -\langle N|\bar{s}s|N\rangle)/2  \nonumber \\
D_S \!\!\!&=\!\!\! & (\langle p|\bar{u}u |p\rangle (C. I.) - 
2\langle p|\bar{d}d |p\rangle
(C. I.))/2 + (\langle N|\bar{s}s|N\rangle - \langle p|\bar{d}d |p\rangle 
(D. I.))/2
\end{eqnarray}
are strongly affected by the large D. I. part. We see from Table 2
that both $D_S$ and $F_S$ compare favorably with the
phenomenological values obtained from the SU(3) breaking pattern of
the octet baryon masses with either linear
\cite{mmp87,oka96} or quadratic mass relations \cite{gas81}.
This agreement is significantly improved from the valence
quark model which predicts $F_S < 1$ and $D_S = 0$ and also those of
the {\it C. I.} alone \cite{mmp87,oka96}.
The latter yields $F_S = 0.91(13)$ and $D_S = - 0.28(10)$ which are
only half of the phenomenological values \cite{mmp87,oka96,gas81}.
This again underscores the importance of the sea-quark contributions.

 \begin{table}[ht]
\caption{Scalar contents, $\sigma_{\pi N}$, $F_S$, and $D_S$ in
comparison with phenomenology and quark model (QM). The
17.7 MeV in the last column is determined with the quark mass from the
 lattice calculation.}
\begin{tabular}{lllll}
 \multicolumn{1}{c}{} &\multicolumn{1}{c}{C. I.}&\multicolumn{1}{c}
 {C. I + D. I.}
 & \multicolumn{1}{c} {Phenomenology} & \multicolumn{1}{c} {QM}\\
 \hline
 $\langle p|\bar{u}u + \bar{d}d|p\rangle$ & 3.02(9) & 8.43(24) & & 
 $\leq$ 3 \\
 $\langle p|\bar{u}u - \bar{d}d|p\rangle$   & 0.63(9) &  & & $\leq$ 1 \\
 $\langle N|\bar{s}s|N\rangle$  & 1.53(7) &  & & 0 \\
 $F_S$ & 0.91(13) & 1.51(12) & 1.52 \cite{mmp87,oka96} --- 1.81\cite{gas81}
  & $\leq$ 1 \\
 $D_S$ & -0.28(10) & -0.88(28)& -0.52\cite{mmp87,oka96} --- -0.57\cite{gas81} 
  & 0  \\
  $\sigma_{\pi N}$ & 17.8(5) MeV & 49.7(2.6) MeV & 45 MeV \cite{gls91}
   & $\le 17.7$ MeV \\
  \hline
 \end{tabular}
 \end{table}

Next, we address the effect of the cloud quarks in the C. I.. Similar to
the ratio $R_A$ in the axial-vector case, we plot the ratio
\begin{equation}    \label{R_S}
R_S =\frac{g_S^{I =0}}{g_S^{I=1}} = \frac{\langle p|\bar{u}u + 
\bar{d}d|p\rangle}{\langle p|\bar{u}u  - \bar{d}d|p\rangle}
= \frac{(\langle p|\bar{u}u + \bar{d}d|p\rangle)(C. I.) +
(\langle p|\bar{u}u + \bar{d}d|p\rangle)(D. I.)}
{\langle p|\bar{u}u  - \bar{d}d|p\rangle}
\end{equation}
as a function of the quark mass in Fig. 7.

\begin{figure}[h]
\setlength{\unitlength}{0.240900pt}
\ifx\plotpoint\undefined\newsavebox{\plotpoint}\fi
\sbox{\plotpoint}{\rule[-0.200pt]{0.400pt}{0.400pt}}%
\begin{picture}(1500,900)(0,0)
\font\gnuplot=cmr10 at 10pt
\gnuplot
\sbox{\plotpoint}{\rule[-0.200pt]{0.400pt}{0.400pt}}%
\put(176.0,113.0){\rule[-0.200pt]{4.818pt}{0.400pt}}
\put(154,113){\makebox(0,0)[r]{2}}
\put(1416.0,113.0){\rule[-0.200pt]{4.818pt}{0.400pt}}
\put(176.0,216.0){\rule[-0.200pt]{4.818pt}{0.400pt}}
\put(154,216){\makebox(0,0)[r]{3}}
\put(1416.0,216.0){\rule[-0.200pt]{4.818pt}{0.400pt}}
\put(176.0,318.0){\rule[-0.200pt]{4.818pt}{0.400pt}}
\put(154,318){\makebox(0,0)[r]{4}}
\put(1416.0,318.0){\rule[-0.200pt]{4.818pt}{0.400pt}}
\put(176.0,421.0){\rule[-0.200pt]{4.818pt}{0.400pt}}
\put(154,421){\makebox(0,0)[r]{5}}
\put(1416.0,421.0){\rule[-0.200pt]{4.818pt}{0.400pt}}
\put(176.0,524.0){\rule[-0.200pt]{4.818pt}{0.400pt}}
\put(154,524){\makebox(0,0)[r]{6}}
\put(1416.0,524.0){\rule[-0.200pt]{4.818pt}{0.400pt}}
\put(176.0,627.0){\rule[-0.200pt]{4.818pt}{0.400pt}}
\put(154,627){\makebox(0,0)[r]{7}}
\put(1416.0,627.0){\rule[-0.200pt]{4.818pt}{0.400pt}}
\put(176.0,729.0){\rule[-0.200pt]{4.818pt}{0.400pt}}
\put(154,729){\makebox(0,0)[r]{8}}
\put(1416.0,729.0){\rule[-0.200pt]{4.818pt}{0.400pt}}
\put(176.0,832.0){\rule[-0.200pt]{4.818pt}{0.400pt}}
\put(154,832){\makebox(0,0)[r]{9}}
\put(1416.0,832.0){\rule[-0.200pt]{4.818pt}{0.400pt}}
\put(226.0,113.0){\rule[-0.200pt]{0.400pt}{4.818pt}}
\put(226,68){\makebox(0,0){0}}
\put(226.0,812.0){\rule[-0.200pt]{0.400pt}{4.818pt}}
\put(428.0,113.0){\rule[-0.200pt]{0.400pt}{4.818pt}}
\put(428,68){\makebox(0,0){0.1}}
\put(428.0,812.0){\rule[-0.200pt]{0.400pt}{4.818pt}}
\put(630.0,113.0){\rule[-0.200pt]{0.400pt}{4.818pt}}
\put(630,68){\makebox(0,0){0.2}}
\put(630.0,812.0){\rule[-0.200pt]{0.400pt}{4.818pt}}
\put(831.0,113.0){\rule[-0.200pt]{0.400pt}{4.818pt}}
\put(831,68){\makebox(0,0){0.3}}
\put(831.0,812.0){\rule[-0.200pt]{0.400pt}{4.818pt}}
\put(1033.0,113.0){\rule[-0.200pt]{0.400pt}{4.818pt}}
\put(1033,68){\makebox(0,0){0.4}}
\put(1033.0,812.0){\rule[-0.200pt]{0.400pt}{4.818pt}}
\put(1234.0,113.0){\rule[-0.200pt]{0.400pt}{4.818pt}}
\put(1234,68){\makebox(0,0){0.5}}
\put(1234.0,812.0){\rule[-0.200pt]{0.400pt}{4.818pt}}
\put(1436.0,113.0){\rule[-0.200pt]{0.400pt}{4.818pt}}
\put(1436,68){\makebox(0,0){0.6}}
\put(1436.0,812.0){\rule[-0.200pt]{0.400pt}{4.818pt}}
\put(176.0,113.0){\rule[-0.200pt]{303.534pt}{0.400pt}}
\put(1436.0,113.0){\rule[-0.200pt]{0.400pt}{173.207pt}}
\put(176.0,832.0){\rule[-0.200pt]{303.534pt}{0.400pt}}
\put(806,23){\makebox(0,0){{\large\bf $m_q a$}}}
\put(806,877){\makebox(0,0){{\Large\bf $g_s^{I=0}(0)/g_s^{I=1}(0)$}}}
\put(176.0,113.0){\rule[-0.200pt]{0.400pt}{173.207pt}}
\put(1033,627){\makebox(0,0)[r]{CI}}
\put(1077,627){\circle{24}}
\put(364,292){\circle{24}}
\put(463,257){\circle{24}}
\put(653,234){\circle{24}}
\put(1014,220){\circle{24}}
\put(1312,218){\circle{24}}
\put(1055.0,627.0){\rule[-0.200pt]{15.899pt}{0.400pt}}
\put(1055.0,617.0){\rule[-0.200pt]{0.400pt}{4.818pt}}
\put(1121.0,617.0){\rule[-0.200pt]{0.400pt}{4.818pt}}
\put(364.0,284.0){\rule[-0.200pt]{0.400pt}{4.095pt}}
\put(354.0,284.0){\rule[-0.200pt]{4.818pt}{0.400pt}}
\put(354.0,301.0){\rule[-0.200pt]{4.818pt}{0.400pt}}
\put(463.0,255.0){\rule[-0.200pt]{0.400pt}{0.964pt}}
\put(453.0,255.0){\rule[-0.200pt]{4.818pt}{0.400pt}}
\put(453.0,259.0){\rule[-0.200pt]{4.818pt}{0.400pt}}
\put(653.0,233.0){\rule[-0.200pt]{0.400pt}{0.482pt}}
\put(643.0,233.0){\rule[-0.200pt]{4.818pt}{0.400pt}}
\put(643.0,235.0){\rule[-0.200pt]{4.818pt}{0.400pt}}
\put(1014.0,218.0){\rule[-0.200pt]{0.400pt}{0.964pt}}
\put(1004.0,218.0){\rule[-0.200pt]{4.818pt}{0.400pt}}
\put(1004.0,222.0){\rule[-0.200pt]{4.818pt}{0.400pt}}
\put(1312.0,216.0){\rule[-0.200pt]{0.400pt}{0.964pt}}
\put(1302.0,216.0){\rule[-0.200pt]{4.818pt}{0.400pt}}
\put(1302.0,220.0){\rule[-0.200pt]{4.818pt}{0.400pt}}
\sbox{\plotpoint}{\rule[-0.500pt]{1.000pt}{1.000pt}}%
\put(1033,582){\makebox(0,0)[r]{3}}
\multiput(1055,582)(20.756,0.000){4}{\usebox{\plotpoint}}
\put(1121,582){\usebox{\plotpoint}}
\put(176,216){\usebox{\plotpoint}}
\multiput(176,216)(20.756,0.000){61}{\usebox{\plotpoint}}
\put(1436,216){\usebox{\plotpoint}}
\sbox{\plotpoint}{\rule[-0.200pt]{0.400pt}{0.400pt}}%
\put(1033,537){\makebox(0,0)[r]{Sea + CI}}
\put(1077,537){\circle*{24}}
\put(364,697){\circle*{24}}
\put(463,541){\circle*{24}}
\put(653,364){\circle*{24}}
\put(1014,233){\circle*{24}}
\put(1334,218){\circle*{24}}
\put(1055.0,537.0){\rule[-0.200pt]{15.899pt}{0.400pt}}
\put(1055.0,527.0){\rule[-0.200pt]{0.400pt}{4.818pt}}
\put(1121.0,527.0){\rule[-0.200pt]{0.400pt}{4.818pt}}
\put(364.0,655.0){\rule[-0.200pt]{0.400pt}{20.476pt}}
\put(354.0,655.0){\rule[-0.200pt]{4.818pt}{0.400pt}}
\put(354.0,740.0){\rule[-0.200pt]{4.818pt}{0.400pt}}
\put(463.0,511.0){\rule[-0.200pt]{0.400pt}{14.695pt}}
\put(453.0,511.0){\rule[-0.200pt]{4.818pt}{0.400pt}}
\put(453.0,572.0){\rule[-0.200pt]{4.818pt}{0.400pt}}
\put(653.0,348.0){\rule[-0.200pt]{0.400pt}{7.468pt}}
\put(643.0,348.0){\rule[-0.200pt]{4.818pt}{0.400pt}}
\put(643.0,379.0){\rule[-0.200pt]{4.818pt}{0.400pt}}
\put(1014.0,227.0){\rule[-0.200pt]{0.400pt}{2.650pt}}
\put(1004.0,227.0){\rule[-0.200pt]{4.818pt}{0.400pt}}
\put(1004.0,238.0){\rule[-0.200pt]{4.818pt}{0.400pt}}
\put(1334.0,213.0){\rule[-0.200pt]{0.400pt}{2.409pt}}
\put(1324.0,213.0){\rule[-0.200pt]{4.818pt}{0.400pt}}
\put(1324.0,223.0){\rule[-0.200pt]{4.818pt}{0.400pt}}
\end{picture}
\caption{The ratio $R_S$ in Eq. (\ref{R_S}) is plotted against the
quark mass. The $\bullet$ shows the full result with both the C. I. and
the D. I. The $\circ$ indicates the C. I. result only. The solid line
is non-relativistic quark model prediction of 3.}
\end{figure}
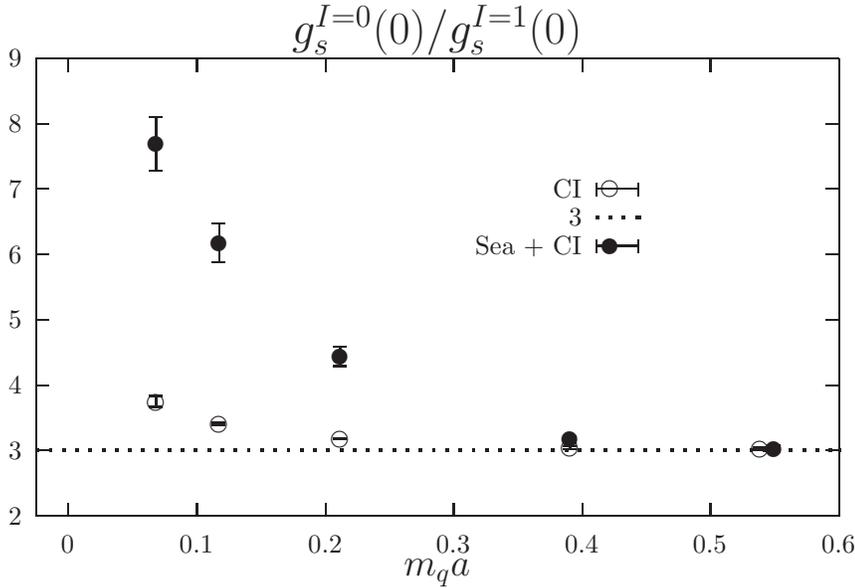

\newpage
\clearpage
The dotted line is the valence quark model prediction of 3 for both
the non-relativistic and relativistic cases. Again for heavy quarks
(i.e. $\kappa \geq 0.133$ or $ma \geq 0.4$ in Fig. 7), we see that 
the ratio $R_S$ is 3 irrespective whether the D. I. is included	
(shown as $\bullet$ in Fig. 7) or not (C. I. alone is indicated as
$\circ$). As for light quarks, the full result
(C. I. + D. I.) is much larger than 3 largely due to the large
sea contribution in the D. I. (Table 2 lists the results at
the chiral limit).  Even for the C. I. alone, 
$R_S$ still overshoots 3. As we shall see later, this is caused
by the cloud quarks. As we remove these cloud quarks in the Z-graphs in
an approximation, $R_S$ will become much closer to 3, the $SU(6)$ limit.

\subsubsection{Magnetic Moments of the Nucleon}

After having established the importance of the sea and cloud effects in
the axial and scalar matrix elements, one
should naturally ask what happens to vector current matrix elements,
especially the neutron to proton magnetic
moment ratio $\mu_n/\mu_p$. How much will the sea and cloud affect the 
ratio and in what way? After all, the $\mu_n/\mu_p$ ratio was well
predicted by the valence picture -- a celebrated
defining success of the $SU(6)$ symmetry. 
 
I has been known for some time that a nontrivial sea-quark contribution
to baryon magnetic moments is essential to reproducing the
experimental moments~\cite{lei92,lei95,lei96}.
It turns out that the individual sea contribution of each flavor is
not small~\cite{lei96,dlw98}. A direct lattice calculation of the strangeness
magnetic form factor is $G_M^s(0) = - 0.36 \pm 0.20 $~\cite{dlw98}
\footnote{Note this is defined via the vector current $\bar{s}\gamma_{\mu}s$ 
without the electric charge}. Although the central value of our lattice result
differs in sign from that of the SAMPLE experiment which has 
$G^s_M(Q^2=0.1$GeV$^2)= +0.23\pm 0.37\pm 0.15\pm 0.19 $ from the
elastic parity-violating electron scattering~\cite{sample97}, 
they are consistent within errors. The $u$ and $d$ contributions are
$\sim$ 80\% larger, $G_M^{u,d}(0) (D. I.) = -0.65 \pm 0.30$. However, their
net contribution to the proton and neutron magnetic moment
\begin{equation}  \label{mudi}
\mu (D. I.) = (2/3 G_M^u(0) (D. I.) -1/3 G_M^d(0) (D. I.) -1/3 G_M^s(0))
\mu_N  = -0.097 \pm 0.037 \mu_N 
\end{equation}
becomes smaller due the cancellation
of the quark charges of $u, d,$ and $s$. This small $SU(6)$-breaking sea quark 
effect is further nullified by the cloud effect. 

\begin{figure}[h]
\setlength{\unitlength}{0.240900pt}
\ifx\plotpoint\undefined\newsavebox{\plotpoint}\fi
\sbox{\plotpoint}{\rule[-0.200pt]{0.400pt}{0.400pt}}%
\begin{picture}(1500,900)(0,0)
\font\gnuplot=cmr10 at 10pt
\gnuplot
\sbox{\plotpoint}{\rule[-0.200pt]{0.400pt}{0.400pt}}%
\put(176.0,113.0){\rule[-0.200pt]{4.818pt}{0.400pt}}
\put(154,113){\makebox(0,0)[r]{-0.75}}
\put(1416.0,113.0){\rule[-0.200pt]{4.818pt}{0.400pt}}
\put(176.0,257.0){\rule[-0.200pt]{4.818pt}{0.400pt}}
\put(154,257){\makebox(0,0)[r]{-0.7}}
\put(1416.0,257.0){\rule[-0.200pt]{4.818pt}{0.400pt}}
\put(176.0,401.0){\rule[-0.200pt]{4.818pt}{0.400pt}}
\put(154,401){\makebox(0,0)[r]{-0.65}}
\put(1416.0,401.0){\rule[-0.200pt]{4.818pt}{0.400pt}}
\put(176.0,544.0){\rule[-0.200pt]{4.818pt}{0.400pt}}
\put(154,544){\makebox(0,0)[r]{-0.6}}
\put(1416.0,544.0){\rule[-0.200pt]{4.818pt}{0.400pt}}
\put(176.0,688.0){\rule[-0.200pt]{4.818pt}{0.400pt}}
\put(154,688){\makebox(0,0)[r]{-0.55}}
\put(1416.0,688.0){\rule[-0.200pt]{4.818pt}{0.400pt}}
\put(176.0,832.0){\rule[-0.200pt]{4.818pt}{0.400pt}}
\put(154,832){\makebox(0,0)[r]{-0.5}}
\put(1416.0,832.0){\rule[-0.200pt]{4.818pt}{0.400pt}}
\put(226.0,113.0){\rule[-0.200pt]{0.400pt}{4.818pt}}
\put(226,68){\makebox(0,0){0}}
\put(226.0,812.0){\rule[-0.200pt]{0.400pt}{4.818pt}}
\put(428.0,113.0){\rule[-0.200pt]{0.400pt}{4.818pt}}
\put(428,68){\makebox(0,0){0.1}}
\put(428.0,812.0){\rule[-0.200pt]{0.400pt}{4.818pt}}
\put(630.0,113.0){\rule[-0.200pt]{0.400pt}{4.818pt}}
\put(630,68){\makebox(0,0){0.2}}
\put(630.0,812.0){\rule[-0.200pt]{0.400pt}{4.818pt}}
\put(831.0,113.0){\rule[-0.200pt]{0.400pt}{4.818pt}}
\put(831,68){\makebox(0,0){0.3}}
\put(831.0,812.0){\rule[-0.200pt]{0.400pt}{4.818pt}}
\put(1033.0,113.0){\rule[-0.200pt]{0.400pt}{4.818pt}}
\put(1033,68){\makebox(0,0){0.4}}
\put(1033.0,812.0){\rule[-0.200pt]{0.400pt}{4.818pt}}
\put(1234.0,113.0){\rule[-0.200pt]{0.400pt}{4.818pt}}
\put(1234,68){\makebox(0,0){0.5}}
\put(1234.0,812.0){\rule[-0.200pt]{0.400pt}{4.818pt}}
\put(1436.0,113.0){\rule[-0.200pt]{0.400pt}{4.818pt}}
\put(1436,68){\makebox(0,0){0.6}}
\put(1436.0,812.0){\rule[-0.200pt]{0.400pt}{4.818pt}}
\put(176.0,113.0){\rule[-0.200pt]{303.534pt}{0.400pt}}
\put(1436.0,113.0){\rule[-0.200pt]{0.400pt}{173.207pt}}
\put(176.0,832.0){\rule[-0.200pt]{303.534pt}{0.400pt}}
\put(806,23){\makebox(0,0){{\large\bf $m_q a$}}}
\put(806,877){\makebox(0,0){{\Large\bf $\mu_n/\mu_p$}}}
\put(176.0,113.0){\rule[-0.200pt]{0.400pt}{173.207pt}}
\put(1134,746){\makebox(0,0)[r]{CI}}
\put(1178,746){\circle{24}}
\put(226,498){\circle{24}}
\put(365,493){\circle{24}}
\put(463,470){\circle{24}}
\put(661,435){\circle{24}}
\put(1021,426){\circle{24}}
\put(1318,398){\circle{24}}
\put(1156.0,746.0){\rule[-0.200pt]{15.899pt}{0.400pt}}
\put(1156.0,736.0){\rule[-0.200pt]{0.400pt}{4.818pt}}
\put(1222.0,736.0){\rule[-0.200pt]{0.400pt}{4.818pt}}
\put(226.0,435.0){\rule[-0.200pt]{0.400pt}{30.594pt}}
\put(216.0,435.0){\rule[-0.200pt]{4.818pt}{0.400pt}}
\put(216.0,562.0){\rule[-0.200pt]{4.818pt}{0.400pt}}
\put(365.0,424.0){\rule[-0.200pt]{0.400pt}{33.244pt}}
\put(355.0,424.0){\rule[-0.200pt]{4.818pt}{0.400pt}}
\put(355.0,562.0){\rule[-0.200pt]{4.818pt}{0.400pt}}
\put(463.0,406.0){\rule[-0.200pt]{0.400pt}{30.594pt}}
\put(453.0,406.0){\rule[-0.200pt]{4.818pt}{0.400pt}}
\put(453.0,533.0){\rule[-0.200pt]{4.818pt}{0.400pt}}
\put(661.0,378.0){\rule[-0.200pt]{0.400pt}{27.703pt}}
\put(651.0,378.0){\rule[-0.200pt]{4.818pt}{0.400pt}}
\put(651.0,493.0){\rule[-0.200pt]{4.818pt}{0.400pt}}
\put(1021.0,418.0){\rule[-0.200pt]{0.400pt}{4.095pt}}
\put(1011.0,418.0){\rule[-0.200pt]{4.818pt}{0.400pt}}
\put(1011.0,435.0){\rule[-0.200pt]{4.818pt}{0.400pt}}
\put(1318.0,383.0){\rule[-0.200pt]{0.400pt}{6.986pt}}
\put(1308.0,383.0){\rule[-0.200pt]{4.818pt}{0.400pt}}
\put(1308.0,412.0){\rule[-0.200pt]{4.818pt}{0.400pt}}
\sbox{\plotpoint}{\rule[-0.500pt]{1.000pt}{1.000pt}}%
\put(1134,701){\makebox(0,0)[r]{-2/3}}
\multiput(1156,701)(20.756,0.000){4}{\usebox{\plotpoint}}
\put(1222,701){\usebox{\plotpoint}}
\put(176,353){\usebox{\plotpoint}}
\multiput(176,353)(20.756,0.000){61}{\usebox{\plotpoint}}
\put(1436,353){\usebox{\plotpoint}}
\sbox{\plotpoint}{\rule[-0.200pt]{0.400pt}{0.400pt}}%
\put(1134,656){\makebox(0,0)[r]{Expt.~~-0.685}}
\put(1156.0,656.0){\rule[-0.200pt]{15.899pt}{0.400pt}}
\put(176,300){\usebox{\plotpoint}}
\put(176.0,300.0){\rule[-0.200pt]{303.534pt}{0.400pt}}
\put(1134,611){\makebox(0,0)[r]{Sea + CI}}
\put(1178,611){\circle*{24}}
\put(226,291){\circle*{24}}
\put(365,418){\circle*{24}}
\put(463,429){\circle*{24}}
\put(661,429){\circle*{24}}
\put(1156.0,611.0){\rule[-0.200pt]{15.899pt}{0.400pt}}
\put(1156.0,601.0){\rule[-0.200pt]{0.400pt}{4.818pt}}
\put(1222.0,601.0){\rule[-0.200pt]{0.400pt}{4.818pt}}
\put(226.0,185.0){\rule[-0.200pt]{0.400pt}{51.312pt}}
\put(216.0,185.0){\rule[-0.200pt]{4.818pt}{0.400pt}}
\put(216.0,398.0){\rule[-0.200pt]{4.818pt}{0.400pt}}
\put(365.0,360.0){\rule[-0.200pt]{0.400pt}{27.703pt}}
\put(355.0,360.0){\rule[-0.200pt]{4.818pt}{0.400pt}}
\put(355.0,475.0){\rule[-0.200pt]{4.818pt}{0.400pt}}
\put(463.0,386.0){\rule[-0.200pt]{0.400pt}{20.958pt}}
\put(453.0,386.0){\rule[-0.200pt]{4.818pt}{0.400pt}}
\put(453.0,473.0){\rule[-0.200pt]{4.818pt}{0.400pt}}
\put(661.0,401.0){\rule[-0.200pt]{0.400pt}{13.731pt}}
\put(651.0,401.0){\rule[-0.200pt]{4.818pt}{0.400pt}}
\put(651.0,458.0){\rule[-0.200pt]{4.818pt}{0.400pt}}
\end{picture}
\caption{The ratio of the neutron to proton magnetic moment ratio 
$\mu_n/\mu_p$ is plotted against the 
quark mass. The $\circ$ indicates the C. I. result only and the $\bullet$ 
shows the full result with both the C. I. and
the D. I. The solid line is valence quark model prediction of - 2/3 and the
dashed line is the experimental result of - 0.685.} 
\end{figure}
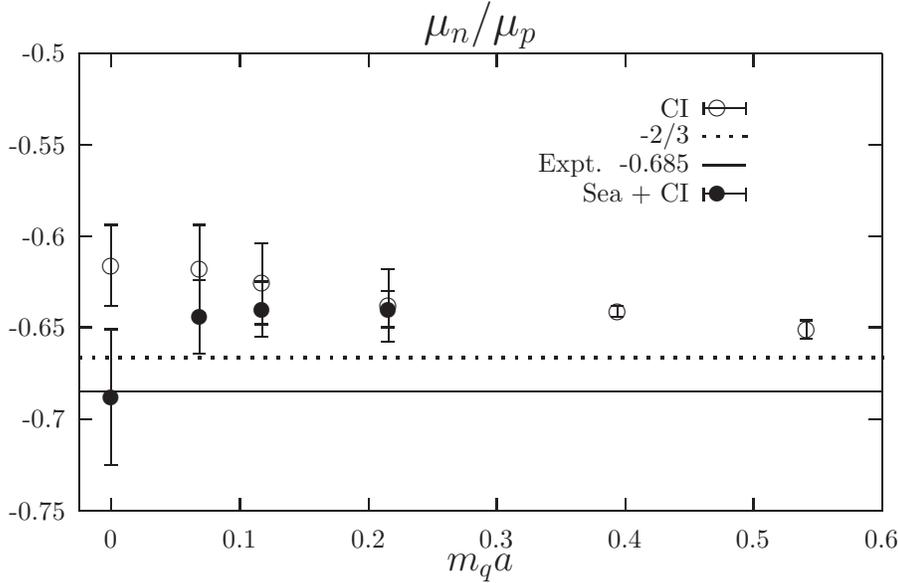

This is illustrated in
Fig. 8, where the neutron to proton magnetic moment ratio is plotted against
the quark mass. We see in Fig. 8 that for heavy quarks in the charm region
($m_q a$ at 0.55 corresponds to $m_q \sim 1$ GeV), the ratio from the
C. I. is close to the $SU(6)$ prediction of -2/3. This is quite reasonable, as 
this is in the non-relativistic regime where one expects $SU(6)$ to work well
as in the case of $g_A$ and $g_S$ in the preceeding sections.
When the quark mass comes down to the strange
region (i.e. $m_qa =0.07$), the ratio becomes less negative. Extrapolated to 
the chiral limit, the ratio is $ - 0.616 \pm 0.022$ which deviates from the 
$SU(6)$ prediction by $\sim$ 8 \%. We understand this deviation as mainly due 
to the cloud quark effect in the Z-graphs; this will be verified later. 
Now as the sea quark contribution $\mu (D. I.)$ in Eq. (\ref{mudi})
is added to the C. I., we find that it tends to counter the cloud effect and
bring the ratio back closer to the $SU(6)$ prediction.
At the chiral limit, the $\mu_n/\mu_p$ ratio then comes down to 
$-0.68 \pm 0.04$ which is quite consistent with the experimental value 
of 0.685. We should point out that the $\mu_n/\mu_p$ ratio for the
full result ($\bullet$) is more negative at the chiral limit compared
with those at other $m_qa$ has to do with the fact that different 
extrapolations are used for the C. I. and the D. I.. The C. I. employs
the linear quark mass extrapolation, as do other observables for the
C. I.~\cite{dll95,dll96,ldd94a}; whereas, the D. I. uses the $\sqrt{m_q}$ 
dependence for the chiral extrapolation~\cite{dlw98} as in the
scalar case to reflect the non-analytic behavior in 
chiral loops~\cite{gss88,dll96}. From this analysis,
we see that although the individual $G_M^u(0) (D. I.), G_M^d(0) (D. I.)$,
and $G_M^s(0)$ are large, their net contribution
$\mu (D. I.)$ in Eq. (\ref{mudi}) is much smaller
because of the partial cancellation due to the quark charges of u, d, and s.
The net sea contribution is further cancelled by the cloud effect
to bring the $\mu_n/\mu_p$ ratio close to the experimental value.
Barring any known symmetry principle yet to
surface, this cancellation is probably accidental and in stark contrast
with the $\pi N \sigma$ term and flavor-singlet $g_A^0$ where the
cloud and sea effects add up to enhance the $SU(6)$ breaking~\cite{dll95,dll96}
as detailed in the preceeding sections.

So it seems that 
the success of the $SU(6)$ prediction of the neutron to proton magnetic moment 
ratio is fortuitous due to these two cancellations and
as a result it has not revealed, and may even has helped conceal, the
importance of the cloud and sea quark effects over the years.

\subsection{Form Factors and Meson Dominance}   

In all the above ratios we considered, i.e. $R_A, R_S$, and $\mu_n/\mu_p$,
the $SU(6)$ breaking due to the cloud in the Z-graphs is at the level
of 10 -- 20\% which is relatively small compared with, say, the sea quark
effect in $R_S$. However, its effect is large in the nucleon
form factors and has been a subject of wide interest.

It is well known from the $e^+ e^- \longrightarrow \pi\pi$ experiment that
the process is dominated by the $\rho$ meson pole. Extending this
$\rho$ meson dominance of the photon coupling to the space-like part of
the pion electric form factor $F_{\pi}(q^2 < 0)$, it 
accounts for 95\% of the pion charge radius and is the basis of the
`vector dominance'~\cite{sak66}. This is supported by the lattice
studies of the pion electric form factor~\cite{ww85,dww89,aw97}. It is found
that the calculated form factor for different quark masses can be well fitted 
with a monopole form $1/(1 + q^2/m_{\rho}^2)$, with $m_{\rho}$ being 
the lattice $\rho$ mass for the corresponding quark mass. 
Similarly, the dipole form of the nucleon electromagnetic and
axial form factors is interpreted as the product of two monopoles.
For example, the isovector part of the nucleon Dirac form factor can
be written as~\cite{bmw86} 
\begin{figure}[h]
\hspace*{2.0in}
\setlength{\unitlength}{0.01pt}
\begin{picture}(45000,20000)
\qbezier(-4800, 8800)(-4200, 9400)(-3600, 8800)
\qbezier(-3600, 8800)(-3000, 8200)(-2400, 8800)
\qbezier(-2400, 8800)(-1800, 9400)(-1200, 8800)
\qbezier(-1200, 8800)(-600,  8200)(0000, 8800)
\put(0000, 8800){\circle*{1200}}
\qbezier(6500, 3800)(3400, 8800)(6500,13800)
\put(5000, 8800){\circle*{1200}}
\put(-5000, 7800){$\gamma$}
\put(0000,10500){$\frac{1}{f_\rho}$}
\put(2300, 7000){$\rho$}
\put(6000, 8200){$g_{\rho NN}(q^2)$}
\put(5530, 5800){\vector(-1,3){200}}
\put(5500,11600){\vector(1,3){200}}
\put(7000,15800){$N$}
\put(7000, 3300){$N$}
\linethickness{0.3cm}
\put(0000, 8800){\line(1,0){5000}}
%
\end{picture}
\caption{The schematic diagram which depicts the photon coupling to
the nucleon goes through the $\rho$ meson in a vector dominance picture.}
\end{figure}
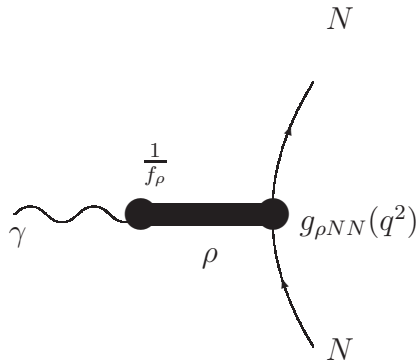
\newpage
\clearpage
\begin{equation}  \label{f1ff}
F_1^V(q^2) = \frac{1}{2}(F_1^p(q^2) - F_1^n(q^2)) = \frac{1}{2}\frac{1}
{1 - q^2/m_{\rho}^2} \frac{g_{\rho NN}(q^2)}{f_{\rho}}
\end{equation}
to reflect that the dominating process is the photon coupling to the 
 $\rho$ meson which in turn couples to the nucleon as shown in Fig. 9.
  
One monopole in $F_1^V(q^2)$ is the $\rho$ meson propagator, and the other 
one is $g_{\rho NN}(q^2) = f_{\rho}\frac{\Lambda^2 - m_{\rho}^2}
{\Lambda^2 -q^2}$ to parametrize the $\rho NN$ vertex (see Fig. 9).
By the same token, the isovector axial form factor with axial meson
dominance takes the form
\begin{equation}  \label{gaff}
g_A^3(q^2) = \frac{g_A^3(0)}{1 - q^2/m_{a_1}^2} g_{a_1 NN}(q^2)
\end{equation}
where $g_{a_1 NN}(q^2)$ is the $a_1 NN$ vertex and can be parametrized with
a monopole form. The isovector pseudoscalar form factor should reflect
the pion pole for small $q^2 $ and has the form
\begin{equation}   \label{gpff}
g_P^3(q^2) = \frac{g_P^3(0)}{1 - q^2/m_{\pi}^2} \frac{g_{\pi NN}(q^2)}
{g_{\pi NN}(0)}
\end{equation}
where $g_{\pi NN}(q^2)$ is the $\pi NN$ form factor.
\begin{figure}[h]
\input{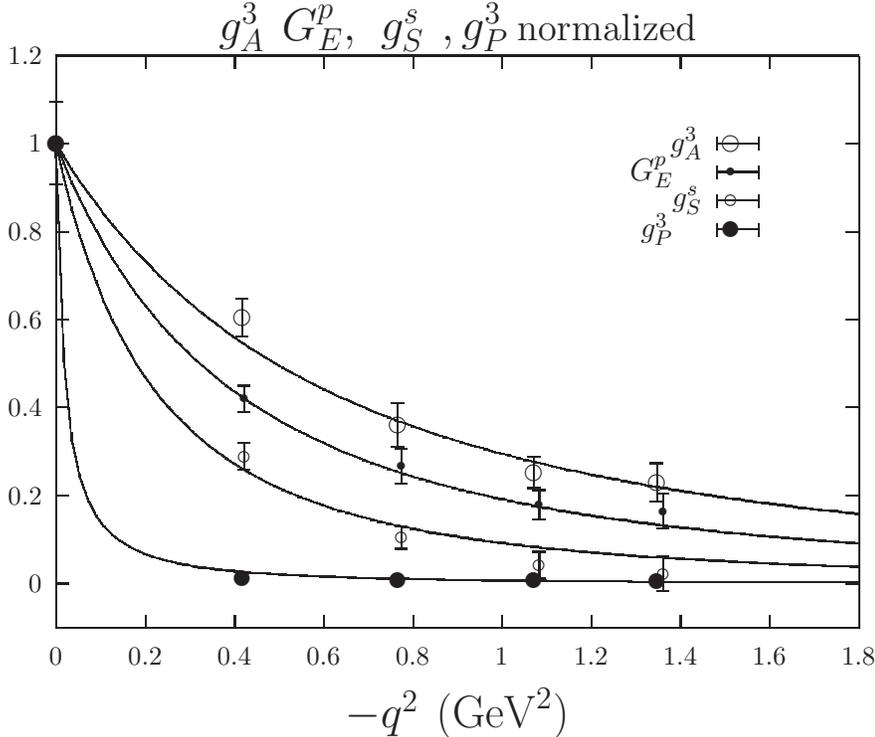}
\caption{The isovector axial form factor $g_A^3(q^2)$, the proton electric form
factor $G_E^p(q^2)$, the strangeness scalar form factor $g_S^s(q^2)$,
and the isovector pseudoscalar form factor $g_P^3(q^2)$ are plotted as
a function of - $q^2$.}
\end{figure}

\newpage
\clearpage
 Thus one of the major differences of the various form factors of the nucleon 
is reflected in the mass of the meson which dominates the  
matrix element in the t-channel for the specific current. We plot in the 
following the isovector axial form factor $g_A^3(q^2)$, the proton electric form 
factor $G_E^p(q^2)$, the strangeness scalar form factor $g_S^s(q^2)$
~\cite{dll96}, and the isovector pseudoscalar form factor $g_P^3(q^2)$
~\cite{dll95} in Fig. 10. 

We see that, since $g_A^3(q^2)$ and $g_P^3(q^2)$ involve only
the C. I. and $G_E^p(q^2)$ is dominated by the C. I.~\cite{dlw98}, their
different behaviors in $q^2$ reflect the $\rho, a_1$ and $\pi$ propagators 
in the cloud which serve as the intermediate states in the meson dominance 
picture as depicted in Fig. 11. 

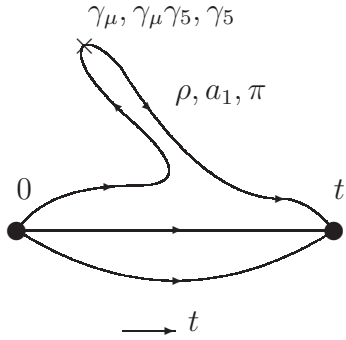
\begin{figure}
\hspace*{2.0in}
\setlength{\unitlength}{0.01pt}
\begin{picture}(45000,20000)
\put(-10000, 8800){\circle*{700}}
\put(-10000, 8800){\line(1,0){12000}}
\put(-4000,  8800){\vector(1,0){300}}
\put(2000, 8800){\circle*{700}}
\qbezier(-10000, 8800)(-4000, 5000)(2000, 8800)
\qbezier(-10000, 8800)(-8900,10500)(-5500,10500)
\qbezier(-5500,10500)(-3000,10500)(-5500,12800)
\qbezier(-5500,12800)(-8000,15000)(-7500,15800)
\qbezier(-7500,15800)(-7000,16000)(-6100,15000)
\qbezier(-6100,15000)(-3000,10000)(0000,10000)
\qbezier(0000,10000)(1000,10000)(2000, 8800)
\put(-4000, 6900){\vector(1,0){300}}
\put(-6000, 5000){\vector(1,0){2000}}
\put(-6500,10450){\vector(1,0){200}}
\put(-6200,13500){\vector(-1,1){200}}
\put(-5070,13500){\vector(1,-1){200}}
\put(0000,10000){\vector(1,0){200}}
\put(-7900,15500){$\times$}
\put(-7300,16800){{\bf $\gamma_\mu,\gamma_\mu\gamma_5,\gamma_5$}}
\put(-4000,13800){$\rho,a_1,\pi$}
\put(-3500, 5000){$t$}
\put(-10000,10000){$0$}
\put(2000,10000){$t$}
%
\end{picture}
\caption{The quark line diagram for the C. I. which illustrates the
meson dominance picture with different intermediate meson state corresponding
to the respective probing current.}
\end{figure}

\newpage
\clearpage
Given that experimentally 
$m_{a_1} = 1230 {\rm MeV} > m_{\rho} = 769{\rm MeV} > m_{\pi} = 140{\rm MeV}$
and if one 
assumes that $g_{\rho NN}(q^2), g_{a_1 NN}(q^2)$ and $g_{\pi NN}(q^2)$ have
a similar form in $q^2$, the fact that $g_P^3(q^2)$ falls off faster than
$G_E^p(q^2)$ which in turn falls off faster than $g_A^3(q^2)$ is then
to be expected. We should mention in passing that both $G_E^p(q^2)$ and
$g_A^3(q^2)$ shown in Fig. 10 from the lattice calculation~\cite{ldd94a,dlw98} 
agree with the experiments within $\sim$ 6\%. This is a clear manifestation
of the cloud quark effect through the meson cloud. We can attempt to
define the meson-nucleon vertex by dividing the form factors in Eqs.
(\ref{f1ff}), (\ref{gaff}), and (\ref{gpff}) by their respective meson
propagators. These are plotted in Fig. 12. We see that the resulting
$g_{\rho NN}(q^2), g_{a_1 NN}(q^2)$ and $g_{\pi NN}(q^2)$ extracted this
way are much closer to each other than those in Fig. 10. We should mention
that the monopole fit of $g_{\pi NN}(q^2)$ gives $g_{\pi NN}(0)
= 12.2 \pm 2.3$ which checks out the Goldberger-Treiman relation~\cite{ldd95}.

\begin{figure}[h]
\input{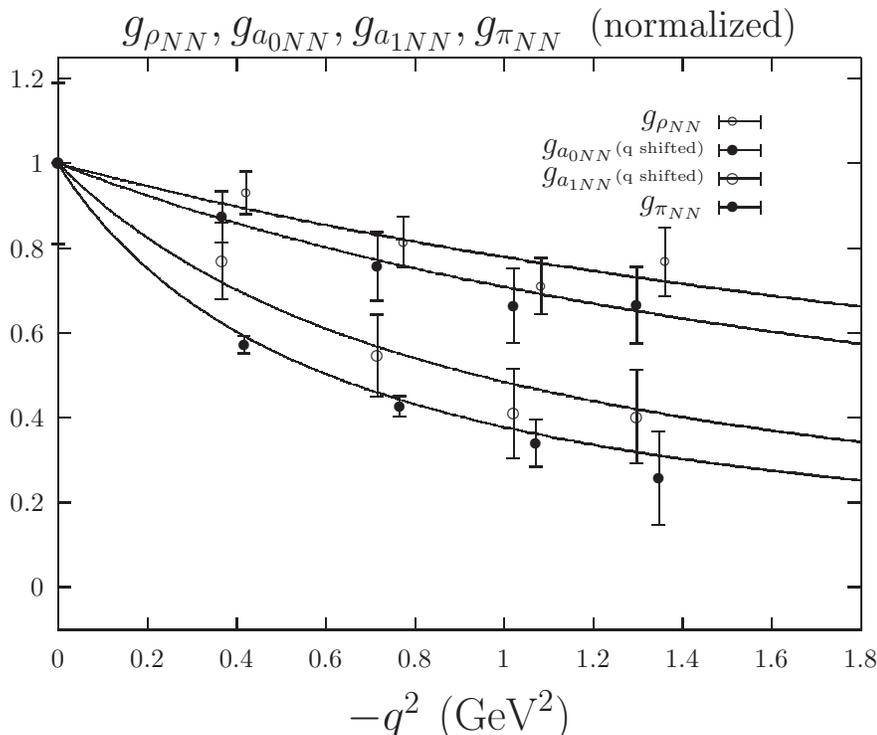}
\caption{The meson-nucleon-nucleon vertices $g_{\rho NN}(q^2), g_{a_1 NN}(q^2)$,
$g_{\pi NN}(q^2)$, and $g_{a_0 NN}(q^2)$ deduced from the EM form factors,	
$g_A^3(q^2), g_P^3(q^2)$, and the isovector scalar form factor $g_S^3(q^2)$
are plotted as a function of  - $q^2$.}
\end{figure}

Also plotted in Fig. 10 is the strangeness scalar form factor $g_S^s(q^2)$ 
which is from the D. I. (Fig. 3(c)). It is very soft and has been interpreted
as due to the $K \bar{K}$ intermediate state as depicted in Fig. 13(a)~\cite
{dll96}. The D. I. with u or d quarks are even softer~\cite{dll96} and are 
consistent with the dispersion analysis of $\pi\pi$ intermediate state in
chiral perturbation theory ($\chi$ PT)~\cite{gls91}. This appears to be the 
source of the pion and kaon loops in $\chi$ PT~\cite{gl82} which are 
responsible for the non-analytic contribution of $m_q^{3/2}$ or $m_{\pi}^3$ 
and $m_K^3$ in hadron masses (see Fig. 13(b)).

\newpage
\clearpage
\begin{figure}[h]
\hspace*{2.0in}
\setlength{\unitlength}{0.01pt}
\begin{picture}(45000,20000)
\put(-10000, 8800){\circle*{700}}
\put(-10000, 8800){\line(1,0){12000}}
\put(-4000,  8800){\vector(1,0){300}}
\put(2000, 8800){\circle*{700}}
\qbezier(-10000, 8800)(-4000, 5000)(2000, 8800)
\qbezier(-10000, 9000)(-8000, 9000)(-5500,15000)
\qbezier(-5500,15000)( 500,26500)(-1900,15000)
\qbezier(-1900,15000)(-2800, 8800)( 2000, 9000)
\qbezier(-4000,15000)(-6500, 6000)(-2500,15000)
\qbezier(-4000,15000)(-1000,23000)(-2500,15000)
\put(-4000,68900){\vector(1,0){300}}
\put(-6000, 5000){\vector(1,0){2000}}
\put(-6100,13500){\vector(1,3){200}}
\put(-2000,13500){\vector(0,-1){200}}
\put(-3950,15000){\vector(-1,-3){200}}
\put(-2900,14000){\vector(1,3){200}}
\put(-2550,18700){$\times$}
\put(-2500,21500){{\bf $\bar{S}S$}}
\put(-7000,15000){{\bf $K$}}
\put(-500,15000){{\bf $\bar{K}$}}
\put(-5300,13000){$\bar{s}$}
\put(-4000,13000){$s$}
\put(-3500, 5000){$t$}
\put(-10000,10000){$0$}
\put(2000,10000){$t$}
\put(-4500, 3000){$(a)$}
%
\put(10000, 8800){\line(1,0){15000}}
\put(17500, 8800){\vector(1,0){200}}
\put(11000, 8800){\vector(1,0){200}}
\put(24000, 8800){\vector(1,0){200}}
\qbezier[100](13000, 8800)(17500,18000)(22000, 8800)
\put(11000, 7000){$N$}
\put(16000, 7000){$\Lambda,~\Sigma$}
\put(23500, 7000){$N$}
\put(17000,14000){$K$}
\put(16500, 3000){$(b)$}
\end{picture}
\caption{(a) The quark line diagram which illustrates the $K \bar{K}$ 
intermediate state which dominates the form factor $g_S^s(q^2)$.
(b) The kaon loop diagram in chiral perturbation theory.}
\input{fig14.tex}
\caption{The neutron electric form factor $G_E^n(q^2)$ together with the
fit to the experimental result (solid line). The $\circ$ indicates the C. I.
contribution and the $\bullet$ shows the full result with both the C. I. and
the D. I.}
\end{figure}

\newpage
\clearpage
This nonlinear dependence on
the quark masses
or $m_{\pi}^2$ has been observed prominently in hadron masses with
dynamical fermions in lattice simulations~\cite{ceh96}. This illustrates the
sea quark effect in hadron masses and form factors.
The neutron charge form factor in the strict $SU(6)$ quark model would be
identically zero, since the positively charged $u$ quark and the negatively
charged $d$ quarks have the same spatial wave function.
Thus, the small positive $G_E^n(q^2)$ signals the effects of the cloud and
the sea without the contamination of the valence part like in other quantities. 
We present the lattice calculation of $G_E^n(q^2)$~\cite{dlw98}
in Fig. 14 together with the experimental result. It is seen that both
the cloud from the C. I. and the sea from the D. I. are positive and
their contributions are similar in size. 


\subsubsection{Density and Size of Nucleon}
Now, we can look at the time averaged radial density distribution of the 
nucleon due to different current probes. Define the time averaged 
density distribution as
\begin{equation}
\rho(r) = N \frac{1}{(2\pi)^2} \int dt d^4 q e^{i(\vec{q}\cdot\vec{r} -
q_0 t)} F(q^2) = N \frac{1}{(2\pi)^{3/2}} \int d^4 q \delta(q_0) 
e^{i\vec{q}\cdot\vec{r}} F(q^2)
\end{equation}
where N is the normalization factor so that $\int d^3 r \rho(r) = 1$.
We plot the pseudoscalar density $\rho_P (r)$, the scalar strangeness
density $\rho_{S}^s(r)$, the electric charge density $\rho_c(r)$, and
the axial current density $\rho_{A}(r)$ so obtained from $g_P^3(q^2),	
g_S^s(q^2), G_E^p(q^2)$, and $g_A^3(q^2)$ in Fig. 15. 

\begin{figure}[h]
\includegraphics{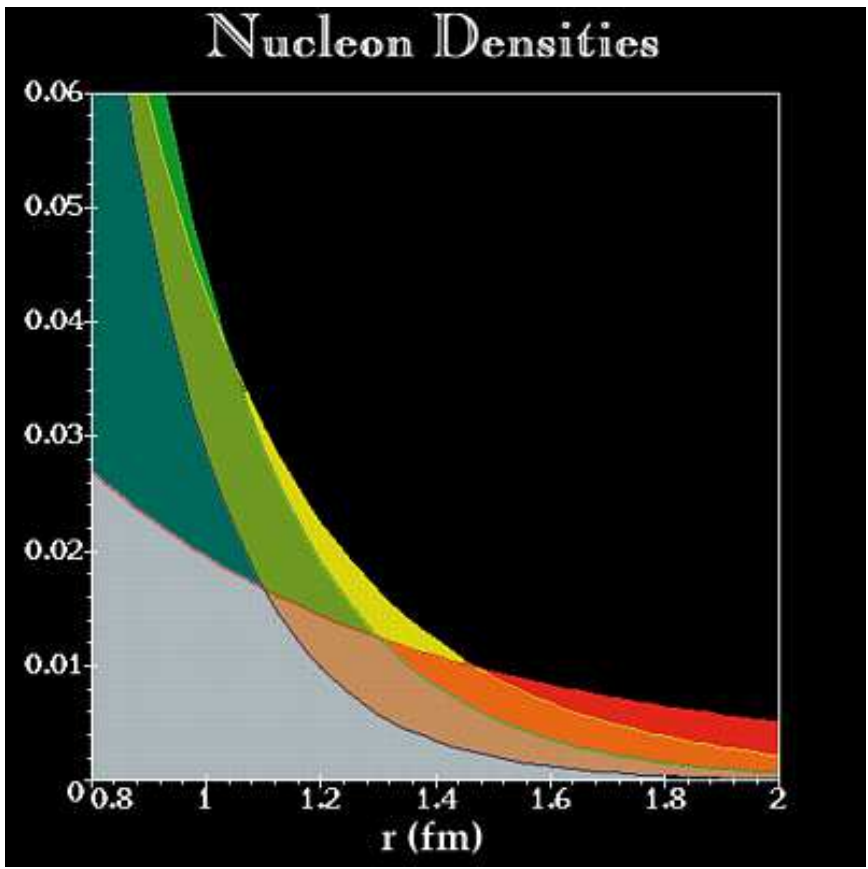}
\caption{The normalized pseudoscalar density $\rho_P (r)$ (in red), the scalar 
strangeness density $\rho_{S}^s(r)$ (in yellow), the electric charge density 
$\rho_c(r)$ (in green), 
and the axial current density $\rho_{A}(r)$ (in blue) are plotted as a 
function of the radial distance from the center.} 
\end{figure}

We see that 
$\rho_P (r)$ has the longest range. This is presumable due to the
the pion cloud which dominates the pseudoscalar channel and  has the
longest Compton wavelength of all hadrons. The next longest is the scalar 
strangeness density $\rho_{S^s}(r)$ which seems to reflect the 
$K\bar{K}$ meson intermediate state in Fig. 13(a) which would correspond to
the kaon loop in chiral perturbation theory as shown in Fig. 13(b). 
Then comes the electric charge density in the proton $\rho_c(r)$ which
is well known and has been frequently used to extract the size of the
nucleon. Finally, the one with the smallest size is the axial
current density $\rho_{A}(r)$ which reflects the small Compton wavelength
of the $a_1$ meson.

Now what is the size of the nucleon? As seen from Fig. 15, it is in
the eyes of the beholder. In other words, it depends on what probe is
used to measure it. It ranges from 3.56(3) fm for the pseudoscalar density,
1.06(9) fm for the strangeness density, 0.797(29) fm for the proton charge 
density, to 0.627(29) fm for the axial current density, indeed a large 
variation.

We see that even though the clouds in the C. I. do not break the $SU(6)$ 
symmetry as much as the seas in the D. I. for the scalar and axial currents,
they afford a large variation in hadron form factors and sizes. Short of
these meson clouds, the valence quark model simply is not capable of
delineating the richness of the various form factors.  
A model like the skyrmion, on the other hand, is capable of detailing 
the G-T relation~\cite{anw83},	
the meson dominance of the nucleon form factors~\cite{brw86,mkw87}, 
negative square charge
radius of the neutron~\cite{anw83}, etc. All these are achieved via the 
ingredient of the meson clouds.

\section{Valence QCD}

   After having examined the roles of the dynamical quark degree of freedom,
we come back to the original question of what approximation to QCD
the valence quark model represents. As illustrated in Fig. 5, the
sea is only involved in the D. I. part of the three-point function
and thus can be isolated as far as its contribution is concerned. On the
other hand, we have stressed in Sec. \ref{dof} that the cloud and valence 
contributions are lumped in the C. I. in Fig. 5 and cannot be 
separated a posteriori. Thus to single out the valence effects and
compare to the valence quark model requires an approximation to QCD. This
can be achieved by forcefully eliminating pair creation and annihilation
by decoupling the quark from the antiquark. In other words, we want to
eliminate all the Z-graphs such as a typical one illustrated in Fig. 16. 

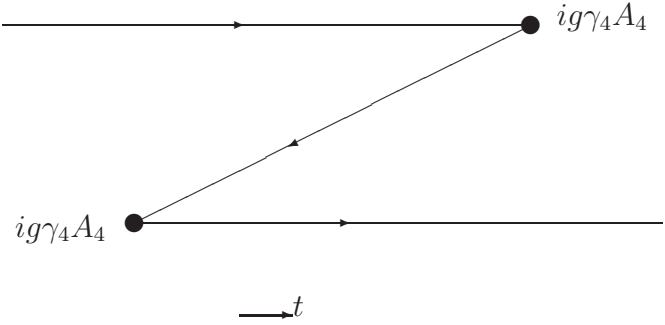
\begin{figure}[h]
\hspace*{1.5in}\setlength{\unitlength}{0.01pt}
\begin{picture}(45000,14000)
\put(-10000,12500){\line(1,0){20000}}
\put(10000,12500){\circle*{700}}
\put(10000,12500){\line(-2,-1){15000}}
\put(-5000, 5000){\circle*{700}}
\put(-5000,5000){\line(1,0){20000}}
\put(-1000,12500){\vector(1,0){200}}
\put( 3000, 5000){\vector(1,0){200}}
\put( 1000, 8000){\vector(-2,-1){200}}
\put(11000,12500){$i g \gamma_4 A_4$}
\put(-9500,4500){$i g \gamma_4 A_4$}
\put(-1000, 1500){\vector(1,0){2000}}
\put( 1000, 1500){$t$}
%
\end{picture}
\caption{A typical Z-graph as a diagram in the time-ordered perturbation.}
\end{figure}
We introduce
valence QCD (VQCD), a theory which is designed to achieve this goal.  

First of all, we shall introduce the particle field $u$ and the antiparticle 
field $v$ in lieu of the Dirac field $\Psi$ in the valence QCD lagrangian 
\begin{equation}   \label{vqcd}
{\cal L}_{VQCD} = - \frac{1}{4} F_{\mu\nu}F_{\mu\nu} -
\bar{u}[\frac{\gamma_4 +1}{2}D_4 + \vec{\gamma}\cdot\vec{D} + m]u 
- \bar{v}[\frac{\gamma_4 -1}{2}D_4 + \vec{\gamma}\cdot\vec{D} + m]v.
\end{equation}
Comparing with the QCD lagrangian, the valence version has changed the
$\gamma_4$ into $\frac{\gamma_4 +1}{2}$ for the particle field u and
$\frac{\gamma_4 -1}{2}$ for the antiparticle field v. We note that the
u and v fields do not couple. Now we want to prove that the propagator
of the u field only propagates forward in time and does not zig-zag in
the time direction to generate particle-antiparticle pairs. 
The propagator $S_u(x,y;A)$ satisfies the equation
\begin{equation} 
 - (\frac{\gamma_4 +1}{2}D_4 + \vec{\gamma}\cdot\vec{D} + m)S_u(x,y;A)
 = \delta(x-y).
\end{equation}
This can be cast in the integral representation with
the static propagator $S_u^0(x,y;A_4)$  as the bare part of the
solution~\cite{ef81}. The static propagator $S_u^0(x,y;A_4)$ satisfies the 
following equation where there is no propagation in the spatial direction
\begin{equation}
- (\frac{\gamma_4 +1}{2}D_4 + m)S_u^4(x,y;A_4)
 = \delta(x-y).
\end{equation}

\newpage
\clearpage
It is easy to write down the formal solution for $S_u^0(x,y;A_4)$
\begin{eqnarray}  \label{S0}
S_u^0(x,y;A_4) &=& - \theta(x_4 - y_4) e^{- m (x_4 - y_4)}\frac{1 + \gamma_4}{2}
P\left[ \begin{array}{c}x \\ y \end{array} \right]\delta(\vec{x} - \vec{y}) 
\nonumber \\
& &-\frac{\delta(x_4 - y_4)}{m} e^{- m (y_4 - x_4)}\frac{1 - \gamma_4}{2}
P\left[ \begin{array}{c}x \\ y \end{array} \right]\delta(\vec{x} - \vec{y})
\end{eqnarray}
where $P\left[ \begin{array}{c}x \\ y  \end{array} \right]\equiv
P e^{ig \int_{y_4}^{x_4} dz_4 A_4}$ is the path-ordered parallel transport
factor in the time direction. We see that the usual antiparticle propagation in QCD which involves the
$\theta(y_4 - x_4)$ is now replaced with $\delta(x_4 - y_4)$ in the 
second term. Thus, $S_u^0(x,y;A_4)$ is the static particle propagator which
moves forward in time only. Now the full propagator $S_u(x,y;A)$
can be represented in an integral equation in terms of $S_u^0(x,y;A_4)$
\begin{equation}
S_u(x,y;A) =  S_u^0(x,y;A_4) + \int d^4z S_u^0(x,z;A_4) \vec{\gamma}\cdot
\vec{D} S_u(z,y;A).
\end{equation}
The kernel $\vec{\gamma}\cdot\vec{D}$ is responsible for hopping in the
spatial direction. The full solution can be obtained by substituting $S_u$ 
with $S_u^0$ iteratively leading to a hopping expansion series
\begin{eqnarray}  \label{hop_ser}
&&S_u(x,y;A) =  S_u^0(x,y;A_4) + \int_{y_4}^{x_4} dz_4 
\int d^3z 
S_u^0(x,z;A_4) \vec{\gamma}\cdot\vec{D} S_u^0(z,y;A_4)~~~~~~~~~~~~~~~~~~~
\nonumber \\
&+&\int_{z_4}^{x_4}dz'_4 
\int_{y_4}^{z'_4} dz_4 \int d^3z' d^3z S_u^0(x,z';A_4) 
\vec{\gamma}\cdot\vec{D} S_u^0(z',z;A_4) \vec{\gamma}\cdot\vec{D} 
S_u^0(z,y;A_4)+\cdots 
\end{eqnarray}
It is clear from this expansion that the time integration
variables $z'_4, z_4, ...$ are sequenced between $x_4$ and $y_4$ due to
$\theta$ and $\delta$ functions in Eq. (\ref{S0}). A typical term in the
series is shown graphically in Fig. 17.

\begin{figure}[h]
\hspace*{1.5in}
\setlength{\unitlength}{0.01pt}
\begin{picture}(45000,14000)
\qbezier(-10000, 6000)(0000,11000)(10000, 6000)
\qbezier( 10000, 6000)(20000,1000)(30000, 6000)
\put(-10000, 6000){\circle*{700}}
\put(-10000, 4000){y}
\put(30000, 6000){\circle*{700}}
\put(30000, 4000){x}
\put(00000, 8500){\vector(1,0){200}}
\put(20000, 3500){\vector(1,0){200}}
\put(10000, 6000){\vector(2,-1){200}}
\put(27000, 4730){\vector(3,1){200}}
\put(-6000, 7400){$\times$}
\put(-6000, 8500){$Z_4$}
\put(-7000, 5900){$-\vec{\gamma}\cdot\vec{D}$}
\put( 4000, 7700){$\times$}
\put( 4000, 8800){$Z_3$}
\put( 2000, 6200){$-\vec{\gamma}\cdot\vec{D}$}
\put(13000, 4300){$\times$}
\put(13000, 5400){$Z_2$}
\put(11000, 2500){$-\vec{\gamma}\cdot\vec{D}$}
\put(23000, 3500){$\times$}
\put(23000, 5000){$Z_1$}
\put(21000, 2000){$-\vec{\gamma}\cdot\vec{D}$}
%
\end{picture}
\caption{A term in the hopping expansion series in Eq. (\ref{hop_ser}) is
graphically presented.}
\end{figure}
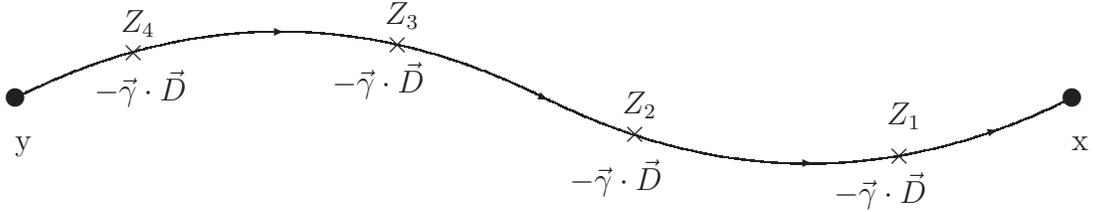

From this we see that there is no time backward propagation in $S_u(x,y; A)$.
Therefore, there is no pair creation or annihilation in the particle
propagator, albeit it still propagates forward and backward in the spatial
direction. Similarly, one can show that the antiparticle propagator
$S_v(x,y; A)$ moves only in the time backward direction.

Although there is no pair creation or annihilation in VQCD, there are still
quark loops in the spatial direction which could lead to non-trivial
dynamical effect via the fermion determinant. Since we want to emulate the
valence quark model, the sea quark degree of freedom needs to be removed
also. Therefore, we will not include the fermion determinant in the
calculation. In other words, VQCD is inherently ``quenched''.


\subsection{Pauli Spinor Representation}

   In the process of replacing the fermion field $\Psi$ in QCD by
two Dirac spinors $u$ and $v$ in valence QCD (VQCD), we seem to have doubled 
the degrees of freedom. It turns out that half of the degrees of freedom in
$u$ and $v$ are not dynamical and thus can be integrated over. As a result,
VQCD can be represented by Pauli spinor fields. To prove this, we first look at
the particle part of the fermion action from the VQCD lagrangian in 
Eq. (\ref{vqcd}) and write it in terms of the upper and lower components
\begin{eqnarray}      
S_u \!\!\! &=\!\!\!& - \int d^4x (\bar{u}_1 \bar{u}_2) \left( \begin{array}{cc}
D_4 + m  & i\vec{\sigma}\cdot\vec{D} \\
- i\vec{\sigma}\cdot\vec{D} & m \end{array} \right) 
\left( \begin{array}{c} u_1 \\ u_2 \end{array} \right), \nonumber  \\
 \!\!\!&=\!\!\!& - \int d^4x [\bar{u}_1 (D_4 + m - \frac{(\vec{\sigma}
 \cdot\vec{D})^2}{m}) u_1 \nonumber \\
 \!\!\!& \!\!\!& + (\bar{u}_2 + \bar{u}_1\frac{i(\vec{\sigma}\cdot\vec{D})}{m})
   m (u_2 + \frac{ -i(\vec{\sigma}\cdot\vec{D})}{m} u_1)].
\end{eqnarray}
After changing the field variables $u_1 \rightarrow \chi_1$, and
$u_2 + \frac{ -i(\vec{\sigma}\cdot\vec{D})}{m} u_1 \rightarrow \xi_1$, the
action becomes
\begin{equation}  \label{xi}
S_u = \int d^4x [\bar{\chi}_1 (D_4 + m - \frac{(\vec{\sigma}\cdot\vec{D})^2}
   {m}) \chi_1 + \bar{\xi}_1 m \xi_1].
\end{equation}
Since the $\bar{\xi}_1 m \xi_1$ part has no dynamics and is quadratic, it
can be integrated over, leaving the particle action represented by the 	
Pauli spinor $\chi_1$
\begin{equation}  \label{S_u_Pauli}
S_u = \int d^4x \bar{\chi}_1 (D_4 + m - \frac{\vec{D}^2 + \vec{\sigma}\cdot
   \vec{B}}{m}) \chi_1.
\end{equation}

Similarly, the antiparticle action $S_v$ can also be written in terms of
the Pauli spinor
\begin{equation}  \label{S_v_Pauli}
S_v =  \int d^4x \bar{\chi}_2 ( -D_4 + m - \frac{\vec{D}^2 + \vec{\sigma}\cdot
   \vec{B}}{m}) \chi_2.
\end{equation}
We can redefine the Dirac spinor as $\chi = \left( \begin{array}{c}
\chi_1 \\ \chi_2 \end{array} \right)$ and rewrite the VQCD lagrangian as
\begin{eqnarray}  \label{pauli}
{\cal L}_{VQCD}\!\!\! &\!\!\!= - \frac{1}{4} F_{\mu\nu}F_{\mu\nu}
- \chi (\gamma_4 D_4 + m - \frac{\vec{D}^2 + \vec{\sigma}\cdot
   \vec{B}}{m}) \chi  \nonumber \\
   \!\!\!&\!\!\!= - \frac{1}{4} F_{\mu\nu}F_{\mu\nu}
- (\chi_1 \chi_2) \left( \begin{array}{cc} D_4 + m - \frac{\vec{D}^2 + 
\vec{\sigma}\cdot\vec{B}}{m} & 0 \\ 0 & -D_4 + m - \frac{\vec{D}^2 + 
\vec{\sigma}\cdot\vec{B}}{m} \end{array} \right) \left( \begin{array}{c}
\chi_1 \\ \chi_2 \end{array}\right).
\end{eqnarray}
It is clear from Eq. (\ref{pauli}) that particle field $\chi_1$ and
antiparticle field $\chi_2$ decouple. This also proves that 
${\cal L}_{VQCD}$ in Eq. (\ref{vqcd}) does not double the fermion degrees of
freedom. After integrating out the non-dynamical d.o.f.  it has exactly 
4 propagting spinors as shown in Eq. (\ref{pauli}). It is worthwhile
remarking that the Pauli form of ${\cal L}_{VQCD}$ in Eq. (\ref{pauli})
resembles that of the non-relativistic QCD lagrangian after Foldy-Wouthuysen
transformation. It has the single
time derivative like in the Schr\"{o}edinger action and it contains
the $\frac{\vec{D}^2 + \vec{\sigma}\cdot\vec{B}}{m}$ term, much like the
non-relativistic expansion. However, we should stress that the Pauli form
of VQCD is {\it not} a non-relativistic or other expansion. Its form is
exact. Furthermore, it does not have the spin-orbit,tensor, and Darwin terms as
in non-relativistic QCD.

\subsection{Discrete symmetry}  \label{dis_sym}

   Let's explore the symmetries of VQCD and see if there is any change
from QCD. First we examine the discrete symmetries: the parity, charge, 
and time reversal.

   The gluon part of the VQCD is the same as in QCD and there is no
need to modify the transformation of the gluon field. For parity and time
reversal, the u and v fields transform the same way as $\Psi$ in QCD. Thus
in VQCD
\begin{eqnarray}
P\left( \begin{array}{c} u(x) \\ v(x) \end{array}\right)P^{-1}
= \gamma_4 \left( \begin{array}{c} u(x^P) \\ v(x^P) \end{array}\right)  \\     
T\left( \begin{array}{c} u(x) \\ v(x) \end{array}\right)T^{-1}
=\sigma_2 \left( \begin{array}{c} u(x^T) \\ v(x^T) \end{array}\right)
\end{eqnarray}
where $x^P = (-\vec{x}, x_4), x^T = (\vec{x}, - x_4)$. It is easy to
show that the VQCD action $S_{VQCD} = \int d^4 x {\cal L}_{VQCD}$
is invariant under the above parity and time reversal transformations.

As for the charge transformation, we need to take into account the fact
that u and v are particle and antiparticle fields which should be
transformed into each other under charge transformation. We find that
$S_{VQCD}$ is invariant under the following charge transformation:
\begin{equation}
C\left( \begin{array}{c} u_{\alpha} \\ v_{\alpha} \end{array}\right)C^{-1}      
= \left( \begin{array}{c} (\gamma_2)_{\alpha\beta}v_{\beta}^{\dagger} \\ 
(\gamma_2)_{\alpha\beta}u_{\beta}^{\dagger} \end{array}\right)  
\end{equation}

Thus, with the appropriate definition, VQCD satisfies the C, P, and T 
invariance.
    
\subsection{Continuous Symmetry --- $U(2N_F)$} \label{con_sym}

   Next, we shall address the continuous symmetries. Since the
Dirac structure of the time derivative is modified in VQCD, it is no longer 
Lorentz invariant, although it is still translational invariant. This should be
acceptable for the purpose of our study, i.e. low energy hadron physics near 
the rest frame. After all, the quark model is supposed to be an effective
theory of low energy and small momentum transfer, unlike the parton model	
which addresses different kinetic regimes. 

   Like in QCD, VQCD has the global vector and axial symmetries. It is 
invariant under the U(1) transformation 
\begin{equation}
u \longrightarrow e^{i \alpha} u,  v \longrightarrow e^{i \alpha'} v     
\end{equation}
This leads to conserved vector currents
\begin{equation} \label{cvc}
\partial_{\mu} J_{\mu}^u = 0, \,\,\,\, \partial_{\mu} J_{\mu}^v = 0,
\end{equation}
where the Noether currents associated with these gauge transformations are 
\begin{equation}  \label{vc}
J_{\mu}^u = \bar{u} \left( \begin{array}{c} i\gamma_i \\ i\frac{\gamma_4 +1}{2} 
\end{array} \right) u,  \,\,\,\,
J_{\mu}^v = \bar{v} \left( \begin{array}{c} i\gamma_i \\ i\frac{\gamma_4 -1}{2} 
\end{array} \right) v.
\end{equation}
 
Therefore, the particle and antiparticle are separately conserved. This
is in contrast to the conserved current $J_{\mu} = \bar{\Psi} i\gamma_{\mu}
\Psi$ in QCD where only the difference of the particle and the
antiparticle numbers or the valence number is conserved, i.e.
\begin{equation}
N_v = \int d^3x \bar{\Psi}\gamma_4\Psi = \int \frac{d^3p}{(2\pi)^3}\sum_s
[b_s^{\dagger}(\vec{p}) b_s(\vec{p}) - d_s^{\dagger}(\vec{p}) d_s(\vec{p})]
\end{equation}
where $b^{\dagger}/d^{\dagger}$ and b/d are the creation and annihilation
operators of the particle/antiparticle in QCD. 

 The axial symmetry of VQCD is realized in the $\gamma_5$ transformation
\begin{equation}
u \longrightarrow e^{i \theta \gamma_5} v,
v \longrightarrow e^{i \theta' \gamma_5} u.
\end{equation}
The lagrangian ${\cal L}_{VQCD}$ with $m$ = 0 is invariant under this 
transformation which transposes the u and v terms in the lagrangian. 
As a result, one has the conserved axial currents
\begin{eqnarray}
A_{\mu} & = &i \bar{u} \left( \begin{array}{c} \frac{\gamma_4 +1}{2} \\
\gamma_i \end{array} \right) \gamma_5 v,  \\
A_{\mu}^{\dagger} & = &i \bar{v} \left( \begin{array}{c} \frac{\gamma_4 -1}{2} 
\\ \gamma_i \end{array} \right) \gamma_5 u.
\end{eqnarray}

We should point out that there is no Adler-Bell-Jackiw anomaly~\cite{abj69}
in VQCD. This is so because in VQCD there is no quark loop involving the
time direction, hence there is no triangle diagram to generate the
axial anomaly. With m $\neq$ 0, the axial Ward identities are
\begin{eqnarray} 
\partial_{\mu} A_{\mu} &= &2 m \bar{u} i\gamma_5 v, \label{axi1} \\
\partial_{\mu} A_{\mu}^{\dagger} &=& 2 m \bar{v} i\gamma_5 u. \label{axi2}
\end{eqnarray}

  It is useful to consider the particle field $u$ and antiparticle field $v$ 
like two flavors and define $\zeta = \left( \begin{array}{c} u \\ v \end{array}
\right)$, then the VQCD lagrangian in Eq. (\ref{vqcd}) can be written as
\begin{equation}   \label{vqcd2}
{\cal L}_{VQCD} = - \frac{1}{4} F_{\mu\nu}F_{\mu\nu} -
\bar{\zeta}[\frac{\gamma_4 + \tau_3}{2}D_4 + \vec{\gamma}\cdot\vec{D} + m]\zeta. 
\end{equation}
 At the massless limit, VQCD is invariant under the transformation
\begin{equation}
\zeta \longrightarrow e^{i \alpha I} \zeta, \zeta \longrightarrow 
e^{i \alpha' \tau_3} \zeta, \zeta \longrightarrow e^{i \theta \gamma_5\tau_1} 
\zeta, \zeta \longrightarrow e^{i \theta \gamma_5\tau_2} \zeta
\end{equation}
where $\tau's$ are the Pauli spinor in the two-component $u, v$ space.
The 4 operators $I, \tau_3, \gamma_5 \tau_1$, and $\gamma_5 \tau_2$
are the generators of the $U(2)$ algebra. So, massless VQCD has the $U(2)$ 
vector and axial symmetries in the particle-antiparticle space. 
For degenerate massless $N_F$ flavors, it has the $U(2N_F)$ symmetry.    
It is in contrast to the $SU(N_F)_L \times SU(N_F)_R \times U_V(1)$ 
chiral and $U_V(1)$ symmetry 
of QCD. In VQCD with $N_F$ flavors, the charges $Q^a_{\pm} = \int d^3x 
[\bar{u}\gamma_4 (t^a/2)u \pm \bar{u}\gamma_4\gamma_5 (t^a/2) v]$ do not form 
a complete $SU(N_F) \times SU(N_F)$ algebra because the vector and axial 
current contain 
different fields. This can also be seen from the states. For massless
particle, $u$ satisfies the Dirac equation
\begin{equation}  \label{dirac_u}
[\frac{\gamma_4 +1}{2}D_4 + \vec{\gamma}\cdot\vec{D}] u = 0.	
\end{equation}
$\gamma_5 v$ satisfies the same equation
\begin{equation}
[\frac{\gamma_4 +1}{2}D_4 + \vec{\gamma}\cdot\vec{D}] (\gamma_5 v) = 0.	
\end{equation}
Therefore, $\chi_{\pm} = \frac{1}{2}(u \pm \gamma_5 v)$ is a solution of the
Dirac equation in  Eq. (\ref{dirac_u}), but it has different particle
content, i.e. it is a mixture of particle and antiparticle. As
a result, $\chi_{\pm}$ does not have a definite handedness, it contains
both helicity states. From this we conclude that massless VQCD does not
have $SU(N_F)_L \times SU(N_F)_R$ chiral symmetry as in QCD. Instead, it
has the vector-axial $U(2N_F)$ in the flavor and particle-antiparticle space.
 
\subsection{Zero Quark Mass Limit}
   
   Even though we have explored the axial symmetry of VQCD in Sec. 
(\ref{con_sym}) at the massless limit, there is a concern that the
the zero quark mass limit may be singular. This can be seen from the
Dirac equation in Eq. (\ref{dirac_u}) for a free quark
\begin{equation}
\left( \begin{array}{cc} \partial_4 & \vec{\sigma}\cdot\vec{\partial} \\
\vec{\sigma}\cdot\vec{\partial} &  0 \end{array}\right) 
\left( \begin{array}{c} u_1 \\ u_2 \end{array}\right).	
\end{equation}
This leads to two Laplace equations for the upper and lower
components of the particle field u
\begin{equation}
\bigtriangledown^2 u_1 = 0, \bigtriangledown^2 u_2 = 0.
\end{equation}
There are no time derivatives in these constraint equations and thus
no dynamics~\cite{gol98}. Similarly, ones sees that the $\xi$ field in
Eq. (\ref{xi}) is ill-defined for the m = 0 case.

To address this problem, we consider the following approach. Let's consider 
the fermion part of VQCD lagrangian with a	
small mixture of antiparticle part in the particle action and vice versa 
 
\begin{equation}
 {\cal L}_F = - \bar{u}[ \frac{\gamma_4 +1}{2}D_4 + \epsilon 
 \frac{\gamma_4 -1}{2}D_4 + m + \vec{\gamma}\cdot\vec{D}] u 
- \bar{v}[ \frac{\gamma_4 -1}{2}D_4 + \epsilon 
 \frac{\gamma_4 +1}{2}D_4 + m + \vec{\gamma}\cdot\vec{D}] v
\end{equation}
We will then let both m and $\epsilon$ go to zero.

Let's first consider the free quark case. In this
case, the fermion lagrangian is
\begin{equation}
{\cal L^0}_F = - \bar{u} \left( \begin{array}{cc} \partial_4 + m &
\vec{\sigma}\cdot\vec{\partial} \\ \vec{\sigma}\cdot\vec{\partial}  &
-\epsilon\partial_4 + m \end{array} \right) u
- \bar{v} \left( \begin{array}{cc} \epsilon \partial_4 + m &
\vec{\sigma}\cdot\vec{\partial} \\ \vec{\sigma}\cdot\vec{\partial}  &
-\partial_4 + m \end{array} \right) v.
\end{equation}
This involves two time derivatives. The eigenvalues for $u$ are determined from 
\begin{equation}
det \left( \begin{array}{cc} - E + m & i \vec{\sigma}\cdot\vec{p} \\
i \vec{\sigma}\cdot\vec{p} &  \epsilon E + m \end{array}\right) = 0,
\end{equation}
which are, for small $\epsilon$,
\begin{eqnarray}
E = m + \frac{\vec{p}^2}{m},    \label{fq_dis}  \\
E = - \frac{m (1 - \epsilon)}{\epsilon} - (m + \frac{\vec{p}^2}{m}). 
\label{decou}
\end{eqnarray}
Note in Eq. (\ref{fq_dis}), the kinetic energy term $\vec{p}^2/m$ is
different from $\vec{p}^2/2m$ in the non-relativistic case.
Now if we let $\epsilon$ approaches zero faster than $m$, 
the second branch in Eq. (\ref{decou}) will be decoupled from the physical
spectrum. 
At this limit, a gap between E = p = 0 and  $E = \infty$ is created. This
could pose a problem for perturbation treatment around this axial symmetry 
point. 

However, the situation is modified when the quarks are interacting. 
In this case, the Dirac equation for $u$ is,
\begin{equation}  \label{int_zq}
\left( \begin{array}{cc} D_4 + m &
\vec{\sigma}\cdot\vec{D} \\ \vec{\sigma}\cdot\vec{D}  &
-\epsilon D_4 + m \end{array} \right) 
\left( \begin{array}{c} u_1 \\ u_2 \end{array}\right). 
\end{equation}
One of the coupled equation from Eq. (\ref{int_zq}) is
\begin{equation} \label{int2}
(D_4^2 - \frac{m}{\epsilon} (1 - \epsilon) D_4 - \frac{m^2}{\epsilon}) u_2
= \frac{(\vec{D}^2 + \vec{\sigma}\cdot{B}) u_2 + ig\vec{\sigma}\cdot\vec{E}u_1}
{\epsilon}.
\end{equation}
If we let $m$ and $\epsilon$ approaches zero at the same rate
such that $m/\epsilon = \lambda \gg \lambda_{QCD}$, the left-hand side
of Eq. (\ref{int2}) leads to a constraint equation
\begin{equation}  \label{iq1}
 (\vec{D}^2 + \vec{\sigma}\cdot{B}) u_2 + ig\vec{\sigma}\cdot\vec{E}u_1 = 0.
\end{equation}
The right-hand side leads to two equations, both with linear time dependence
\begin{eqnarray}
D_4 u_2 = 0,  \label{iq2}  \\   
(D_4 - \lambda) u_2 = 0.  \label{iq3} 
\end{eqnarray}
Since $\lambda \gg \lambda_{QCD}$, the solution from Eq. (\ref{iq3}) is
decoupled from the physical system of the hadrons.

Therefore, the Dirac equation for the interacting massless quark with
the $\epsilon$ regulator leads to the following coupled equations,
\begin{eqnarray}
D_4 u_1 + i\vec{\sigma}\cdot\vec{D} u_2 = 0,  \\
D_4 u_2 = 0,
\end{eqnarray}
with Eq. (\ref{iq1}) as a constraint.
This should admit propagating solutions. Similar situation exists for $v$.
Thus, we can approach the interacting massless quark case with the help of the
infrared $\epsilon$ regulator.

\section{Lattice VQCD}

   In order to solve VQCD, we need a lattice version like in QCD. To this
end, we shall devise a lattice VQCD action based on Wilson's action in
QCD. We shall use the following lattice VQCD action
\begin{equation}  \label{lvqcd}
S_{VQCD}^L =  S_G + S_F^u + S_F^v	
\end{equation}
where $S_G$ is Wilson's gauge action which is preserved here and
the quark action $S_F^u$ and the antiquark action $S_F^v$ are
\begin{eqnarray}
S_F^u =  \sum_x [ \bar{u}(x)u(x) - \kappa( \bar{u}(x + a_4) (1 + \gamma_4)
U_4^{\dagger}(x) u(x) + u_0 \bar{u}(x)(1 - \gamma_4)u(x)) \nonumber \\
- \kappa \sum_i (\bar{u}(x+ a_i) (1 + \gamma_i) U_i^{\dagger}(x) u(x) + 
 \bar{u}(x)(1 - \gamma_i)u(x + a_i))]   \label{s_u}  \\
S_F^v =  \sum_x [ \bar{v}(x)v(x) - \kappa( \bar{v}(x) (1 - \gamma_4)
U_4(x) u(x + a_4) + u_0 \bar{v}(x)(1 + \gamma_4)v(x)) \nonumber \\
- \kappa \sum_i (\bar{v}(x+ a_i) (1 + \gamma_i) U_i^{\dagger}(x) v(x) + 
 \bar{v}(x)(1 - \gamma_i)v(x + a_i))]   \label{s_v} 
\end{eqnarray}
where $u_0$ is the tadpole contribution of the gauge link $U_{\mu}$ which
we take to be $(Tr \Box)^{1/4}$~\cite{lm93}. This has the VQCD in 
Eq. (\ref{vqcd}) as the classical continuum limit. 

\subsection{Reflection Positivity and Hermiticity}

   Similar to the continuum case in Sec. (\ref{dis_sym}), lattice VQCD
action in Eq. (\ref{lvqcd}) is invariant under the corresponding lattice
C, P, and T transformations.

   For Euclidean action, it is imperative that it satisfies 
Osterwalder-Schrader reflection positivity~\cite{ow73} in order to allow the 
Euclidean correlations to be continued back to the Minkowski space. We shall
follow the derivation for the Wilson action~\cite{mm94}. To prove reflection
positivity, one needs to show 
\begin{equation}  \label{rp}
\langle (\Theta F) F \rangle \geq 0 
\end{equation}
where F is a function of the fields $\bar{u}, u, \bar{v}, v$, and $U$ on the
positive time part of the lattice and $\Theta$ is the time reflection
operator. We shall consider the `link-reflection' case where the time
reflection is with respect to the t = (0 $\rightarrow$ 1) link. In this
case, $\Theta$ is defined by the transformation  
\begin{eqnarray}
\Theta u_{x,t} &= &\bar{u}_{x,1-t}\gamma_4, \Theta \bar{u}_{x,t} =
\gamma_4 u_{x,1-t}, \\
\Theta v_{x,t} &= &\bar{v}_{x,1-t}\gamma_4, \Theta \bar{v}_{x,t} =
\gamma_4 v_{x,1-t}, \\
\Theta U(x,t_x;yt_y)& =& U^{\dagger}(x,1-t_x;y,1-t_y).
\end{eqnarray}
Let's illustrate the proof with the u part of the action alone. The
proof can be extended similarly to include the $v$ field. Denoting the
field variables in the half-space with positive time $t \geq 1$ by
$u^+, \bar{u}^+, U^+$, and in the other half-space with $t \leq 0$ by
$u^-, \bar{u}^-, U^-$, the VQCD action can be separated into three parts
\begin{equation}
S_{VQCD} = S_+[\bar{u}^+, u^+, U^+] + S_-[\bar{u}^-, u^-, U^-]
          + S_c[\bar{u}^+, u^-],
\end{equation}
where 
\begin{equation} 
S_c [\bar{u}^+, u^-] = -\kappa \sum_{\vec{x}}
[\bar{u}_{\vec{x},1}^+ (1 + \gamma_4) u_{\vec{x},0}^-],
\end{equation}
is the action which connects $S_+$ and $S_-$ and involves links between 
$t = 0$ and $t = 1$.
Here are have used the temporal gauge. Now since
\begin{equation}  
\Theta S_+[\bar{u}^+, u^+, U^+] = S_+^{\dagger}[\Theta \bar{u}^+,
\Theta u^+, \Theta U^+] = S_-[\bar{u}^-, u^-, U^-],
\end{equation}       
the integral in Eq. (\ref{rp}) is then
\begin{eqnarray}   \label{rp_1}
\langle (\Theta F) F \rangle = Z^{-1} \int dU d\bar{u}^+ du^+
e^{-S_+[\bar{u}^+, u^+, U^+]} F[\bar{u}^+, u^+, U^+] \nonumber \\
\cdot \int d(\Theta \bar{u}^+) d(\Theta u^+) e^{-S_+^{\dagger}
[\Theta \bar{u}^+, \Theta u^+, \Theta U^+]} F^{\dagger}[\Theta \bar{u}^+,
\Theta u^+, \Theta U^+]  \nonumber \\
e^{ - \kappa \sum_{\vec{x}} \bar{u}_{\vec{x},1}^+ (\gamma_4 +1) 
(\Theta \bar{u}_{\vec{x},1}^+)}.
\end{eqnarray}

Consider the Taylor expansion of the last exponential in Eq. (\ref{rp_1})
\begin{equation}
1 - \kappa \sum_{\vec{x}} \bar{u}_{\vec{x},1}^+ (\gamma_4 +1) 
(\Theta \bar{u}_{\vec{x},1}^+) + ...
\end{equation}
The only terms that survive the Grassmann integration are the first
two terms and, with a diagonal representation of $\gamma_4$, they give
semi-positive definite contributions to  $\langle (\Theta F) F \rangle$.
Extension to include the v field is straightforward and thus  
the reflection positivity for the VQCD action is proved.

In constructing meson propagators, the usual practice is to first
invert the quark matrix to obtain the quark propagator from the source
to all lattice points, i.e. $M^{-1}(x, 0)$. Then the anti-quark propagator
which goes backward in time is obtained through the hermiticity 
relation
\begin{equation}
{M^{-1}}^{\dagger}(0,x) = \gamma_5 M^{-1}(x,0) \gamma_5,
\end{equation}
where $\dagger$ indicates the hermitian conjugation in the color and Dirac
indices. In VQCD, a similar situation exists. In constructing a $q\bar{q}$ 
meson, one needs the quark propagator $M_u^{-1}(x, 0)$ and and the
anti-quark propagator $M_v^{-1}(0, x)$. It turns out that the hermiticity 
relation
\begin{equation}
{M_v^{-1}}^{\dagger}(0,x) = \gamma_5 M_u^{-1}(x,0) \gamma_5,
\end{equation}
still exists so that one can obtain the anti-quark propagator from the
quark to construct a meson propagator as before.

\subsection{Free Quark Propagator}

   It is useful to understand the free quark spectrum and its residue at the
pole for the lattice VQCD and see how different it is from the Wilson and
the continuum ones. The inverse of the free quark propagator of VQCD in
momentum space is
\begin{equation}
{S_F^u}^{-1}(p) = 1 - \kappa (1 -\gamma_4) - \kappa (1 + \gamma_4) e^{- i p_4a}
- \kappa \sum_i [ ( 1 + \gamma_i) e^{ - i p_ia} + (1 - \gamma_i) e^{i p_ia}],
\end{equation}
where $a$ is the lattice spacing.
We can compute the propagator in discrete time $t = n_ta$
\begin{equation}
S_F^u(t, \vec{p}) = \int_{-\pi}^{\pi} \frac{d p_4}{2 \pi} S_F^u(p) e^{i p_4t}
\end{equation}
For $n_t > 0$ and $\vec{p}$ in the 3 direction,
\begin{equation}
S_F^u(t, p_3) = \frac{e^{ - E t}}{B}[A - \kappa e^{Ea} + \kappa(e^{Ea} - 1)
\gamma_4 - 2i\kappa\gamma_3 \sin (p_3a)],
\end{equation}
where 
\begin{eqnarray}  \label{Ea}
A & = & 1 - 5\kappa - 2\kappa \cos(p_3a), \nonumber  \\
B &= & A^2 - \kappa^2 + 4 \kappa^2 \sin^2(p_3a), \nonumber \\
Ea &= & \ln (\frac{B}{2A\kappa - 2 \kappa^2}).
\end{eqnarray}
E is the energy. For small $p_3a$, i.e. $p_3a \ll 1$, 
\begin{equation}  \label{dr}
Ea = \bar{m}a + \frac{ma + 2}{2 ma(ma + 1)} (p_3a)^2,
\end{equation}
where 
\begin{equation}  \label{fq_mass}
\bar{m}a = \ln (\frac{1 - 6\kappa}{2\kappa}), 
\end{equation}
is the free quark mass
which is the same as in the Wilson case and $ma = \frac{1}{2\kappa} - 4$
is the small mass approximation for $\bar{m}a$. 
\begin{figure}[h]
\rotatebox{270}{
\resizebox{10cm}{15cm}{
\includegraphics{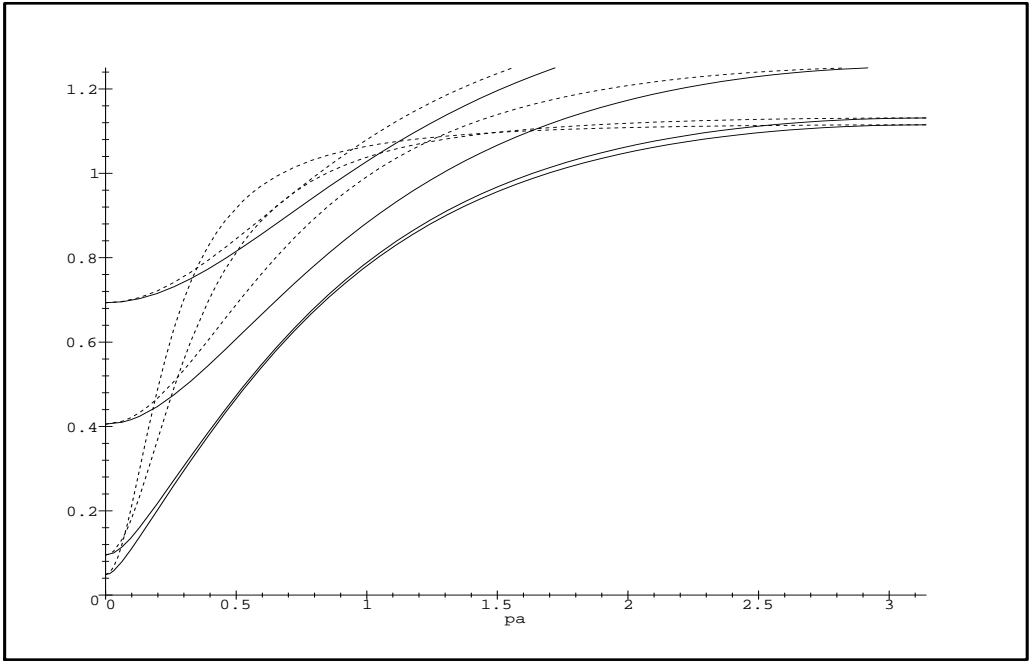}}
}
\caption{The dispersion relations between Ea and $p_3a$ for a free quark are
compared between the Wilson (solid lines) and the lattice VQCD (dashed lines)
version for ma = 0.05, 0.1, 0.5, and 1.0.}
\end{figure}

 We plot in Fig. 18, the dispersion relation of $Ea$ vs. $p_3a$ for a range of	
of $ma$ (ma = 0.05, 0.1, 0.5, and 1.0)
for both the Valence (dashed lines) and the Wilson case (solid lines).
We see that for heavy quarks, i.e. $ma$ = 0.5 and 1, the two curves at the
top are close to each other and they differ at small mass and low momentum.
At small $p_3a$, the behavior of Eq. (\ref{dr}) holds for the valence
case. At $ma = \bar{m}a = 0$, there is a sigularity at $Ea = p_3a = 0$.
For any finite $p_3a$, $Ea = \ln 3$ which resembles the infinite gap in 
the massless free quark situation in the continuum (see Eq. (\ref{fq_dis})).

Finally, we see at zero momentum, the static propagator is
\begin{equation}
S(t > 0, \vec{p} = 0) = \frac{1}{1 - 6 \kappa} e^{ - \bar{m}t}\frac{1 + 
\gamma_4}{2}.
\end{equation}
This is the same as in the Wilson case and the wave function normalization
factor $\frac{1}{1 - 6\kappa}$ is also the same. To convert lattice 
matrix elements of local currents with bilinear quark fields, e.g.
$\bar{\Psi}(x)\Gamma\Psi(x)$ to the continuum ones, besides the finite lattice 
renormalization one needs to multiply the factor $\frac{1 - 6\kappa u_0} 	
{2\kappa } = u_0 e^{m_qa}$ to take into account the finite mass normalization
due to the Wilson quark action with tadpole improvement~\cite{lm93,ldd94a}. 
Here $u_0 = \frac{1}{8\kappa_c}$ where $\kappa_c$ is the critical $\kappa$
at which point the pion mass is zero and $m_qa = \ln(\frac{1}{2 \kappa u_0}
- 3)$ is the tadpole improved definition of the bare quark mass
in Eq. (\ref{fq_mass}).

\subsection{Lattice Details}

    Before we present our results on VQCD from the Monte Carlo calculation,
we should mention the lattice specifics and the details of the calculation.

We use the same gauge configurations which have been used for the
study of hadron masses, matrix elements and form factors~\cite{ldd94a,ldd95,
dll95,dll96,dlw98} in the quenched approximation. This way we keep the
scale of the lattice spacing unchanged. 
These quenched gauge configurations were generated on a $16^3 \times 24$
lattice at $\beta = 6.0$.  The gauge field was thermalized for $5000$
pseudo-heatbath sweeps from a cold start and 100 configurations separated 
by at least
$1000$ sweeps were used.  Periodic boundary conditions were imposed on
the quark fields in the spatial directions.  In the time direction,
fixed boundary conditions were imposed on the quarks to provide
larger time separations than available with periodic boundary conditions.  
As
long as the time separations $t_1$ and $t -t_1$ in Fig. 5(a) are large enough,
the form factors should not depend on the nucleon interpolation field.
All quark propagators in the quenched approximation were chosen to originate 
from lattice time slice $5$; the secondary nucleon source was fixed at time 
slice $20$ (except for $\kappa = 0.154$ where the quark propagators 
from time slice $3$ to $22$ were used). In the case of VQCD, all quark
propagators originate from time slice $2$ and terminate at time slice $22$ for
the three-point function calculation. We also averaged over the directions of 
equivalent lattice momenta in each configuration; this reduces error bars.

We have verified that the time separation is sufficient so that
there is a plateau for the quark bilinear current insertion at time slices 
$t_1$ after the nucleon ground state is achieved. 
The quenched approximation part is done for the
lightest quarks with $\kappa = 0.154, 0.152, 0.148$, and 0.140, and
$\vec{q\,}^2 a^2$ up to $4(2\pi/L)^2$. The nucleon masses $M_N a$ for 
$\kappa = 0.154, 0.152$, and 0.148 are 0.731(11), 0.883(9), and
1.153(7) respectively. The corresponding pion masses $m_{\pi} a$ are
0.375(4), 0.487(3), and 0.679(3). Extrapolating the nucleon and 
pion masses to the chiral limit where we determine $\kappa_c = 
0.15672(4)$
and the nucleon mass at the chiral limit to be 0.536(13). Using
the nucleon mass to set the scale which we believe to be 
appropriate for studying nucleon properties \cite{ldd94a,ldd95,dll95},
the lattice spacing $a^{-1} = 1.75(2)$ GeV is determined. The
three $\kappa's$ then correspond to quark masses of about 120, 205, 
and 370 MeV respectively.

Since we use the same guage configurations for VQCD, the lattice spacing
is the same as that in the quenched approximation. This is certainly 
obvious if we choose the string tension or the glueball mass to set the
scale. Using the physical nucleon mass to set the scale in the quenched
approximation opens up the question as to what extend the fermion determinant
effects are implicitly inlcuded. It is shown~\cite{sw97} that the quenched
approximation can be viewed as including the leading terms in the loop
expansion of the fermion determinant which are commensurate with the size of
loops in the gauge action. This leads to a shift in $\beta$ or the coupling
constant. However, when the infinite volume and continuum limits are
taken~\cite{bcs93}, the scales due to hadron masses and the string tension
are consistent. Since we are not at the infinite volume and continuum
limits, the scale set by the nucleon mass differs from that by the
string tension by $\sim$ 20\%. This is the caveat that we should be aware of.
Nevertheless, whatever scale we decide to choose, the lattice spacing is
the same in the following VQCD calculation as that in our quenched 
approximation. 

The determination of $\kappa_c$ which corresponds to the zero quark mass
will be discussed in the next section. 
To determine the finite quark mass,
we shall use the tadpole improved form of the lattice free-quark mass
in Eq. (\ref{fq_mass}), i.e.
\begin{equation}
m_q a = \ln (\frac{1}{2 \kappa u_0} - 3) = \ln (\frac{4 \kappa_c}{\kappa} - 3),
\end{equation}
where we have used $u_0 = \frac{1}{8\kappa_c}$.

\subsection{Pion Decay Constant, Pion Mass and $\kappa_c$}  \label{pi_sym}

Pion decay constant $f_{\pi}$ plays an essential role in understanding
low-energy chiral dynamics and chiral symmetry breaking in QCD. It sets the
scale for chiral perturbation, and relates the Goldstone boson mass to
the quark mass through the Gell-Mann-Oakes-Renner's relation~\cite{gor68}
\begin{equation}
f_{\pi}^2 m_{\pi}^2 = - (m_u + m_d) \langle \bar{q}q\rangle,
\end{equation}
where $\langle \bar{q}q\rangle$ is the quark condensate, which is the order
parameter for chiral symmetry breaking. In VQCD, it is not clear if 
there is a corresponding relation, or more importantly, if the $U(2N_f)$
symmetry is broken to generate Goldstone bosons. We can, however,
look for clues from the pion decay matrix element with the axial current.
In QCD, the pion decay constant is defined by
\begin{equation}
\langle 0|A_{\mu}(x)|\pi(p)\rangle = i f_{\pi}p_{\mu} e^{ip \cdot x}.
\end{equation}
Applying axial identity from Eq. (\ref{axi2}) to the zero-momentum pion state,
we obtain in VQCD
\begin{eqnarray}  \label{pi_m}
\langle 0|\partial_4 A_4^{\dagger}(x)|\pi(0)\rangle &=& - m_{\pi}^2 
f_{\pi}(m_{\pi}) e^{- m_{\pi} t} \nonumber \\
  &= & 2 m \langle 0|\bar{v} i \gamma_5 u|\pi(0)\rangle e^{- m_{\pi} t},
\end{eqnarray}
where m is the quark mass. From this, we find
\begin{equation} \label{quark_mass}
- m_{\pi}^2\frac{f_{\pi}(m_{\pi})}{\langle 0|\bar{v} i \gamma_5 u|\pi(0)\rangle} 
= 2 m .
\end{equation}

It is clear from Eq. (\ref{quark_mass}) that as long as the ratio 
$f_{\pi}(m_{\pi})/\langle 0|\bar{v} i \gamma_5 u|\pi(0)\rangle$ does not
diverge as fast as $1/m_{\pi}^2$ when the quark mass approaches zero, the
pion mass will go to zero at massless quark limit. Furthermore, if
the pion decay constant $f_{\pi}$ is not zero at the massless limit, it
would signal spontaneous axial symmetry breaking with the pion as
the Goldstone boson. $f_{\pi}, m_{\pi}$ and the current quark mass $m$ 
from lattice VQCD are calculated in the
standard way by fitting the following two-point functions to
\begin{eqnarray}
\langle \sum_{\vec{x}} [ i \bar{v}\frac{\gamma_4 -1}{2} 
\gamma_5 u (\vec{x}, t)] P(0,0)\rangle_{\,\,
\stackrel{\longrightarrow}{t\,>>\,a}}\,\, \frac{f_{\pi} m_{\pi} Z_{\pi}}
{2 m_{\pi}} e^{- m_{\pi}t}, \label{f_pi}\\
\langle \sum_{\vec{x}} [ i \partial_t \bar{v}\frac{\gamma_4 -1}{2} 
\gamma_5 u (\vec{x}, t)] P(0,0)\rangle_{\,\,
\stackrel{\longrightarrow}{t\,>>\,a}}\,\, \frac{2 m Z_{\pi}^2}
{2 m_{\pi}} e^{- m_{\pi}t}, \label{q_m}\\
\langle \sum_{\vec{x}} 
P(\vec{x}, t) P(0,0)\rangle_{\,\,
\stackrel{\longrightarrow}{t\,>>\,a}}\,\, \frac{Z_{\pi}^2}{2m_{\pi}} 
e^{- m_{\pi}t}.
\end{eqnarray}
Here $P$ is the pseudoscalar interpolation field $\bar{u}i\gamma_5 v$ and
$Z_{\pi}= \langle \pi|P|0\rangle$ is the wave function overlap.
We use the local current for the axial current in Eq. (\ref{f_pi})
for the lattice calculation. There
are finite lattice renormalizations associated with the operators in
these matrix elements. We have not calculated them, but we expect the
multiplicative renormalization constants $Z_A$ and $Z_P$ for the axial and
pseudoscalar operators to be of order 1, as in the quenched approximation.
Our results presented below are subject to this caveat.

    With Wilson-type fermions, one needs to find out where the $\kappa_c$ is 
for zero quark mass. To determine $\kappa_c$, we plot the dimensionless pion
mass, the current quark mass, and the pion decay constant in Fig. 19 as a 
function of $m_qa = \ln (4\kappa_c/\kappa -3)$ where $\kappa_c$ is to be 
determined from the extrapolation. We see that the pion mass is very 
linear in the range of the
quark mass that we calculated. At the same time, the pion decay constant
$f_{\pi}$ behaves like $1/m_{\pi}$ in this range. Since from Eq. (\ref{f_pi})
and Eq. (\ref{q_m}) we have the relation
\begin{equation}  \label{fpi2}
f_{\pi}(m_{\pi}) m_{\pi}^2 = 2 m Z_{\pi},
\end{equation}
\begin{figure}[h]
\input{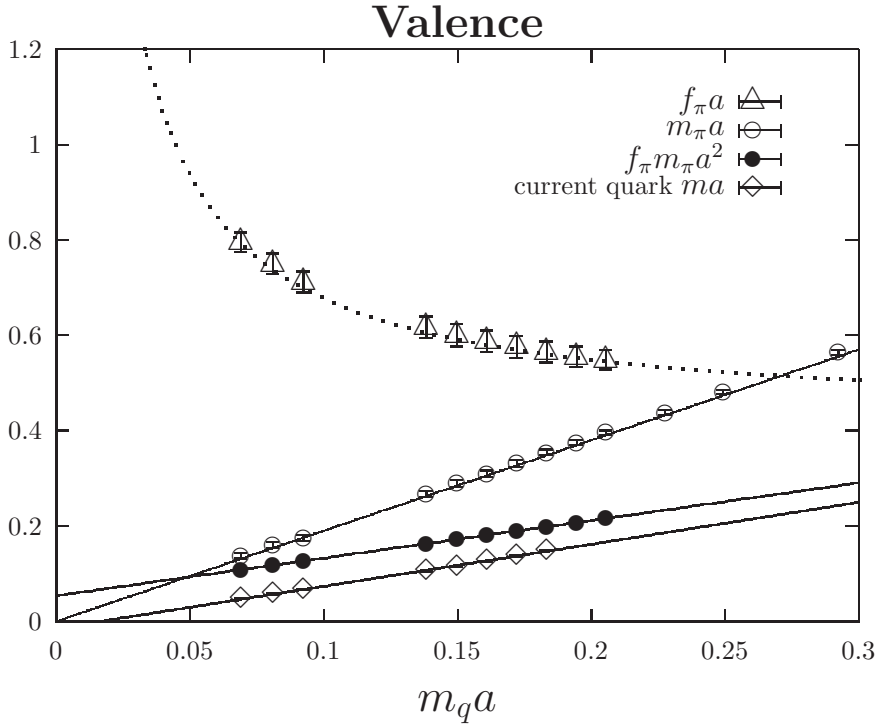}
\caption{The dimensionless pion mass $m_{\pi}a$, the current quark mass $ma$,
and the pion decay constant $f_{\pi}a$ are shown as a function of
$m_qa = \ln (4\kappa_c/\kappa -3)$.
The solid lines represent the linear extrapolation with respect to $m_qa$.}
\end{figure}
if the linear behavior between $m_{\pi}$ and m persists all the way
down to zero quark mass and $Z_{\pi}$ remains constant, then $f_{\pi}$
will diverge like $1/m_{\pi}$ or 1/m. Alternatively, at smaller quark mass
than we calculated here, $f_{\pi}$ could conceivably turn flat and in this
case $m_{\pi}$ will fall off like $\sqrt{m}$ as in QCD with a constant 
$Z_{\pi}$.
Unfortunately, using conjugate gradient to invert the quark matrix, we have 
encountered critical slowing down. The smallest quark mass we run at 
$\kappa =0.162$ already takes 5,000 iterations to converge. It is impractical 
for us to go down
any further. Short of theoretical guidance and numerical evidence, we 
extrapolate the pion mass to zero both linearly and quadratically with 
respect to $m_qa = \ln(4\kappa_c/\kappa -3)$ with $\kappa = 
0.162, 0.1615, 0.1610, 0.1590, 0.1585, 0.1580, 0.1575, 0.1570, 0.1565,
0.1560, 0.155, \\
0.154,  0.152$, and 0.148.
We found that $\kappa_c = 0.1649(10)$ ($\chi^2 = 0.002$ with 14 data points)
for the linear dependence and $\kappa_c = 0.1636(19)$ ($\chi^2 = 0.04$ from
the three largest $\kappa's$) for the quadratic dependence. 
We plot in Fig. 20 the quadratic fit of $m_{\pi}$ as a funciton of $m_qa$
with $\kappa_c$ determined from the linear $m_{\pi}$ fit. We see that
the $\kappa_c$ point from the quadratic fit crosses the abscissa at 
$m_qa = 0.031$, however, its error bar overlaps with that from the linear 
$m_{\pi}$ fit. Also plotted in Figs. 19 and 20 is the current quark mass 
$ma$ from Eq. (\ref{q_m}) as a function of $m_qa$. Extrapolating the quark 
mass linearly with respect to 
$m_qa$, we obtain $\kappa_c = 0.1642(9)$ ($\chi^2 =3.5$ from the 
first 8 $\kappa's$).  We have not used
covariance matrix in these extrapolations.
We see that the current quark mass $ma$ from Eqs. (\ref{quark_mass}) and
(\ref{q_m}) crosses the abscissa at $m_qa = 0.017$. This is consistent with
with that extrapolated from the pion mass, either linearly or quadratically. 
The $\kappa_c$ so obtained overlaps with both
of the above two $\kappa_c's$ within errors. It is gratifying to know that
different definitions of $\kappa_c$ agree. On the other hand, it does not
differentiate the two scenarios of the pion mass dependence on the quark mass. 
We shall use the linear extrapolaiton with $\kappa_c = 0.1649(10)$ to
define zero quark mass in this study. Also plotted in Fig. 20 is 
$f_{\pi}m_{\pi}^2 a^3$. We see that it is 
\begin{figure}[h]
\input{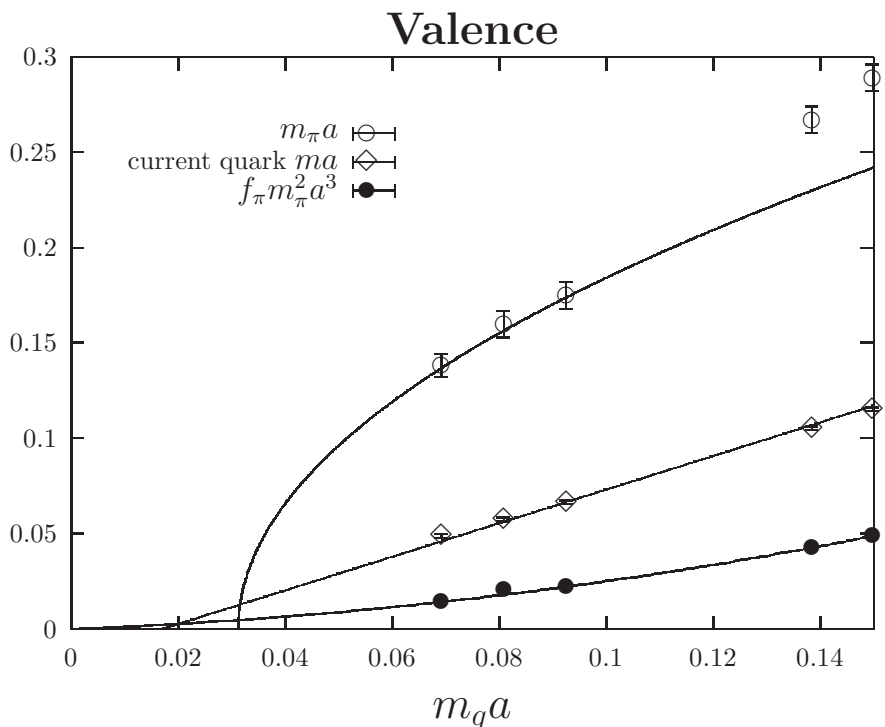}
\caption{The dimensionless pion mass $m_{\pi}a$,
$f_{\pi}m_{\pi}^2a^3$, and current quark mass $ma$ are shown shown as a
function of $m_qa = \ln (4\kappa_c/\kappa -3)$ with $\kappa_c$ determined
from the linearly fit of $m_{\pi}$ with reepect to $m_qa$.}
\end{figure}

\newpage
\clearpage
\noindent quite linear in the range of
quark mass that we have considered. This confirms that
$f_{\pi}m_{\pi}^2 \propto m_q$ or equivalently Eq. (\ref{fpi2}), since
we have just shown in Fig. 20 that $m_qa$ and $ma$ are linearly related.
We should stress that we still do not know the behavior of
the pion mass and pion decay constant when the quark mass is small. But,
at least we can say that $f_{\pi}$ is non-zero (divergent or not)
and $m_{\pi}$ approaches zero
at the massless quark limit. This we take to be the evidence that there is
a spontaneous axial symmetry breaking with the two pions $\bar{u}i\gamma_5 v$	
and $\bar{v}i\gamma_5 u$ as the Goldstone bosons for each flavor.

\section{$SU(6)$ Relations}

   We shall first examine the ratios $R_A$ in Eq. (\ref{R_A}),
$R_S$ in Eq. (\ref{R_S}) and the neutron to proton magnetic moment ratio
$\mu^n/\mu^p$ to check their $SU(6)$ relations as compared to
those in full QCD, albeit with quenched approximation. 
In VQCD, the sea quark contributions (e.g. Fig. 5(b)) are scrapped. We will
only consider the connected insertions (e.g. Fig. 4(a) and 5(a)). 

\subsection{$R_A$ and $F_A/D_A$}

   The same 100 gauge configurations used for the earlier $R_A$ calculation
in Sec.~\ref{g_A} are used for the VQCD case. Since in VQCD, there is only
the C. I., the $R_A$ ratio becomes
\begin{equation}  \label{VR_A}
R_A =\frac{g_A^0}{g_A^3} (C. I.) 
= \frac{(\Delta u + \Delta d)(C. I.)}{(\Delta u - \Delta d)(C. I.)}. 
\end{equation}
We plot it in Fig. 21 as a function of the dimensionless quark mass $m_qa$
(with $\kappa$ = 0.162, 0.1615, 0.1610, 0.1590, and 0.1585) 
together with the corresponding results from the quenched calculation presented 
earlier in Fig. 6 in Sec.~\ref{g_A}. We see that, even for light quarks in 
the strange region ($m_q a \sim 0.07$), it is much closer to the valence 
prediction of 3/5, in contrast to the QCD calculation with C. I. alone.

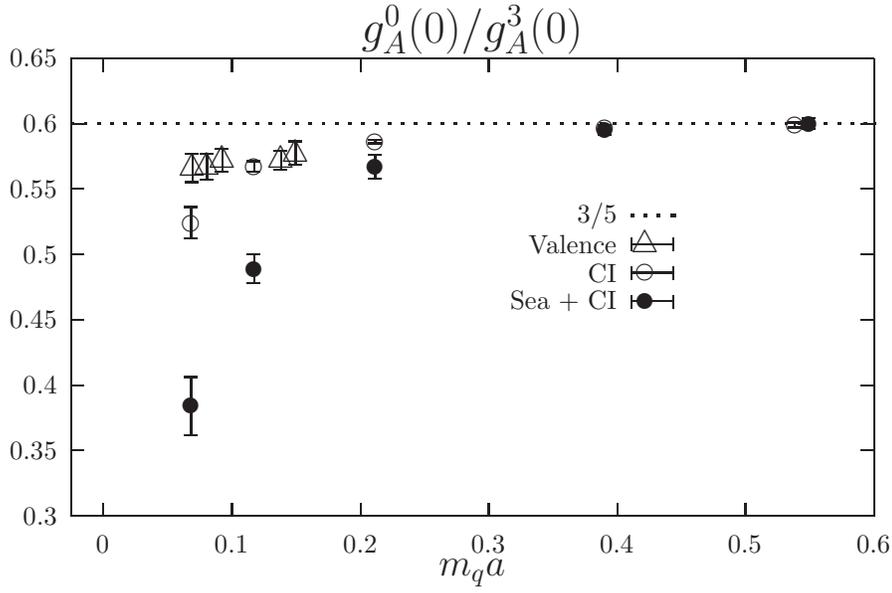
\begin{figure}[h]
\setlength{\unitlength}{0.240900pt}
\ifx\plotpoint\undefined\newsavebox{\plotpoint}\fi
\sbox{\plotpoint}{\rule[-0.200pt]{0.400pt}{0.400pt}}%
\begin{picture}(1500,900)(0,0)
\font\gnuplot=cmr10 at 10pt
\gnuplot
\sbox{\plotpoint}{\rule[-0.200pt]{0.400pt}{0.400pt}}%
\put(176.0,113.0){\rule[-0.200pt]{4.818pt}{0.400pt}}
\put(154,113){\makebox(0,0)[r]{0.3}}
\put(1416.0,113.0){\rule[-0.200pt]{4.818pt}{0.400pt}}
\put(176.0,216.0){\rule[-0.200pt]{4.818pt}{0.400pt}}
\put(154,216){\makebox(0,0)[r]{0.35}}
\put(1416.0,216.0){\rule[-0.200pt]{4.818pt}{0.400pt}}
\put(176.0,318.0){\rule[-0.200pt]{4.818pt}{0.400pt}}
\put(154,318){\makebox(0,0)[r]{0.4}}
\put(1416.0,318.0){\rule[-0.200pt]{4.818pt}{0.400pt}}
\put(176.0,421.0){\rule[-0.200pt]{4.818pt}{0.400pt}}
\put(154,421){\makebox(0,0)[r]{0.45}}
\put(1416.0,421.0){\rule[-0.200pt]{4.818pt}{0.400pt}}
\put(176.0,524.0){\rule[-0.200pt]{4.818pt}{0.400pt}}
\put(154,524){\makebox(0,0)[r]{0.5}}
\put(1416.0,524.0){\rule[-0.200pt]{4.818pt}{0.400pt}}
\put(176.0,627.0){\rule[-0.200pt]{4.818pt}{0.400pt}}
\put(154,627){\makebox(0,0)[r]{0.55}}
\put(1416.0,627.0){\rule[-0.200pt]{4.818pt}{0.400pt}}
\put(176.0,729.0){\rule[-0.200pt]{4.818pt}{0.400pt}}
\put(154,729){\makebox(0,0)[r]{0.6}}
\put(1416.0,729.0){\rule[-0.200pt]{4.818pt}{0.400pt}}
\put(176.0,832.0){\rule[-0.200pt]{4.818pt}{0.400pt}}
\put(154,832){\makebox(0,0)[r]{0.65}}
\put(1416.0,832.0){\rule[-0.200pt]{4.818pt}{0.400pt}}
\put(226.0,113.0){\rule[-0.200pt]{0.400pt}{4.818pt}}
\put(226,68){\makebox(0,0){0}}
\put(226.0,812.0){\rule[-0.200pt]{0.400pt}{4.818pt}}
\put(428.0,113.0){\rule[-0.200pt]{0.400pt}{4.818pt}}
\put(428,68){\makebox(0,0){0.1}}
\put(428.0,812.0){\rule[-0.200pt]{0.400pt}{4.818pt}}
\put(630.0,113.0){\rule[-0.200pt]{0.400pt}{4.818pt}}
\put(630,68){\makebox(0,0){0.2}}
\put(630.0,812.0){\rule[-0.200pt]{0.400pt}{4.818pt}}
\put(831.0,113.0){\rule[-0.200pt]{0.400pt}{4.818pt}}
\put(831,68){\makebox(0,0){0.3}}
\put(831.0,812.0){\rule[-0.200pt]{0.400pt}{4.818pt}}
\put(1033.0,113.0){\rule[-0.200pt]{0.400pt}{4.818pt}}
\put(1033,68){\makebox(0,0){0.4}}
\put(1033.0,812.0){\rule[-0.200pt]{0.400pt}{4.818pt}}
\put(1234.0,113.0){\rule[-0.200pt]{0.400pt}{4.818pt}}
\put(1234,68){\makebox(0,0){0.5}}
\put(1234.0,812.0){\rule[-0.200pt]{0.400pt}{4.818pt}}
\put(1436.0,113.0){\rule[-0.200pt]{0.400pt}{4.818pt}}
\put(1436,68){\makebox(0,0){0.6}}
\put(1436.0,812.0){\rule[-0.200pt]{0.400pt}{4.818pt}}
\put(176.0,113.0){\rule[-0.200pt]{303.534pt}{0.400pt}}
\put(1436.0,113.0){\rule[-0.200pt]{0.400pt}{173.207pt}}
\put(176.0,832.0){\rule[-0.200pt]{303.534pt}{0.400pt}}
\put(806,23){\makebox(0,0){{\large\bf $m_q a$}}}
\put(806,877){\makebox(0,0){{\Large\bf $g_A^0(0)/g_A^3(0)$}}}
\put(176.0,113.0){\rule[-0.200pt]{0.400pt}{173.207pt}}
\sbox{\plotpoint}{\rule[-0.500pt]{1.000pt}{1.000pt}}%
\put(1033,585){\makebox(0,0)[r]{3/5}}
\multiput(1055,585)(20.756,0.000){4}{\usebox{\plotpoint}}
\put(1121,585){\usebox{\plotpoint}}
\put(176,729){\usebox{\plotpoint}}
\multiput(176,729)(20.756,0.000){61}{\usebox{\plotpoint}}
\put(1436,729){\usebox{\plotpoint}}
\sbox{\plotpoint}{\rule[-0.200pt]{0.400pt}{0.400pt}}%
\put(1033,540){\makebox(0,0)[r]{Valence}}
\put(1077,540){\makebox(0,0){$\triangle$}}
\put(366,659){\makebox(0,0){$\triangle$}}
\put(389,661){\makebox(0,0){$\triangle$}}
\put(413,672){\makebox(0,0){$\triangle$}}
\put(505,672){\makebox(0,0){$\triangle$}}
\put(528,682){\makebox(0,0){$\triangle$}}
\put(1055.0,540.0){\rule[-0.200pt]{15.899pt}{0.400pt}}
\put(1055.0,530.0){\rule[-0.200pt]{0.400pt}{4.818pt}}
\put(1121.0,530.0){\rule[-0.200pt]{0.400pt}{4.818pt}}
\put(366.0,637.0){\rule[-0.200pt]{0.400pt}{10.840pt}}
\put(356.0,637.0){\rule[-0.200pt]{4.818pt}{0.400pt}}
\put(356.0,682.0){\rule[-0.200pt]{4.818pt}{0.400pt}}
\put(389.0,641.0){\rule[-0.200pt]{0.400pt}{9.877pt}}
\put(379.0,641.0){\rule[-0.200pt]{4.818pt}{0.400pt}}
\put(379.0,682.0){\rule[-0.200pt]{4.818pt}{0.400pt}}
\put(413.0,653.0){\rule[-0.200pt]{0.400pt}{8.913pt}}
\put(403.0,653.0){\rule[-0.200pt]{4.818pt}{0.400pt}}
\put(403.0,690.0){\rule[-0.200pt]{4.818pt}{0.400pt}}
\put(505.0,657.0){\rule[-0.200pt]{0.400pt}{6.986pt}}
\put(495.0,657.0){\rule[-0.200pt]{4.818pt}{0.400pt}}
\put(495.0,686.0){\rule[-0.200pt]{4.818pt}{0.400pt}}
\put(528.0,664.0){\rule[-0.200pt]{0.400pt}{8.913pt}}
\put(518.0,664.0){\rule[-0.200pt]{4.818pt}{0.400pt}}
\put(518.0,701.0){\rule[-0.200pt]{4.818pt}{0.400pt}}
\put(1033,495){\makebox(0,0)[r]{CI}}
\put(1077,495){\circle{24}}
\put(364,573){\circle{24}}
\put(463,661){\circle{24}}
\put(653,701){\circle{24}}
\put(1014,723){\circle{24}}
\put(1312,727){\circle{24}}
\put(1055.0,495.0){\rule[-0.200pt]{15.899pt}{0.400pt}}
\put(1055.0,485.0){\rule[-0.200pt]{0.400pt}{4.818pt}}
\put(1121.0,485.0){\rule[-0.200pt]{0.400pt}{4.818pt}}
\put(364.0,549.0){\rule[-0.200pt]{0.400pt}{11.804pt}}
\put(354.0,549.0){\rule[-0.200pt]{4.818pt}{0.400pt}}
\put(354.0,598.0){\rule[-0.200pt]{4.818pt}{0.400pt}}
\put(463.0,653.0){\rule[-0.200pt]{0.400pt}{4.095pt}}
\put(453.0,653.0){\rule[-0.200pt]{4.818pt}{0.400pt}}
\put(453.0,670.0){\rule[-0.200pt]{4.818pt}{0.400pt}}
\put(653.0,698.0){\rule[-0.200pt]{0.400pt}{1.204pt}}
\put(643.0,698.0){\rule[-0.200pt]{4.818pt}{0.400pt}}
\put(643.0,703.0){\rule[-0.200pt]{4.818pt}{0.400pt}}
\put(1014.0,719.0){\rule[-0.200pt]{0.400pt}{1.927pt}}
\put(1004.0,719.0){\rule[-0.200pt]{4.818pt}{0.400pt}}
\put(1004.0,727.0){\rule[-0.200pt]{4.818pt}{0.400pt}}
\put(1312.0,723.0){\rule[-0.200pt]{0.400pt}{1.927pt}}
\put(1302.0,723.0){\rule[-0.200pt]{4.818pt}{0.400pt}}
\put(1302.0,731.0){\rule[-0.200pt]{4.818pt}{0.400pt}}
\put(1033,450){\makebox(0,0)[r]{Sea + CI}}
\put(1077,450){\circle*{24}}
\put(364,286){\circle*{24}}
\put(463,501){\circle*{24}}
\put(653,661){\circle*{24}}
\put(1014,719){\circle*{24}}
\put(1334,729){\circle*{24}}
\put(1055.0,450.0){\rule[-0.200pt]{15.899pt}{0.400pt}}
\put(1055.0,440.0){\rule[-0.200pt]{0.400pt}{4.818pt}}
\put(1121.0,440.0){\rule[-0.200pt]{0.400pt}{4.818pt}}
\put(364.0,240.0){\rule[-0.200pt]{0.400pt}{21.922pt}}
\put(354.0,240.0){\rule[-0.200pt]{4.818pt}{0.400pt}}
\put(354.0,331.0){\rule[-0.200pt]{4.818pt}{0.400pt}}
\put(463.0,479.0){\rule[-0.200pt]{0.400pt}{10.840pt}}
\put(453.0,479.0){\rule[-0.200pt]{4.818pt}{0.400pt}}
\put(453.0,524.0){\rule[-0.200pt]{4.818pt}{0.400pt}}
\put(653.0,643.0){\rule[-0.200pt]{0.400pt}{8.913pt}}
\put(643.0,643.0){\rule[-0.200pt]{4.818pt}{0.400pt}}
\put(643.0,680.0){\rule[-0.200pt]{4.818pt}{0.400pt}}
\put(1014.0,711.0){\rule[-0.200pt]{0.400pt}{3.854pt}}
\put(1004.0,711.0){\rule[-0.200pt]{4.818pt}{0.400pt}}
\put(1004.0,727.0){\rule[-0.200pt]{4.818pt}{0.400pt}}
\put(1334.0,721.0){\rule[-0.200pt]{0.400pt}{4.095pt}}
\put(1324.0,721.0){\rule[-0.200pt]{4.818pt}{0.400pt}}
\put(1324.0,738.0){\rule[-0.200pt]{4.818pt}{0.400pt}}
\end{picture}
\caption{The ratio $R_A$ between isoscalar and isovector $g_A$ in 
VQCD and QCD are plotted against the dimensionless quark mass
$m_q a$ from the strange to the charm region. $\bigtriangleup $ indicates
the VQCD case, $\circ$/$\bullet$ indicates the C. I./ Sea + C. I. in the 
QCD case. The dashed line is the $SU(6)$ prediction of 3/5.}
\end{figure}

This shows that VQCD indeed seems to confirm our expectation of the
valence quarks behavior, i.e. obeying the $SU(6)$ relation. The deviation
from the exact 3/5 prediction in Fig. 21 reflects the fact that are still
spin-spin interaction between the valence quarks as evidenced in the
$\vec{\sigma}\cdot\vec{B}$ term in the VQCD action with Pauli spinors 
in Eq. (\ref{S_u_Pauli}). Its effects, however, appear to be small. 
This also confirms our earlier assertion that the deviation of
the C. I. of $R_A$ in QCD ($\circ$ in Fig. 21) is largely due to the
the cloud quark-antiquark pairs.

With only the C. I., the $F_A/D_A$ ratio is related to VQCD $R_A$ in 
Eq. (\ref{VR_A})
\begin{equation}
\frac{F_A}{D_A} (C. I.) = \frac{1 + R_A}{3 - R_A},
\end{equation}    
From $R_A = 0.566(11)$ for the smallest quark mass ($\kappa = 0.162$), we
obtain $F_A/D_A = 0. 643(4)$. This is larger than
the QCD prediciton of $0.60(2)$ for the \mbox{C. I.} (see Table 1) and 
closer to the valence quark prediction of 2/3.

\newpage
\clearpage
\subsection{$R_S$ and $D_S/F_S$ }

Similar results are obtained for the $R_S$ ratio in VQCD. In this case,
\begin{equation}
R_S = \frac{(\langle p|\bar{u}u + \bar{d}d|p\rangle)(C. I.)}
{(\langle p|\bar{u}u  - \bar{d}d|p\rangle)(C. I.)}.
\end{equation}
Plotted in Fig. 22 are the VQCD results together with those 
from QCD in Fig. 7 as a function of the quark mass $m_q a$. Again, we find
that the ratios for the light quarks are approaching the valence quark
prediction of 3. This again confirms that the deviation of the C. I. result
in QCD is primarily due to the Z-graphs with cloud quarks and antiquarks.
When they are eliminated in VQCD, $R_S$ becomes close to the $SU(6)$ relation.

\begin{figure}[h]
\setlength{\unitlength}{0.240900pt}
\ifx\plotpoint\undefined\newsavebox{\plotpoint}\fi
\sbox{\plotpoint}{\rule[-0.200pt]{0.400pt}{0.400pt}}%
\begin{picture}(1500,900)(0,0)
\font\gnuplot=cmr10 at 10pt
\gnuplot
\sbox{\plotpoint}{\rule[-0.200pt]{0.400pt}{0.400pt}}%
\put(176.0,113.0){\rule[-0.200pt]{4.818pt}{0.400pt}}
\put(154,113){\makebox(0,0)[r]{2}}
\put(1416.0,113.0){\rule[-0.200pt]{4.818pt}{0.400pt}}
\put(176.0,216.0){\rule[-0.200pt]{4.818pt}{0.400pt}}
\put(154,216){\makebox(0,0)[r]{3}}
\put(1416.0,216.0){\rule[-0.200pt]{4.818pt}{0.400pt}}
\put(176.0,318.0){\rule[-0.200pt]{4.818pt}{0.400pt}}
\put(154,318){\makebox(0,0)[r]{4}}
\put(1416.0,318.0){\rule[-0.200pt]{4.818pt}{0.400pt}}
\put(176.0,421.0){\rule[-0.200pt]{4.818pt}{0.400pt}}
\put(154,421){\makebox(0,0)[r]{5}}
\put(1416.0,421.0){\rule[-0.200pt]{4.818pt}{0.400pt}}
\put(176.0,524.0){\rule[-0.200pt]{4.818pt}{0.400pt}}
\put(154,524){\makebox(0,0)[r]{6}}
\put(1416.0,524.0){\rule[-0.200pt]{4.818pt}{0.400pt}}
\put(176.0,627.0){\rule[-0.200pt]{4.818pt}{0.400pt}}
\put(154,627){\makebox(0,0)[r]{7}}
\put(1416.0,627.0){\rule[-0.200pt]{4.818pt}{0.400pt}}
\put(176.0,729.0){\rule[-0.200pt]{4.818pt}{0.400pt}}
\put(154,729){\makebox(0,0)[r]{8}}
\put(1416.0,729.0){\rule[-0.200pt]{4.818pt}{0.400pt}}
\put(176.0,832.0){\rule[-0.200pt]{4.818pt}{0.400pt}}
\put(154,832){\makebox(0,0)[r]{9}}
\put(1416.0,832.0){\rule[-0.200pt]{4.818pt}{0.400pt}}
\put(226.0,113.0){\rule[-0.200pt]{0.400pt}{4.818pt}}
\put(226,68){\makebox(0,0){0}}
\put(226.0,812.0){\rule[-0.200pt]{0.400pt}{4.818pt}}
\put(428.0,113.0){\rule[-0.200pt]{0.400pt}{4.818pt}}
\put(428,68){\makebox(0,0){0.1}}
\put(428.0,812.0){\rule[-0.200pt]{0.400pt}{4.818pt}}
\put(630.0,113.0){\rule[-0.200pt]{0.400pt}{4.818pt}}
\put(630,68){\makebox(0,0){0.2}}
\put(630.0,812.0){\rule[-0.200pt]{0.400pt}{4.818pt}}
\put(831.0,113.0){\rule[-0.200pt]{0.400pt}{4.818pt}}
\put(831,68){\makebox(0,0){0.3}}
\put(831.0,812.0){\rule[-0.200pt]{0.400pt}{4.818pt}}
\put(1033.0,113.0){\rule[-0.200pt]{0.400pt}{4.818pt}}
\put(1033,68){\makebox(0,0){0.4}}
\put(1033.0,812.0){\rule[-0.200pt]{0.400pt}{4.818pt}}
\put(1234.0,113.0){\rule[-0.200pt]{0.400pt}{4.818pt}}
\put(1234,68){\makebox(0,0){0.5}}
\put(1234.0,812.0){\rule[-0.200pt]{0.400pt}{4.818pt}}
\put(1436.0,113.0){\rule[-0.200pt]{0.400pt}{4.818pt}}
\put(1436,68){\makebox(0,0){0.6}}
\put(1436.0,812.0){\rule[-0.200pt]{0.400pt}{4.818pt}}
\put(176.0,113.0){\rule[-0.200pt]{303.534pt}{0.400pt}}
\put(1436.0,113.0){\rule[-0.200pt]{0.400pt}{173.207pt}}
\put(176.0,832.0){\rule[-0.200pt]{303.534pt}{0.400pt}}
\put(806,23){\makebox(0,0){{\large\bf $m_q a$}}}
\put(806,877){\makebox(0,0){{\Large\bf $g_s^{I=0}(0)/g_s^{I=1}(0)$}}}
\put(176.0,113.0){\rule[-0.200pt]{0.400pt}{173.207pt}}
\put(1033,627){\makebox(0,0)[r]{Sea + CI}}
\put(1077,627){\circle*{24}}
\put(364,697){\circle*{24}}
\put(463,541){\circle*{24}}
\put(653,364){\circle*{24}}
\put(1014,233){\circle*{24}}
\put(1334,218){\circle*{24}}
\put(1055.0,627.0){\rule[-0.200pt]{15.899pt}{0.400pt}}
\put(1055.0,617.0){\rule[-0.200pt]{0.400pt}{4.818pt}}
\put(1121.0,617.0){\rule[-0.200pt]{0.400pt}{4.818pt}}
\put(364.0,655.0){\rule[-0.200pt]{0.400pt}{20.476pt}}
\put(354.0,655.0){\rule[-0.200pt]{4.818pt}{0.400pt}}
\put(354.0,740.0){\rule[-0.200pt]{4.818pt}{0.400pt}}
\put(463.0,511.0){\rule[-0.200pt]{0.400pt}{14.695pt}}
\put(453.0,511.0){\rule[-0.200pt]{4.818pt}{0.400pt}}
\put(453.0,572.0){\rule[-0.200pt]{4.818pt}{0.400pt}}
\put(653.0,348.0){\rule[-0.200pt]{0.400pt}{7.468pt}}
\put(643.0,348.0){\rule[-0.200pt]{4.818pt}{0.400pt}}
\put(643.0,379.0){\rule[-0.200pt]{4.818pt}{0.400pt}}
\put(1014.0,227.0){\rule[-0.200pt]{0.400pt}{2.650pt}}
\put(1004.0,227.0){\rule[-0.200pt]{4.818pt}{0.400pt}}
\put(1004.0,238.0){\rule[-0.200pt]{4.818pt}{0.400pt}}
\put(1334.0,213.0){\rule[-0.200pt]{0.400pt}{2.409pt}}
\put(1324.0,213.0){\rule[-0.200pt]{4.818pt}{0.400pt}}
\put(1324.0,223.0){\rule[-0.200pt]{4.818pt}{0.400pt}}
\put(1033,582){\makebox(0,0)[r]{CI}}
\put(1077,582){\circle{24}}
\put(364,292){\circle{24}}
\put(463,257){\circle{24}}
\put(653,234){\circle{24}}
\put(1014,220){\circle{24}}
\put(1312,218){\circle{24}}
\put(1055.0,582.0){\rule[-0.200pt]{15.899pt}{0.400pt}}
\put(1055.0,572.0){\rule[-0.200pt]{0.400pt}{4.818pt}}
\put(1121.0,572.0){\rule[-0.200pt]{0.400pt}{4.818pt}}
\put(364.0,284.0){\rule[-0.200pt]{0.400pt}{4.095pt}}
\put(354.0,284.0){\rule[-0.200pt]{4.818pt}{0.400pt}}
\put(354.0,301.0){\rule[-0.200pt]{4.818pt}{0.400pt}}
\put(463.0,255.0){\rule[-0.200pt]{0.400pt}{0.964pt}}
\put(453.0,255.0){\rule[-0.200pt]{4.818pt}{0.400pt}}
\put(453.0,259.0){\rule[-0.200pt]{4.818pt}{0.400pt}}
\put(653.0,233.0){\rule[-0.200pt]{0.400pt}{0.482pt}}
\put(643.0,233.0){\rule[-0.200pt]{4.818pt}{0.400pt}}
\put(643.0,235.0){\rule[-0.200pt]{4.818pt}{0.400pt}}
\put(1014.0,218.0){\rule[-0.200pt]{0.400pt}{0.964pt}}
\put(1004.0,218.0){\rule[-0.200pt]{4.818pt}{0.400pt}}
\put(1004.0,222.0){\rule[-0.200pt]{4.818pt}{0.400pt}}
\put(1312.0,216.0){\rule[-0.200pt]{0.400pt}{0.964pt}}
\put(1302.0,216.0){\rule[-0.200pt]{4.818pt}{0.400pt}}
\put(1302.0,220.0){\rule[-0.200pt]{4.818pt}{0.400pt}}
\put(1033,537){\makebox(0,0)[r]{Valence}}
\put(1077,537){\raisebox{-.8pt}{\makebox(0,0){$\Diamond$}}}
\put(366,225){\raisebox{-.8pt}{\makebox(0,0){$\Diamond$}}}
\put(389,225){\raisebox{-.8pt}{\makebox(0,0){$\Diamond$}}}
\put(414,224){\raisebox{-.8pt}{\makebox(0,0){$\Diamond$}}}
\put(505,223){\raisebox{-.8pt}{\makebox(0,0){$\Diamond$}}}
\put(528,221){\raisebox{-.8pt}{\makebox(0,0){$\Diamond$}}}
\put(1055.0,537.0){\rule[-0.200pt]{15.899pt}{0.400pt}}
\put(1055.0,527.0){\rule[-0.200pt]{0.400pt}{4.818pt}}
\put(1121.0,527.0){\rule[-0.200pt]{0.400pt}{4.818pt}}
\put(366.0,223.0){\rule[-0.200pt]{0.400pt}{0.964pt}}
\put(356.0,223.0){\rule[-0.200pt]{4.818pt}{0.400pt}}
\put(356.0,227.0){\rule[-0.200pt]{4.818pt}{0.400pt}}
\put(389.0,223.0){\rule[-0.200pt]{0.400pt}{0.964pt}}
\put(379.0,223.0){\rule[-0.200pt]{4.818pt}{0.400pt}}
\put(379.0,227.0){\rule[-0.200pt]{4.818pt}{0.400pt}}
\put(414.0,222.0){\rule[-0.200pt]{0.400pt}{0.964pt}}
\put(404.0,222.0){\rule[-0.200pt]{4.818pt}{0.400pt}}
\put(404.0,226.0){\rule[-0.200pt]{4.818pt}{0.400pt}}
\put(505.0,221.0){\rule[-0.200pt]{0.400pt}{0.964pt}}
\put(495.0,221.0){\rule[-0.200pt]{4.818pt}{0.400pt}}
\put(495.0,225.0){\rule[-0.200pt]{4.818pt}{0.400pt}}
\put(528.0,219.0){\rule[-0.200pt]{0.400pt}{0.964pt}}
\put(518.0,219.0){\rule[-0.200pt]{4.818pt}{0.400pt}}
\put(518.0,223.0){\rule[-0.200pt]{4.818pt}{0.400pt}}
\sbox{\plotpoint}{\rule[-0.500pt]{1.000pt}{1.000pt}}%
\put(1033,492){\makebox(0,0)[r]{3}}
\multiput(1055,492)(20.756,0.000){4}{\usebox{\plotpoint}}
\put(1121,492){\usebox{\plotpoint}}
\put(176,216){\usebox{\plotpoint}}
\multiput(176,216)(20.756,0.000){61}{\usebox{\plotpoint}}
\put(1436,216){\usebox{\plotpoint}}
\end{picture}
\caption{The ratio $R_S$ between isoscalar and isovector scalar charge  
in VQCD and QCD are plotted against the dimensionless quark mass
$m_q a$ from the strange to the charm region. $\bigtriangleup $ indicates
the VQCD case and $\circ$/$\bullet$ indicates the C. I./ Sea + C. I. in the
QCD case. The dashed line is the valence quark prediction of 3.}
\end{figure}
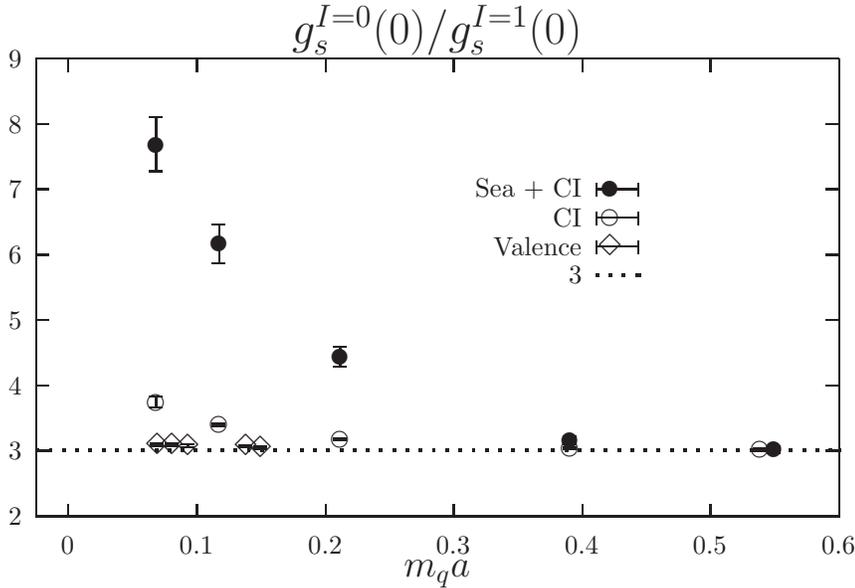

The $D_S/F_S$ ratio in VQCD is
\begin{equation}
D_S/F_S (C. I.) = \frac{3 - R_S}{1 + R_S},
\end{equation} 
From $R_S = 3.086(19)$ for the smallest quark mass ($\kappa = 0.162$), we
obtain $D_S/F_S = - 0.021(4)$ which is close to zero as in the valence quark 
picture  (Table 2) and differs from the lattice QCD calculation of - 0.58(18) 
(Sea + C. I.) and -0.31(11) (C. I.) (see Table 2) by a large margin. 


\subsection{Neutron to Proton Magnetic Moment Ratio}

We next come to the neutron to proton magnetic moment ratio
$\mu^n/\mu^p$. The results of VQCD  (indicated as 
$\bigtriangleup$ in Fig. 23) are very close to the well-known $SU(6)$
value of - 2/3 (the result for the smallest quark mass case is - 0.662(22)), 
indirectly verifying the cloud effects of QCD ($\circ$ for the C. I. 
in Fig. 23) 
which shows a 2.5 $\sigma$ departure from - 2/3 at the chiral limit. If there
is any deviation of the VQCD from - 2/3, it should be due to the residual 
spin-spin interaction between the quarks in the baryon. Given the size of
the error in our present results, we cannot make a definite conclusion on 
this aspect.

\begin{figure}[h]
\setlength{\unitlength}{0.240900pt}
\ifx\plotpoint\undefined\newsavebox{\plotpoint}\fi
\sbox{\plotpoint}{\rule[-0.200pt]{0.400pt}{0.400pt}}%
\begin{picture}(1500,900)(0,0)
\font\gnuplot=cmr10 at 10pt
\gnuplot
\sbox{\plotpoint}{\rule[-0.200pt]{0.400pt}{0.400pt}}%
\put(176.0,113.0){\rule[-0.200pt]{4.818pt}{0.400pt}}
\put(154,113){\makebox(0,0)[r]{-0.75}}
\put(1416.0,113.0){\rule[-0.200pt]{4.818pt}{0.400pt}}
\put(176.0,257.0){\rule[-0.200pt]{4.818pt}{0.400pt}}
\put(154,257){\makebox(0,0)[r]{-0.7}}
\put(1416.0,257.0){\rule[-0.200pt]{4.818pt}{0.400pt}}
\put(176.0,401.0){\rule[-0.200pt]{4.818pt}{0.400pt}}
\put(154,401){\makebox(0,0)[r]{-0.65}}
\put(1416.0,401.0){\rule[-0.200pt]{4.818pt}{0.400pt}}
\put(176.0,544.0){\rule[-0.200pt]{4.818pt}{0.400pt}}
\put(154,544){\makebox(0,0)[r]{-0.6}}
\put(1416.0,544.0){\rule[-0.200pt]{4.818pt}{0.400pt}}
\put(176.0,688.0){\rule[-0.200pt]{4.818pt}{0.400pt}}
\put(154,688){\makebox(0,0)[r]{-0.55}}
\put(1416.0,688.0){\rule[-0.200pt]{4.818pt}{0.400pt}}
\put(176.0,832.0){\rule[-0.200pt]{4.818pt}{0.400pt}}
\put(154,832){\makebox(0,0)[r]{-0.5}}
\put(1416.0,832.0){\rule[-0.200pt]{4.818pt}{0.400pt}}
\put(226.0,113.0){\rule[-0.200pt]{0.400pt}{4.818pt}}
\put(226,68){\makebox(0,0){0}}
\put(226.0,812.0){\rule[-0.200pt]{0.400pt}{4.818pt}}
\put(428.0,113.0){\rule[-0.200pt]{0.400pt}{4.818pt}}
\put(428,68){\makebox(0,0){0.1}}
\put(428.0,812.0){\rule[-0.200pt]{0.400pt}{4.818pt}}
\put(630.0,113.0){\rule[-0.200pt]{0.400pt}{4.818pt}}
\put(630,68){\makebox(0,0){0.2}}
\put(630.0,812.0){\rule[-0.200pt]{0.400pt}{4.818pt}}
\put(831.0,113.0){\rule[-0.200pt]{0.400pt}{4.818pt}}
\put(831,68){\makebox(0,0){0.3}}
\put(831.0,812.0){\rule[-0.200pt]{0.400pt}{4.818pt}}
\put(1033.0,113.0){\rule[-0.200pt]{0.400pt}{4.818pt}}
\put(1033,68){\makebox(0,0){0.4}}
\put(1033.0,812.0){\rule[-0.200pt]{0.400pt}{4.818pt}}
\put(1234.0,113.0){\rule[-0.200pt]{0.400pt}{4.818pt}}
\put(1234,68){\makebox(0,0){0.5}}
\put(1234.0,812.0){\rule[-0.200pt]{0.400pt}{4.818pt}}
\put(1436.0,113.0){\rule[-0.200pt]{0.400pt}{4.818pt}}
\put(1436,68){\makebox(0,0){0.6}}
\put(1436.0,812.0){\rule[-0.200pt]{0.400pt}{4.818pt}}
\put(176.0,113.0){\rule[-0.200pt]{303.534pt}{0.400pt}}
\put(1436.0,113.0){\rule[-0.200pt]{0.400pt}{173.207pt}}
\put(176.0,832.0){\rule[-0.200pt]{303.534pt}{0.400pt}}
\put(806,23){\makebox(0,0){{\large\bf $m_q a$}}}
\put(806,877){\makebox(0,0){{\Large\bf $\mu_n/\mu_p$}}}
\put(176.0,113.0){\rule[-0.200pt]{0.400pt}{173.207pt}}
\put(1134,746){\makebox(0,0)[r]{CI}}
\put(1178,746){\circle{24}}
\put(226,498){\circle{24}}
\put(365,493){\circle{24}}
\put(463,470){\circle{24}}
\put(661,435){\circle{24}}
\put(1021,426){\circle{24}}
\put(1318,398){\circle{24}}
\put(1156.0,746.0){\rule[-0.200pt]{15.899pt}{0.400pt}}
\put(1156.0,736.0){\rule[-0.200pt]{0.400pt}{4.818pt}}
\put(1222.0,736.0){\rule[-0.200pt]{0.400pt}{4.818pt}}
\put(226.0,435.0){\rule[-0.200pt]{0.400pt}{30.594pt}}
\put(216.0,435.0){\rule[-0.200pt]{4.818pt}{0.400pt}}
\put(216.0,562.0){\rule[-0.200pt]{4.818pt}{0.400pt}}
\put(365.0,424.0){\rule[-0.200pt]{0.400pt}{33.244pt}}
\put(355.0,424.0){\rule[-0.200pt]{4.818pt}{0.400pt}}
\put(355.0,562.0){\rule[-0.200pt]{4.818pt}{0.400pt}}
\put(463.0,406.0){\rule[-0.200pt]{0.400pt}{30.594pt}}
\put(453.0,406.0){\rule[-0.200pt]{4.818pt}{0.400pt}}
\put(453.0,533.0){\rule[-0.200pt]{4.818pt}{0.400pt}}
\put(661.0,378.0){\rule[-0.200pt]{0.400pt}{27.703pt}}
\put(651.0,378.0){\rule[-0.200pt]{4.818pt}{0.400pt}}
\put(651.0,493.0){\rule[-0.200pt]{4.818pt}{0.400pt}}
\put(1021.0,418.0){\rule[-0.200pt]{0.400pt}{4.095pt}}
\put(1011.0,418.0){\rule[-0.200pt]{4.818pt}{0.400pt}}
\put(1011.0,435.0){\rule[-0.200pt]{4.818pt}{0.400pt}}
\put(1318.0,383.0){\rule[-0.200pt]{0.400pt}{6.986pt}}
\put(1308.0,383.0){\rule[-0.200pt]{4.818pt}{0.400pt}}
\put(1308.0,412.0){\rule[-0.200pt]{4.818pt}{0.400pt}}
\sbox{\plotpoint}{\rule[-0.500pt]{1.000pt}{1.000pt}}%
\put(1134,701){\makebox(0,0)[r]{-2/3}}
\multiput(1156,701)(20.756,0.000){4}{\usebox{\plotpoint}}
\put(1222,701){\usebox{\plotpoint}}
\put(176,353){\usebox{\plotpoint}}
\multiput(176,353)(20.756,0.000){61}{\usebox{\plotpoint}}
\put(1436,353){\usebox{\plotpoint}}
\sbox{\plotpoint}{\rule[-0.200pt]{0.400pt}{0.400pt}}%
\put(1134,656){\makebox(0,0)[r]{Expt.~~-0.685}}
\put(1156.0,656.0){\rule[-0.200pt]{15.899pt}{0.400pt}}
\put(176,300){\usebox{\plotpoint}}
\put(176.0,300.0){\rule[-0.200pt]{303.534pt}{0.400pt}}
\put(1134,611){\makebox(0,0)[r]{Sea + CI}}
\put(1178,611){\circle*{24}}
\put(226,291){\circle*{24}}
\put(365,418){\circle*{24}}
\put(463,429){\circle*{24}}
\put(661,429){\circle*{24}}
\put(1156.0,611.0){\rule[-0.200pt]{15.899pt}{0.400pt}}
\put(1156.0,601.0){\rule[-0.200pt]{0.400pt}{4.818pt}}
\put(1222.0,601.0){\rule[-0.200pt]{0.400pt}{4.818pt}}
\put(226.0,185.0){\rule[-0.200pt]{0.400pt}{51.312pt}}
\put(216.0,185.0){\rule[-0.200pt]{4.818pt}{0.400pt}}
\put(216.0,398.0){\rule[-0.200pt]{4.818pt}{0.400pt}}
\put(365.0,360.0){\rule[-0.200pt]{0.400pt}{27.703pt}}
\put(355.0,360.0){\rule[-0.200pt]{4.818pt}{0.400pt}}
\put(355.0,475.0){\rule[-0.200pt]{4.818pt}{0.400pt}}
\put(463.0,386.0){\rule[-0.200pt]{0.400pt}{20.958pt}}
\put(453.0,386.0){\rule[-0.200pt]{4.818pt}{0.400pt}}
\put(453.0,473.0){\rule[-0.200pt]{4.818pt}{0.400pt}}
\put(661.0,401.0){\rule[-0.200pt]{0.400pt}{13.731pt}}
\put(651.0,401.0){\rule[-0.200pt]{4.818pt}{0.400pt}}
\put(651.0,458.0){\rule[-0.200pt]{4.818pt}{0.400pt}}
\put(1134,566){\makebox(0,0)[r]{Valence}}
\put(1178,566){\makebox(0,0){$\triangle$}}
\put(366,366){\makebox(0,0){$\triangle$}}
\put(389,363){\makebox(0,0){$\triangle$}}
\put(413,380){\makebox(0,0){$\triangle$}}
\put(505,389){\makebox(0,0){$\triangle$}}
\put(1156.0,566.0){\rule[-0.200pt]{15.899pt}{0.400pt}}
\put(1156.0,556.0){\rule[-0.200pt]{0.400pt}{4.818pt}}
\put(1222.0,556.0){\rule[-0.200pt]{0.400pt}{4.818pt}}
\put(366.0,311.0){\rule[-0.200pt]{0.400pt}{26.499pt}}
\put(356.0,311.0){\rule[-0.200pt]{4.818pt}{0.400pt}}
\put(356.0,421.0){\rule[-0.200pt]{4.818pt}{0.400pt}}
\put(389.0,311.0){\rule[-0.200pt]{0.400pt}{25.054pt}}
\put(379.0,311.0){\rule[-0.200pt]{4.818pt}{0.400pt}}
\put(379.0,415.0){\rule[-0.200pt]{4.818pt}{0.400pt}}
\put(413.0,337.0){\rule[-0.200pt]{0.400pt}{20.958pt}}
\put(403.0,337.0){\rule[-0.200pt]{4.818pt}{0.400pt}}
\put(403.0,424.0){\rule[-0.200pt]{4.818pt}{0.400pt}}
\put(505.0,346.0){\rule[-0.200pt]{0.400pt}{20.717pt}}
\put(495.0,346.0){\rule[-0.200pt]{4.818pt}{0.400pt}}
\put(495.0,432.0){\rule[-0.200pt]{4.818pt}{0.400pt}}
\end{picture}
\caption{The magnetic moment ratio $\mu^n/\mu^p$ between neutron and
proton in VQCD and QCD are plotted against the dimensionless 
quark mass $m_q a$ from the strange to the charm region. $\bigtriangleup $ 
indicates the VQCD case and $\circ$/$\bullet$ indicates the C. I./ Sea + C. I.
in QCD case. The dashed line is the $SU(6)$ prediction of - 2/3.}
\end{figure}
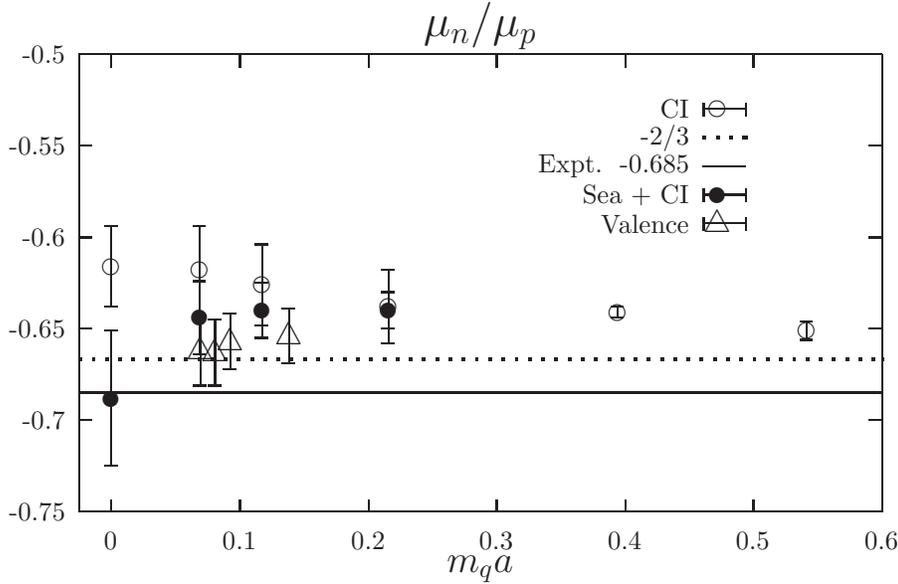

\section{Form Factors}

  Now, we revisit the form factors of the nucleon. We calculated the
isovector-axial form factor $g_A^3(q^2)$, the isoscalar-scalar
form factor $g_S^0(q^2)$, the proton electric form factor $G_E^p(q^2)$
and the isovector pseudoscalar form factor $g_P^3(q^2)$ in VQCD at
$\kappa = 0.162$, which corresponds to the quark mass of $\sim$ 120 MeV. They
are plotted in Fig. 24 as a function of $- q^2$. For comparison, we also
plot in Fig. 25 the corresponding form factors from QCD at $\kappa = 0.154$, 
which is about the same quark mass as in the VQCD case. 

We see that although these form factors in VQCD are still different among 
themselves, the differences are relatively smaller compared to those in 
QCD first of all, and secondly they are overall harder (except for
$g_S^0(q^2)$) , i.e. they fall off slower than
the corresponding ones in QCD. The most dramatic change is the pseudoscalar
form factor where the size as determined by 
\begin{equation}
\langle r^2 \rangle = - 6 \frac{d F(q^2)}{d(- q^2)}|_{q^2 = 0},
\end{equation}
is reduced by about a factor of two. This is consistent with the pseudoscalar
meson-dominance picture in Fig. 11, where the pseudoscalar form factor in QCD 
is dominated by the pion which in turn couples to
the baryon through the $\pi NN$ vertex. Yet, this meson `cloud'
is removed in VQCD by prohibiting the pair creation. In this case, the current 
couples directly to the quarks and consequently the $\langle r^2 \rangle$ 
of the hadron becomes smaller. To a lesser extent, similar situations happen 
in the vector and axial channels. This is again an indirect way of 
visualizing the effects of the meson clouds in the C. I. of QCD. 

\begin{figure}[h]
\input{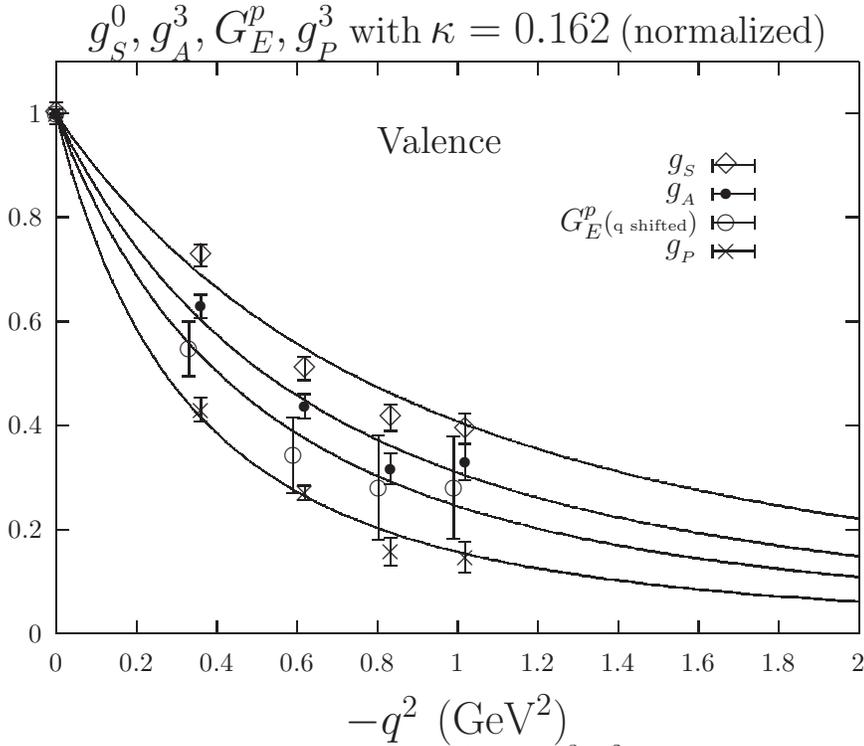}
\caption{The isovector axial form factor $g_A^3(q^2)$, the isoscalar scalar
form factor $g_S^0(q^2)$, the proton electric form factor $G_E^p(q^2)$
and the isovector pseudoscalar form factor $g_P^3(q^2)$ in VQCD at
$\kappa = 0.162$ which corresponds to the quark mass of $\sim$ 120 MeV are
plotted in terms of $-q^2$. They are normalized at $q^2 = 0$ to 1 in order
to compare their $q^2$ dependence.}
\end{figure}

\newpage
\clearpage
\begin{figure}[h]
\input{fig25.tex}
\caption{For comparison, the same $g_A^3(q^2)$, $g_S^0(q^2)$, $G_E^p(q^2)$,
and $g_P^3(q^2)$ in QCD at $\kappa = 0.154$ which is at about the same
quark mass, i.e. $\sim$ 120 MeV are plotted in terms of $-q^2$. They are 
also normalized to 1 at $q^2 = 0$.}
\setlength{\unitlength}{0.240900pt}
\ifx\plotpoint\undefined\newsavebox{\plotpoint}\fi
\sbox{\plotpoint}{\rule[-0.200pt]{0.400pt}{0.400pt}}%
\begin{picture}(1500,1080)(0,0)
\font\gnuplot=cmr10 at 10pt
\gnuplot
\sbox{\plotpoint}{\rule[-0.200pt]{0.400pt}{0.400pt}}%
\put(176.0,377.0){\rule[-0.200pt]{303.534pt}{0.400pt}}
\put(176.0,166.0){\rule[-0.200pt]{4.818pt}{0.400pt}}
\put(154,166){\makebox(0,0)[r]{-0.04}}
\put(1416.0,166.0){\rule[-0.200pt]{4.818pt}{0.400pt}}
\put(176.0,272.0){\rule[-0.200pt]{4.818pt}{0.400pt}}
\put(154,272){\makebox(0,0)[r]{-0.02}}
\put(1416.0,272.0){\rule[-0.200pt]{4.818pt}{0.400pt}}
\put(176.0,377.0){\rule[-0.200pt]{4.818pt}{0.400pt}}
\put(154,377){\makebox(0,0)[r]{0}}
\put(1416.0,377.0){\rule[-0.200pt]{4.818pt}{0.400pt}}
\put(176.0,483.0){\rule[-0.200pt]{4.818pt}{0.400pt}}
\put(154,483){\makebox(0,0)[r]{0.02}}
\put(1416.0,483.0){\rule[-0.200pt]{4.818pt}{0.400pt}}
\put(176.0,589.0){\rule[-0.200pt]{4.818pt}{0.400pt}}
\put(154,589){\makebox(0,0)[r]{0.04}}
\put(1416.0,589.0){\rule[-0.200pt]{4.818pt}{0.400pt}}
\put(176.0,695.0){\rule[-0.200pt]{4.818pt}{0.400pt}}
\put(154,695){\makebox(0,0)[r]{0.06}}
\put(1416.0,695.0){\rule[-0.200pt]{4.818pt}{0.400pt}}
\put(176.0,800.0){\rule[-0.200pt]{4.818pt}{0.400pt}}
\put(154,800){\makebox(0,0)[r]{0.08}}
\put(1416.0,800.0){\rule[-0.200pt]{4.818pt}{0.400pt}}
\put(176.0,906.0){\rule[-0.200pt]{4.818pt}{0.400pt}}
\put(154,906){\makebox(0,0)[r]{0.1}}
\put(1416.0,906.0){\rule[-0.200pt]{4.818pt}{0.400pt}}
\put(176.0,1012.0){\rule[-0.200pt]{4.818pt}{0.400pt}}
\put(154,1012){\makebox(0,0)[r]{0.12}}
\put(1416.0,1012.0){\rule[-0.200pt]{4.818pt}{0.400pt}}
\put(190.0,113.0){\rule[-0.200pt]{0.400pt}{4.818pt}}
\put(190,68){\makebox(0,0){0}}
\put(190.0,992.0){\rule[-0.200pt]{0.400pt}{4.818pt}}
\put(328.0,113.0){\rule[-0.200pt]{0.400pt}{4.818pt}}
\put(328,68){\makebox(0,0){0.2}}
\put(328.0,992.0){\rule[-0.200pt]{0.400pt}{4.818pt}}
\put(467.0,113.0){\rule[-0.200pt]{0.400pt}{4.818pt}}
\put(467,68){\makebox(0,0){0.4}}
\put(467.0,992.0){\rule[-0.200pt]{0.400pt}{4.818pt}}
\put(605.0,113.0){\rule[-0.200pt]{0.400pt}{4.818pt}}
\put(605,68){\makebox(0,0){0.6}}
\put(605.0,992.0){\rule[-0.200pt]{0.400pt}{4.818pt}}
\put(744.0,113.0){\rule[-0.200pt]{0.400pt}{4.818pt}}
\put(744,68){\makebox(0,0){0.8}}
\put(744.0,992.0){\rule[-0.200pt]{0.400pt}{4.818pt}}
\put(882.0,113.0){\rule[-0.200pt]{0.400pt}{4.818pt}}
\put(882,68){\makebox(0,0){1}}
\put(882.0,992.0){\rule[-0.200pt]{0.400pt}{4.818pt}}
\put(1021.0,113.0){\rule[-0.200pt]{0.400pt}{4.818pt}}
\put(1021,68){\makebox(0,0){1.2}}
\put(1021.0,992.0){\rule[-0.200pt]{0.400pt}{4.818pt}}
\put(1159.0,113.0){\rule[-0.200pt]{0.400pt}{4.818pt}}
\put(1159,68){\makebox(0,0){1.4}}
\put(1159.0,992.0){\rule[-0.200pt]{0.400pt}{4.818pt}}
\put(1298.0,113.0){\rule[-0.200pt]{0.400pt}{4.818pt}}
\put(1298,68){\makebox(0,0){1.6}}
\put(1298.0,992.0){\rule[-0.200pt]{0.400pt}{4.818pt}}
\put(1436.0,113.0){\rule[-0.200pt]{0.400pt}{4.818pt}}
\put(1436,68){\makebox(0,0){1.8}}
\put(1436.0,992.0){\rule[-0.200pt]{0.400pt}{4.818pt}}
\put(176.0,113.0){\rule[-0.200pt]{303.534pt}{0.400pt}}
\put(1436.0,113.0){\rule[-0.200pt]{0.400pt}{216.569pt}}
\put(176.0,1012.0){\rule[-0.200pt]{303.534pt}{0.400pt}}
\put(806,-22){\makebox(0,0){{\Large\bf $-q^2$~(GeV$^2$)}}}
\put(806,1057){\makebox(0,0){{\Large\bf $G_E^n(q^2)$}}}
\put(176.0,113.0){\rule[-0.200pt]{0.400pt}{216.569pt}}
\put(1228,959){\makebox(0,0)[r]{valence}}
\put(1272,959){\circle*{24}}
\put(190,377){\circle*{24}}
\put(447,457){\circle*{24}}
\put(638,515){\circle*{24}}
\put(797,467){\circle*{24}}
\put(935,726){\circle*{24}}
\put(1250.0,959.0){\rule[-0.200pt]{15.899pt}{0.400pt}}
\put(1250.0,949.0){\rule[-0.200pt]{0.400pt}{4.818pt}}
\put(1316.0,949.0){\rule[-0.200pt]{0.400pt}{4.818pt}}
\put(190,377){\usebox{\plotpoint}}
\put(180.0,377.0){\rule[-0.200pt]{4.818pt}{0.400pt}}
\put(180.0,377.0){\rule[-0.200pt]{4.818pt}{0.400pt}}
\put(447.0,393.0){\rule[-0.200pt]{0.400pt}{30.594pt}}
\put(437.0,393.0){\rule[-0.200pt]{4.818pt}{0.400pt}}
\put(437.0,520.0){\rule[-0.200pt]{4.818pt}{0.400pt}}
\put(638.0,436.0){\rule[-0.200pt]{0.400pt}{38.062pt}}
\put(628.0,436.0){\rule[-0.200pt]{4.818pt}{0.400pt}}
\put(628.0,594.0){\rule[-0.200pt]{4.818pt}{0.400pt}}
\put(797.0,362.0){\rule[-0.200pt]{0.400pt}{50.830pt}}
\put(787.0,362.0){\rule[-0.200pt]{4.818pt}{0.400pt}}
\put(787.0,573.0){\rule[-0.200pt]{4.818pt}{0.400pt}}
\put(935.0,525.0){\rule[-0.200pt]{0.400pt}{96.842pt}}
\put(925.0,525.0){\rule[-0.200pt]{4.818pt}{0.400pt}}
\put(925.0,927.0){\rule[-0.200pt]{4.818pt}{0.400pt}}
\put(1228,914){\makebox(0,0)[r]{quenched}}
\put(1272,914){\circle{24}}
\put(190,378){\circle{24}}
\put(478,531){\circle{24}}
\put(720,473){\circle{24}}
\put(932,494){\circle{24}}
\put(1123,621){\circle{24}}
\put(1250.0,914.0){\rule[-0.200pt]{15.899pt}{0.400pt}}
\put(1250.0,904.0){\rule[-0.200pt]{0.400pt}{4.818pt}}
\put(1316.0,904.0){\rule[-0.200pt]{0.400pt}{4.818pt}}
\put(190.0,377.0){\usebox{\plotpoint}}
\put(180.0,377.0){\rule[-0.200pt]{4.818pt}{0.400pt}}
\put(180.0,378.0){\rule[-0.200pt]{4.818pt}{0.400pt}}
\put(478.0,467.0){\rule[-0.200pt]{0.400pt}{30.594pt}}
\put(468.0,467.0){\rule[-0.200pt]{4.818pt}{0.400pt}}
\put(468.0,594.0){\rule[-0.200pt]{4.818pt}{0.400pt}}
\put(720.0,399.0){\rule[-0.200pt]{0.400pt}{35.653pt}}
\put(710.0,399.0){\rule[-0.200pt]{4.818pt}{0.400pt}}
\put(710.0,547.0){\rule[-0.200pt]{4.818pt}{0.400pt}}
\put(932.0,330.0){\rule[-0.200pt]{0.400pt}{79.015pt}}
\put(922.0,330.0){\rule[-0.200pt]{4.818pt}{0.400pt}}
\put(922.0,658.0){\rule[-0.200pt]{4.818pt}{0.400pt}}
\put(1123.0,377.0){\rule[-0.200pt]{0.400pt}{117.318pt}}
\put(1113.0,377.0){\rule[-0.200pt]{4.818pt}{0.400pt}}
\put(1113.0,864.0){\rule[-0.200pt]{4.818pt}{0.400pt}}
\end{picture}
\caption{The neutron electric form factor $G_E^n(q^2)$ for VQCD ($\bullet$) at
$\kappa = 0.162$ is compared with the QCD result ($\circ$)
at $\kappa = 0.154$.}
\end{figure}

\newpage
\clearpage
We also plot the neutron electric form factor $G_E^n(q^2)$ for VQCD at	
$\kappa =0.162$ and its counter part (C. I. QCD at $\kappa = 0.154$) in 
Fig. 26. We see that these two results are comparable in size and indicate
that there are still some spin-spin correlation between the quarks in VQCD
which breaks the $SU(6)$ symmetry.

\subsection{Goldberger-Treiman Relation and Vector Dominance}

There are several interesting aspects to observe in VQCD. Since the
axial Ward identities in Eqs. (\ref{axi1}) and (\ref{axi2}) are associated
with the current involving both the u and v fields, they are applicable only
to meson states with the creation and annihilation of quark-antiquark pairs.
Thus, the identities are useful in addressing the relation of the `pion' mass
and decay constant with the quark mass as PCAC is in QCD. On the other hand,
it does not apply to baryons where only the quarks are involved. For 
example, the pseudoscalar current matrix element between the nucleon states 
does not have the pion pole as evidenced in Fig. 24. Consequently, 
there is no Goldberger-Treiman relation in VQCD.
Conversely, the conserved vector current in Eq. (\ref{cvc}) between the
baryons and meson states leads to separately conserved quark and 
antiquark numbers. This entails the three-point function calculation as
illustrated in Fig. 5(a). Yet, it does not apply to situations involving
quark-antiquark creations or annihilations. For example, the vector meson
decay constant (in the continuum) is determined by the matrix element of the 
vector current between the vector meson state and the vacuum
\begin{equation}  \label{f_v}
\langle 0|V_{\mu}|v(k)\rangle = \frac{m_v^2 \epsilon_{\mu} e^{-ik\cdot x}}
{f_v},
\end{equation}
where $\epsilon_{\mu}$ is the polarization vector, $m_v$ is the vector-meson
mass, and $f_v$ is the decay constant. Due to the vector current conservation
$\partial_{\mu} V_{\mu} = 0$, the transversality condition $k_{\mu}\cdot
\epsilon_{\mu} = 0$ is derived in QCD. However in VQCD, the CVC relation in
Eq. (\ref{cvc}) does not apply to Eq. (\ref{f_v}) because the conserved
vector current in Eq. (\ref{vc}) does not have the pair annihilation term
$\bar{v}\gamma_{\mu}u$. Similarly, there is no vector dominance in the pion and nucleon
EM form factors. As discussed in the preceeding section, there should be
no meson dominance in form factors in VQCD. 

More generally, one can say that there
is neither crossing relation, dispersion relation, nor unitarity in VQCD.
But these features, or the lack of them, are shared by the valence quark
model that we are trying to emulate.

\subsection{Quark Condensate and Symmetry Breaking}

The quark condensate $\langle\bar{\Psi}\Psi\rangle$ in QCD is the order
parameter of chiral symmetry breaking. In VQCD, there are two quark 
condensates $\langle \bar{u}u\rangle$ and $\langle \bar{v}v\rangle$. Although
VQCD eliminates the quark loops in the time direction, it still permits
quark loops in the spatial direction. One expects that   
\begin{equation}
|\langle \bar{u}u\rangle| = |\langle \bar{v}v\rangle| < 
|\langle\bar{\Psi}\Psi\rangle|
\end{equation}
due to the fact that $\langle \bar{u}u\rangle$ and $\langle \bar{v}v\rangle$
have null measure in the time direction. Unfortunately there is an 
subtraction in the lattice calculation of quark condensate due to a
contact term for the Wilson action~\cite{bmm85}. Short of a precise control of
this subtraction, we are not able to have a fair comparison between 
$\langle \bar{u}u\rangle$ and $\langle\bar{\Psi}\Psi\rangle$.

Nevertheless,
$\langle \bar{u}u\rangle$ and $\langle \bar{v}v\rangle$ are nonzero which
indicates that, to the extent that they serve as the order parameter of 
axial symmetry breaking as suggested by the existence of the Goldstone
bosons and non-vanishing $f_{\pi}$ in the last section, the symmetry breaking
seems to be weaker than in QCD. It is shown in a Schwinger-Dyson equation
study~\cite{mr97} recently that the pseudoscalar meson grows either linearly or
as the square root of the quark mass depending on whether it is large or small
compared to a scale set by the quark condensate $\langle\bar{\Psi}\Psi\rangle$.
The linear dependence 
we see between the pion mass and the quark mass in Fig. 19 may mean
that the quark masses we are calculating are still larger than the scale
set by the quark condensate  $\langle \bar{u}u\rangle$ and 
$\langle \bar{v}v\rangle$ and the quadratic pion mass dependence of the
quark mass may yet to set in at smaller quark masses. Either way, the
nonzero $\langle \bar{u}u\rangle$ and $\langle \bar{v}v\rangle$ supports
the spontaneous axial symmetry breaking scenario with two Goldstone pions
as alluded to in section \ref{pi_sym}.


\section{Hadron Spectroscopy}  \label{had_spc}

 To further explore the consequences of the valence approximation,
we study the hadron masses. Since hadron masses entail calculations
of two-point functions, the sea quarks do not appear explicitly as they
do in three-point functions (see Fig. 5(b)). The only exception is
the flavor-singlet meson where the D. I. (Fig. 4(b)) is part of the
meson propagator. The implicit sea quark effects in the loops which manifest
themselves through the fermion determinant are known to affect the
scaling~\cite{sw97}, the topological susceptibility, phase transition, 
the $\eta'$ mass, and the slope of the hadron mass with respect to the 
quark mass~\cite{dll96,ceh96}. 

Here, we shall concentrate on the effects of the cloud quarks on hadron masses
which are practically unknown. We first plot in Fig. 27 the masses of
$\Delta, N, \rho$, and $\pi$ as a function of the quark mass 
$ma = \ln (4\kappa_c/\kappa -3)$ on
our lattice with quenched approximation. We see that the hyper-fine
splittings between the $\Delta$ and N, and the $\rho$ and $\pi$
grow when the quark mass approaches the chiral limit as expected.  

In the
infinite volume and continuum limits, it is found~\cite{bcs93} that using 
$m_{\rho}$ to set the scale, the $K^{\*}$, $\Phi$ mesons and the
octet and decuplet baryon masses are all within about 6\% of the experimental
results.
  
\begin{figure}[h]
\setlength{\unitlength}{0.240900pt}
\ifx\plotpoint\undefined\newsavebox{\plotpoint}\fi
\sbox{\plotpoint}{\rule[-0.200pt]{0.400pt}{0.400pt}}%
\begin{picture}(1500,1350)(0,0)
\font\gnuplot=cmr10 at 10pt
\gnuplot
\sbox{\plotpoint}{\rule[-0.200pt]{0.400pt}{0.400pt}}%
\put(220.0,113.0){\rule[-0.200pt]{4.818pt}{0.400pt}}
\put(198,113){\makebox(0,0)[r]{0}}
\put(1416.0,113.0){\rule[-0.200pt]{4.818pt}{0.400pt}}
\put(220.0,293.0){\rule[-0.200pt]{4.818pt}{0.400pt}}
\put(198,293){\makebox(0,0)[r]{0.2}}
\put(1416.0,293.0){\rule[-0.200pt]{4.818pt}{0.400pt}}
\put(220.0,473.0){\rule[-0.200pt]{4.818pt}{0.400pt}}
\put(198,473){\makebox(0,0)[r]{0.4}}
\put(1416.0,473.0){\rule[-0.200pt]{4.818pt}{0.400pt}}
\put(220.0,653.0){\rule[-0.200pt]{4.818pt}{0.400pt}}
\put(198,653){\makebox(0,0)[r]{0.6}}
\put(1416.0,653.0){\rule[-0.200pt]{4.818pt}{0.400pt}}
\put(220.0,832.0){\rule[-0.200pt]{4.818pt}{0.400pt}}
\put(198,832){\makebox(0,0)[r]{0.8}}
\put(1416.0,832.0){\rule[-0.200pt]{4.818pt}{0.400pt}}
\put(220.0,1012.0){\rule[-0.200pt]{4.818pt}{0.400pt}}
\put(198,1012){\makebox(0,0)[r]{1}}
\put(1416.0,1012.0){\rule[-0.200pt]{4.818pt}{0.400pt}}
\put(220.0,1192.0){\rule[-0.200pt]{4.818pt}{0.400pt}}
\put(198,1192){\makebox(0,0)[r]{1.2}}
\put(1416.0,1192.0){\rule[-0.200pt]{4.818pt}{0.400pt}}
\put(220.0,113.0){\rule[-0.200pt]{0.400pt}{4.818pt}}
\put(220,68){\makebox(0,0){0}}
\put(220.0,1262.0){\rule[-0.200pt]{0.400pt}{4.818pt}}
\put(463.0,113.0){\rule[-0.200pt]{0.400pt}{4.818pt}}
\put(463,68){\makebox(0,0){0.05}}
\put(463.0,1262.0){\rule[-0.200pt]{0.400pt}{4.818pt}}
\put(706.0,113.0){\rule[-0.200pt]{0.400pt}{4.818pt}}
\put(706,68){\makebox(0,0){0.1}}
\put(706.0,1262.0){\rule[-0.200pt]{0.400pt}{4.818pt}}
\put(950.0,113.0){\rule[-0.200pt]{0.400pt}{4.818pt}}
\put(950,68){\makebox(0,0){0.15}}
\put(950.0,1262.0){\rule[-0.200pt]{0.400pt}{4.818pt}}
\put(1193.0,113.0){\rule[-0.200pt]{0.400pt}{4.818pt}}
\put(1193,68){\makebox(0,0){0.2}}
\put(1193.0,1262.0){\rule[-0.200pt]{0.400pt}{4.818pt}}
\put(1436.0,113.0){\rule[-0.200pt]{0.400pt}{4.818pt}}
\put(1436,68){\makebox(0,0){0.25}}
\put(1436.0,1262.0){\rule[-0.200pt]{0.400pt}{4.818pt}}
\put(220.0,113.0){\rule[-0.200pt]{292.934pt}{0.400pt}}
\put(1436.0,113.0){\rule[-0.200pt]{0.400pt}{281.612pt}}
\put(220.0,1282.0){\rule[-0.200pt]{292.934pt}{0.400pt}}
\put(45,697){\makebox(0,0){{\Large\bf $M a$}}}
\put(828,-22){\makebox(0,0){{\Large\bf $m_q a$}}}
\put(828,1327){\makebox(0,0){{\Large\bf Quenched}}}
\put(220.0,113.0){\rule[-0.200pt]{0.400pt}{281.612pt}}
\put(415,1183){\makebox(0,0)[r]{{\large\bf $\Delta$}}}
\put(459,1183){\makebox(0,0){$\triangle$}}
\put(554,845){\makebox(0,0){$\triangle$}}
\put(791,956){\makebox(0,0){$\triangle$}}
\put(1251,1174){\makebox(0,0){$\triangle$}}
\put(437.0,1183.0){\rule[-0.200pt]{15.899pt}{0.400pt}}
\put(437.0,1173.0){\rule[-0.200pt]{0.400pt}{4.818pt}}
\put(503.0,1173.0){\rule[-0.200pt]{0.400pt}{4.818pt}}
\put(554.0,817.0){\rule[-0.200pt]{0.400pt}{13.490pt}}
\put(544.0,817.0){\rule[-0.200pt]{4.818pt}{0.400pt}}
\put(544.0,873.0){\rule[-0.200pt]{4.818pt}{0.400pt}}
\put(791.0,935.0){\rule[-0.200pt]{0.400pt}{9.877pt}}
\put(781.0,935.0){\rule[-0.200pt]{4.818pt}{0.400pt}}
\put(781.0,976.0){\rule[-0.200pt]{4.818pt}{0.400pt}}
\put(1251.0,1148.0){\rule[-0.200pt]{0.400pt}{12.527pt}}
\put(1241.0,1148.0){\rule[-0.200pt]{4.818pt}{0.400pt}}
\put(1241.0,1200.0){\rule[-0.200pt]{4.818pt}{0.400pt}}
\put(415,1138){\makebox(0,0)[r]{{\large\bf $N$}}}
\put(459,1138){\circle*{24}}
\put(554,777){\circle*{24}}
\put(791,906){\circle*{24}}
\put(1251,1144){\circle*{24}}
\put(437.0,1138.0){\rule[-0.200pt]{15.899pt}{0.400pt}}
\put(437.0,1128.0){\rule[-0.200pt]{0.400pt}{4.818pt}}
\put(503.0,1128.0){\rule[-0.200pt]{0.400pt}{4.818pt}}
\put(554.0,762.0){\rule[-0.200pt]{0.400pt}{6.986pt}}
\put(544.0,762.0){\rule[-0.200pt]{4.818pt}{0.400pt}}
\put(544.0,791.0){\rule[-0.200pt]{4.818pt}{0.400pt}}
\put(791.0,895.0){\rule[-0.200pt]{0.400pt}{5.300pt}}
\put(781.0,895.0){\rule[-0.200pt]{4.818pt}{0.400pt}}
\put(781.0,917.0){\rule[-0.200pt]{4.818pt}{0.400pt}}
\put(1251.0,1135.0){\rule[-0.200pt]{0.400pt}{4.095pt}}
\put(1241.0,1135.0){\rule[-0.200pt]{4.818pt}{0.400pt}}
\put(1241.0,1152.0){\rule[-0.200pt]{4.818pt}{0.400pt}}
\put(415,1093){\makebox(0,0)[r]{{\large\bf $\rho$}}}
\put(459,1093){\circle{24}}
\put(554,537){\circle{24}}
\put(791,612){\circle{24}}
\put(1251,761){\circle{24}}
\put(437.0,1093.0){\rule[-0.200pt]{15.899pt}{0.400pt}}
\put(437.0,1083.0){\rule[-0.200pt]{0.400pt}{4.818pt}}
\put(503.0,1083.0){\rule[-0.200pt]{0.400pt}{4.818pt}}
\put(554.0,531.0){\rule[-0.200pt]{0.400pt}{3.132pt}}
\put(544.0,531.0){\rule[-0.200pt]{4.818pt}{0.400pt}}
\put(544.0,544.0){\rule[-0.200pt]{4.818pt}{0.400pt}}
\put(791.0,607.0){\rule[-0.200pt]{0.400pt}{2.409pt}}
\put(781.0,607.0){\rule[-0.200pt]{4.818pt}{0.400pt}}
\put(781.0,617.0){\rule[-0.200pt]{4.818pt}{0.400pt}}
\put(1251.0,757.0){\rule[-0.200pt]{0.400pt}{2.168pt}}
\put(1241.0,757.0){\rule[-0.200pt]{4.818pt}{0.400pt}}
\put(1241.0,766.0){\rule[-0.200pt]{4.818pt}{0.400pt}}
\put(415,1048){\makebox(0,0)[r]{{\large\bf $\pi$}}}
\put(459,1048){\raisebox{-.8pt}{\makebox(0,0){$\Box$}}}
\put(554,451){\raisebox{-.8pt}{\makebox(0,0){$\Box$}}}
\put(791,550){\raisebox{-.8pt}{\makebox(0,0){$\Box$}}}
\put(1251,724){\raisebox{-.8pt}{\makebox(0,0){$\Box$}}}
\put(437.0,1048.0){\rule[-0.200pt]{15.899pt}{0.400pt}}
\put(437.0,1038.0){\rule[-0.200pt]{0.400pt}{4.818pt}}
\put(503.0,1038.0){\rule[-0.200pt]{0.400pt}{4.818pt}}
\put(554.0,446.0){\rule[-0.200pt]{0.400pt}{2.650pt}}
\put(544.0,446.0){\rule[-0.200pt]{4.818pt}{0.400pt}}
\put(544.0,457.0){\rule[-0.200pt]{4.818pt}{0.400pt}}
\put(791.0,546.0){\rule[-0.200pt]{0.400pt}{2.168pt}}
\put(781.0,546.0){\rule[-0.200pt]{4.818pt}{0.400pt}}
\put(781.0,555.0){\rule[-0.200pt]{4.818pt}{0.400pt}}
\put(1251.0,720.0){\rule[-0.200pt]{0.400pt}{1.686pt}}
\put(1241.0,720.0){\rule[-0.200pt]{4.818pt}{0.400pt}}
\put(1241.0,727.0){\rule[-0.200pt]{4.818pt}{0.400pt}}
\end{picture}
\caption{The dimensionless $\Delta, N, \rho$, and $\pi$ masses in quenched
QCD are plotted 
as a function of the quark mass $m_qa = \ln (4\kappa_c/\kappa -3)$. The pion
mass is proportional to $\sqrt{m_qa}$, while the others are extrapolated to
the chiral limit with a linear m dependence.}
\end{figure}
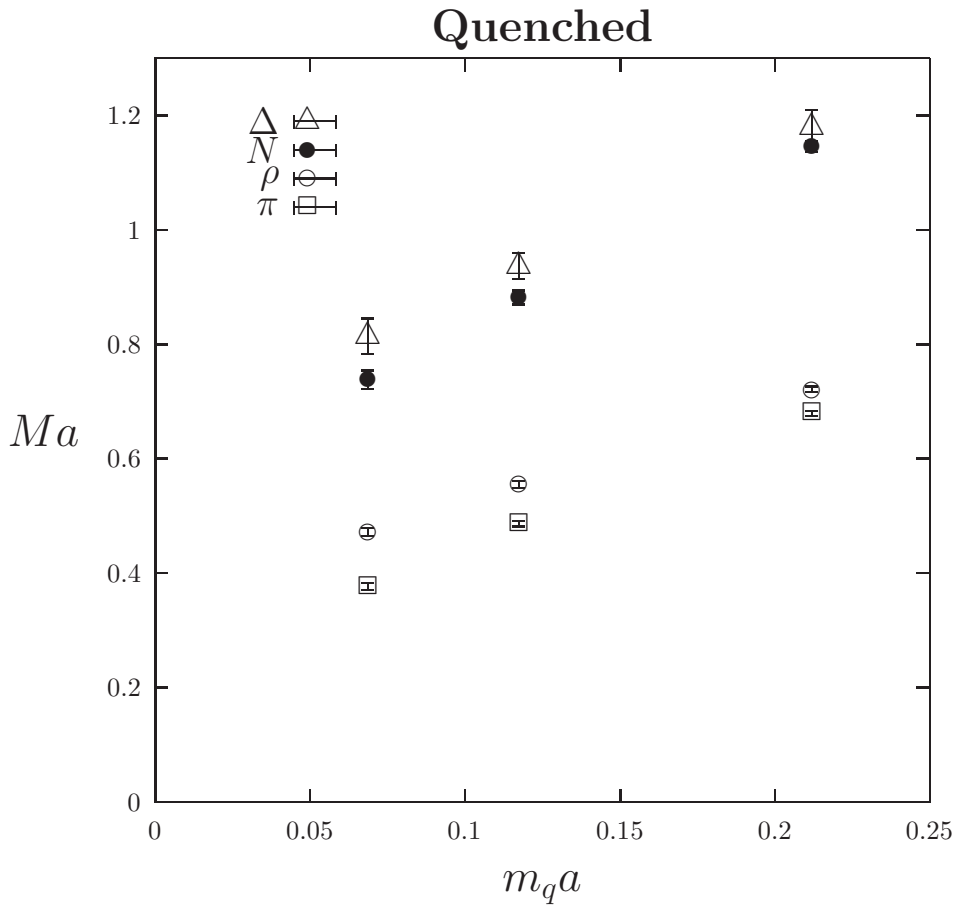    

Next we plot in Fig. 28 the masses of
$\Delta, N, \rho$, and $\pi$ as a function of the quark mass 
$m_qa = \ln (4\kappa_c/\kappa -3)$ ($\kappa_c = 0.1649$ in this case) 
on the same quenched lattice. It is to our surprise that the truncation
of the Z-graphs appears to have a dramatic effect on all these meson and 
baryon masses. 

\newpage
\clearpage
\begin{figure}[h]
\input{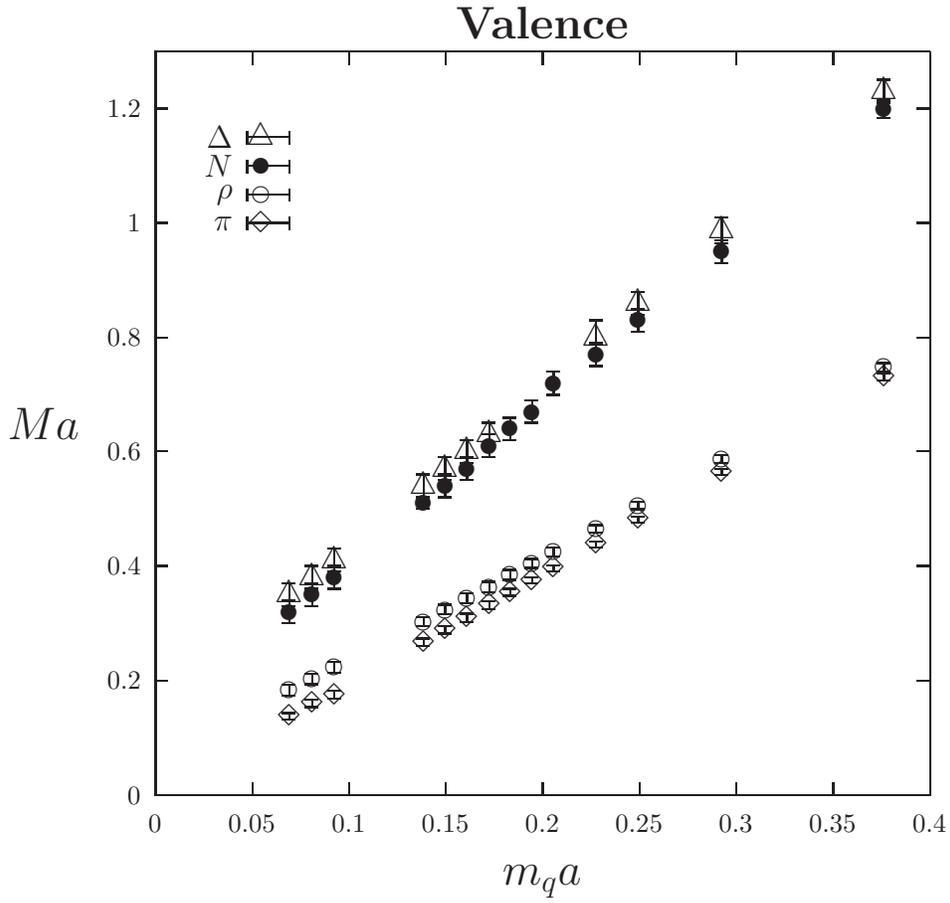}
\caption{The dimensionless $\Delta, N, \rho$, and $\pi$ masses in VQCD are
plotted as a function of the quark mass $m_qa = \ln (4\kappa_c/\kappa -3)$.
All the masses are extrapolated to the zero quark mass limit with a linear
$m_q$ dependence.}
\end{figure}

\newpage
\clearpage
First of all, we notice that the $\Delta$ and the nucleon are on top of
each other within errors all the way down to the smallest quark mass 
around the strange quark range. Thus, the hyper-fine splitting is largely 
gone in VQCD. This is true also between the $\rho$ and $\pi$.
Extrapolating to the zero quark mass limit, the $\rho$ mass $m_{\rho}a$ is 
0.054(8). With $a^{-1} = 1.75$ GeV, $m_{\rho} = 95(14)$ MeV in VQCD. This is
a factor of 6.4 smaller than that in the quenched approximation which
gives $m_{\rho}a = 0.343(6)$. Secondly, we see that $\Delta$, N, and $\rho$
all dropped their masses greatly compared to those in QCD (Fig. 27). The
extrapolated $m_{\Delta} a = 0.102(14), m_{N} a = 0.074(11)$ at the zero 
quark mass limit. They are much smaller than their corresponding values of
0.638(41) and 0.536(13) in the quenched QCD calculation in Fig 27. Furthermore, the
hyper-fine splitting between $\Delta$ and N is now 49(7) MeV which is
$\sim 3.7$ times smaller than our quenched result of 179(11) MeV~\footnote
{Our quenched result is smaller than the experimental $\Delta$-N splitting
of 298 MeV mainly due to the fact that our results are not in the infinite
volume and continuum limits. It is shown that when these limits are taken,
the octet and decuplet baryons are within 6\% of the experimental values
~\cite{bcs93}.}.
 
For a more direct comparison to see how the nucleon and $\Delta$ masses drop, 
we plot the
nucleon and $\Delta$ masses in VQCD and quenched QCD on the same figure.
We see in Fig. 29 that the nucleon moves down from VQCD to quenched QCD case
by about a constant amount 
$\sim 0.4 $ which is about 700 MeV. Similarly, the $\Delta$ mass drops
by the same amount for heavier quarks. For quarks around the strange, it
drops further to meet the nucleon. We also plot the $\rho$ and $\pi$ masses
in VQCD and quenched QCD on Fig. 30. Analogous to the N - $\Delta$ situation, 
the vector meson drops by about
an equal amount $\sim 0.31$ or 537 MeV; whereas, the pseudoscalar meson 
drops about 0.22 or 380 MeV in the strange region and
approaches zero in both the quenched QCD and VQCD cases.

\newpage
\clearpage
\begin{figure}[h]
\input{fig29.tex}
\caption{The dimensionless N and $\Delta$ masses from QCD are compared with
those in VQCD as a function of the quark mass.}
\input{fig30.tex}
\caption{The dimensionless $\pi$ and $\rho$ masses from QCD are compared
with those in VQCD as a function of the quark mass.}
\end{figure}

\newpage
\clearpage
Shown in Fig. 31 are the $a_1$ and $a_0$ mesons calculated in quenched QCD
and VQCD. We see that both mesons come down in mass from QCD to VQCD 
by a large amount. $a_1$ appears to be degenerate with $\pi$ and $\rho$
in VQCD over the range of the quark mass in Fig. 31 and Fig. 28. However,
we cannot be certain about this point, especially in view of the fact that
the errors on $a_1$ for the three lightest quarks are quite large. 
$a_0$, on the other hand, seems to be higher than the pion in this
range of the quark mass. 

\begin{figure}[h]
\input{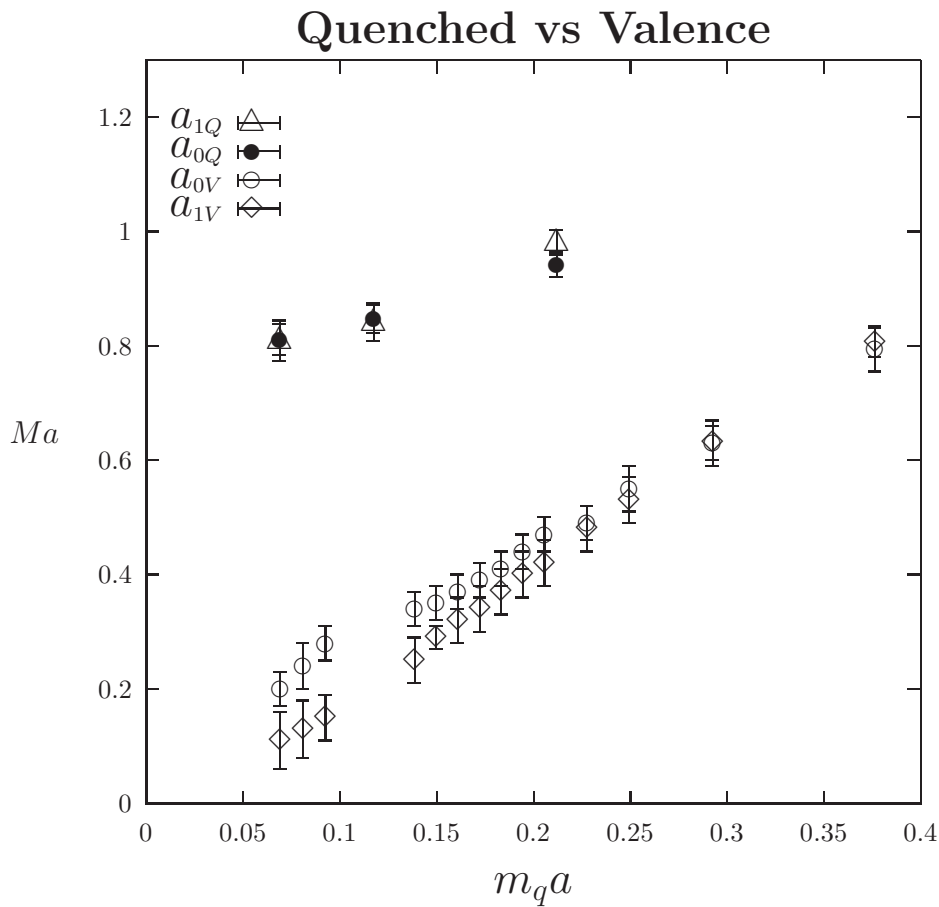}
\caption{The dimensionless $a_1$ and $a_0$ masses from QCD are compared with
those in VQCD as a function of the quark mass.}
\end{figure}    

\newpage
\clearpage
\subsection{$\pi \rho$ correlation}

   There is a concern about the near degeneracy of N and $\Delta$ and
$\rho$ and $\pi$. Since VQCD is not Lorentz invariant, is it possible that
the interpolation field $\bar{u}\gamma_i v$ intended for generating $\rho$
may project onto a $\pi$ state as well and causes the vector correlation 
function $\langle \bar{v}\gamma_i^{\dagger} u (t) \bar{u}\gamma_i v (0)
\rangle$ have
the same asymptotic fall off as that of the pseudoscalar correlation function
$\langle \bar{v}\gamma_5^{\dagger} u (t) \bar{u}\gamma_5 v (0)\rangle$ ? 
To check this possibility, we calculate the cross correlation functions
among the vector (V), pseudoscalar (P), axial (A), and scalar (S) 
interpolation fields. We see in Fig. 32 the cross correlation functions 
$\langle V(t) P(0) \rangle, \langle A(t) P(0) \rangle$, and
$\langle S(t) P(0) \rangle$ have essentially zero overlap over the range
of the Euclidean time t. Also shown in Fig. 33 are the cross correlation
functions $\langle A(t) V(0) \rangle, \langle S(t) V(0) \rangle$,
and $\langle P(t) V(0) \rangle$. Again the overlaps are zero and the 
signals are cleaner. 

\begin{figure}[h]
\input{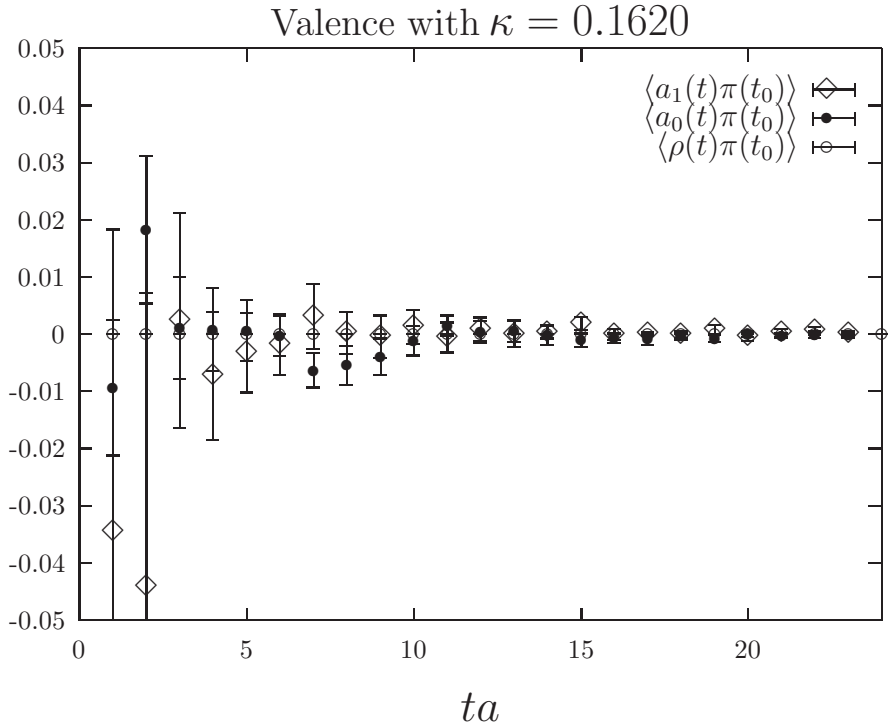}
\caption{The cross correlation functions between the pseudoscalar (P) and
the vector (V), the axial (A) and the scalar (S) channels are shown as a 
function of the time separation ta.}
\end{figure}    

For comparison, we also plot the correlation function in the
pseudoscalar and vector channels $\langle P(t) P(0)\rangle$ and
$\langle V(t) V(0)\rangle$ for both the quenched and the VQCD cases
in Fig. 34. We see that they are indeed much larger than the cross
correlation functions in Figs. 32 and 33.

\newpage
\clearpage
\begin{figure}[h]
\input{fig33.tex}
\caption{The cross correlation functions between the vector (V) and
the axial (A), the scalar (S), and the pseudoscalar (P) channels are shown as a
function of the time separation $ta$.}
\[
\input{fig34_1.tex}
\hspace{-0.25in}\input{fig34_2.tex}
\]
\caption{(a) The direct correlation functions $\langle P(t) P(0)\rangle$
for both the quenched and VQCD as a function of the time separation ta.
(b) The same as in (a) for $\langle V(t) V(0)\rangle$.}
\end{figure}    

\newpage
\clearpage
\subsection{Origin of Hyper-fine Splitting}

We see that the hyper-fine splitting between the $\Delta$ and nucleon
has largely disappeared in the light quark sector when we remove the cloud 
quark and antiqurk in the form of the Z-graphs. This is rather mysterious 
in that according to the usual lore, the hyper-fine splitting is primarily
due to the color-magnetic coupling induced spin-spin interaction between
the quarks~\cite{dgg75,cjj74}. This color-magnetic coupling is related
to the spatial motion of the quarks which should not be affected by the
truncation of the Z-graphs which only constraints the quark motion in the
time direction. Indeed, this color-magnetic coupling is explicitly
shown as $\vec{\sigma}\cdot\vec{B}$ in the Pauli spinor representation of
VQCD in Eqs. (\ref{S_u_Pauli}) and (\ref{pauli}). 
Furthermore, it is this $\vec{\sigma}\cdot\vec{B}$ term which is fully
responsible for the hyperfine splitting between $\Upsilon$ and $\eta_b$ 
in the heavy quark system. The latter is proven by the lattice 
QCD calculation with the non-relativistic QCD action containing such
a term in the form of $\vec{\sigma}\cdot\vec{B}/2M_b$~\cite{shi97,ses98,
dav98}. 

This raises a question as to how effective the color spin interaction is
as far as the hyper-fine splitting is concerned in the light hadron 
spectroscopy. The same question has been raised
by Glozman and Riska~\cite{gr95,gr96}. Upon studying the negative parity and
positive parity excitations of the N, $\Delta$ and $\Lambda$ spectra,
they found that the reverse ordering of the positive and negative
parity resonances of N and particularly $\Delta$ from those of
the $\Lambda$ cannot be accommodated with the color-spin structure of 
the pairwise interaction $\lambda^c_i \cdot \lambda^c_j \vec{\sigma}_i 
\cdot \vec{\sigma}_j$, rather it is consistent with the flavor-spin structure
$\lambda^F_i \cdot \lambda^F_j \vec{\sigma}_i \cdot \vec{\sigma}_j$.
This is so because flavor-spin structure of $\Lambda$ is different
from that of N and $\Delta$. Interpreting this as due to Goldstone boson
exchange, they can fit the low-lying baryon spectrum with
a confinement potential in addition and also the magnetic moments of the baryon 
octets by taking into account of the meson exchange currents~\cite{gr96,gpp96}. 

A similar problem was encountered in searching for scalar diquark
clustering in lattice hadron form factors~\cite{lei93}.
Significant scalar diquark clustering is predicted in quark models
which rely on the hyperfine interaction of the OGEP to split the
N and $\Delta$.  While significant mass splitting is seen in the
lattice simulations of Ref.~\cite{lei93,ldw92,lwd91}
there is no evidence of scalar diquark clustering.  This results leads
one to look for other sources of hyperfine splitting that do not
necessarily lead to clustering in the wave function, such as meson
exchange~\cite{lei93}.

Furthermore, it is well known from the light baryon spectrum that the
spin-orbit interaction is much weaker~\cite{dal82,gs76,ik78,ik79,lw83} than 
that which accompanies
the spin-spin interaction in the one-gluon-exchange picture~\cite{dgg75}.
This is problematic for the gluon-exchange picture if it is to explain
both the heavy quarkonia which require the spin-orbit interaction and
the light baryons which require a much weaker one. However, 
Goldstone boson exchange does not have the spin-orbit interaction between
the light quarks and hence has no problem in this regard.

This Goldstone boson exchange picture appears to be quite consistent with
what we find in VQCD. The flavor-nonsinglet meson exchange between  
the quarks is represented by the Z-graph depicted in Fig. 32. Since all
the Z-graphs are removed in VQCD, there will be no meson exchanges 
between the quarks as a result. This can explain why the hyper-fine 
splitting between $\Delta$ and nucleon is greatly reduced in VQCD (Fig. 29).
But this does not answer the question as to why the color-magnetic 
coupling induced spin-spin interaction is not as effective in light
baryons as in heavy quarkonium. While we don't have a strong evidence
for it, we note that one aspect of the light quark may contribute to the
difference. Unlike the heavy quarks, the propagator of the light quarks in the
background gauge field can fluctuate into color-singlet meson clouds leading
to meson dominance in various form factors (see Fig. 11). The range of
fluctuation depends on the Compton wavelength of the meson. The longest 
range is in the pseudoscalar channel with the pion cloud as evidenced in the
softness of the pseudoscalar form factor of the nucleon (Figs. 10 and 15).
By the same token, Goldstone boson exchange between the quarks in
Fig. 35 can have a range commensurate with its Compton wavelength. On the
other hand, the range of one gluon exchange is limited since the gluon is
confined. If the range of Goldstone boson exchange is longer than
the gluon confinement scale, the hyper-fine interaction from Goldstone
boson exchange is likely to to more effective than that from the 
color-magnetic coupling. In other words, the light quarks in the baryon
have larger separations than those between the quarks and antiquarks in 
heavy quarkonia and this could be the reason for the diminished 
color-magnetic coupling in light baryon due to the limited range of the 
confined gluons.

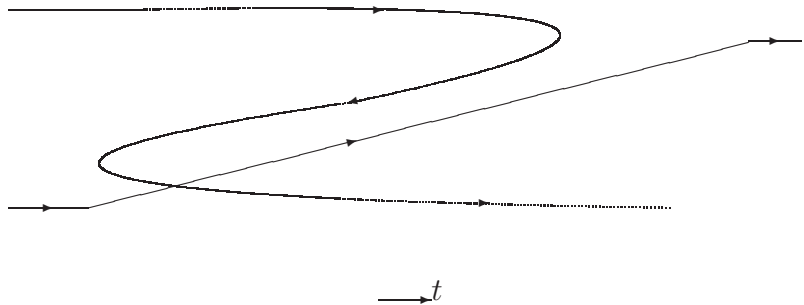
\begin{figure}[h]
\hspace{2.0in}\setlength{\unitlength}{0.01pt}
\begin{picture}(45000,14000)
\qbezier(0000,12500)(27000,13000)(8000,9000)
\qbezier(8000,9000)(-16000,5500)(20000,5000)
\put(-5000,12500){\line(1,0){5000}}
\put(9000,12500){\vector(1,0){200}}
\put(13000, 5180){\vector(1,0){200}}
\put(8000, 9000){\vector(-4,-1){200}}
\put(9000, 1500){\vector(1,0){2000}}
\put(-5000,5000){\line(1,0){3000}}
\put(-2000,5000){\line(4,1){25000}}
\put(23000,11300){\line(1,0){2000}}
\put(-3500,5000){\vector(1,0){200}}
\put(8000,7500){\vector(4,1){200}}
\put(24000,11300){\vector(1,0){200}}
\put(11000, 1500){$t$}
%
\end{picture}
\caption{The meson exchange between quarks in the baryon is depicted as a 
Z-graph. The anti-quark produced in the Z-graph forms a meson with another
quark in this case which is `exchanged' between the two quarks.}
\end{figure}

\subsection{Origin of Dynamical Quark Mass}

  Another significant feature of the VQCD spectroscopy in Figs. 28, 29, 30,
and 31 is that all the hadron masses drop substantially from 
their counterparts in QCD (including pion at finite quark mass).
For example, the nucleon moves down from 940 MeV in QCD (we used this
to fix the scale) to 130(19) MeV in VQCD; $\Delta$ drops from   
1117(72) MeV to 179(25) MeV; and $\rho$ drops from 600(11) MeV to
95 (14) MeV. It is well known that chiral symmetry breaking leads to
a dynamical quark mass related to the quark condensate~\cite{njl61},
 in addition to the existence of Goldstone bosons. This can be seen from 
\begin{equation}
\langle \bar{\Psi}\Psi\rangle = \langle \bar{\Psi}_L\Psi_R + \bar{\Psi}_R
\Psi_L \rangle
\end{equation}
which mixes the left- and right-handed quarks and has the effect of
a dynamical mass as a result of the chiral symmetry breaking. 

To the extent that we can interpret the VQCD result as due to the drop of
dynamical or constituent quark mass, we can draw the following conclusions:
\begin{enumerate}
\item
It is usually assumed in valence quark model that constituent quark mass
arises from the dressing of the glue and the sea quark-antiquark pairs.
Since the hard glue dressing in VQCD is expected to be the same as in QCD, 
it is not likely to be responsible for the dropping of hadron masses in VQCD. 
Furthermore, the quenched lattice 
calculations~\cite{bcs93} can reproduce the $\rho, K^{\*}$, $\Phi$ mesons and 
the octet and decuplet baryon masses to within about 6\% of the experimental
results. This is an indication that the quark loops which generate sea 
quark-antiquark are not the primary source for hadron masses either. 
Here we see from our lattice calculation of VQCD that the dynamical 
quark mass actually arises from the `dressing' of the cloud quarks --- 
quark-antiquark pairs in the connected insertion.  
\item
Since we conclude that the hyper-fine splitting in the light baryons is
largely due to the Goldstone boson exchanges, one would expect the
dynamical mass generation of the chiral symmetry breaking to be
prominently realized in the baryon spectroscopy also. In the
chiral symmetry models, the dynamical mass is generally generated
through the $\sigma$ -- the chiral partner of the pion. For example, in
the linear sigma model, the dynamical mass is given~\cite{es84,bes97} as
\begin{equation}
m_{dyn} = \frac{g_{\sigma qq}^2}{- m_{\sigma}^2} \langle \bar{\Psi}\Psi
\rangle_{m_{\sigma}}
\end{equation}
where $m_{\sigma}$ and $g_{\sigma qq}$ and $m_{\sigma}$ are the sigma mass
and its coupling to the quark. This is represented as the sigma-quark
tadpole diagram as illustrated in Fig. 33(a). Similar mechanism exists in
the four-fermion Nambu-Jona-Lasinio \mbox{model~\cite{cbk96}}. In QCD, the
quark-line diagram which corresponds to the sigma-quark tadpole in Fig. 36(a)
would like something in Fig, 36(b) which inevitably involves cloud quarks and
anti-quarks in the Z-graph. 
\item
Interpreting the tadpole Z-graphs involving mesons in the scalar channel as the
ones responsible for the dynamical quark mass generation, it is consistent
with what we observe in VQCD. Dropping Z-graphs in VQCD which include these
tadpoles diminishes the coupling to the quark condensate 
$\langle \bar{\Psi}\Psi\rangle$ and leads to the downturn of all the
hadron masses from QCD. However, there is still a class of tadpole diagrams
which survive. These are the spatial moving quark loops restricted within time 
slices. They may still couple to $\langle \bar{u}u\rangle$ and
$\langle \bar{v}v\rangle$. But since these condensates in VQCD are much
smaller than that in QCD (see Fig. 26), the dynamical mass is much smaller.
This can explain why the masses of $\Delta$, N, and $\rho$ are small 
but non-zero in VQCD (Figs. 28, 29, and 30). 
\end{enumerate}

\begin{figure}[h]
\hspace{2.0in}\setlength{\unitlength}{0.01pt}
\begin{picture}(45000,20000)
\put(-10000, 8000){\line(1,0){12000}}
\put(-8000,8000){\vector(1,0){200}}
\put( 0000,8000){\vector(1,0){200}}
\put(-4000, 13000){\circle{4500}}
\put(-5000, 5000){\vector(1,0){2000}}
\multiput(-4000,8000)(000,300){10}{\circle*{100}}
\put(-3300,9400){$\sigma$}
\put(-3300,15500){$\langle\bar{\psi}\psi\rangle$}
\put(-2800, 5000){$t$}
\put(-5600, 3000){(a)}
\put(10000, 8000){\line(1,0){5500}}
\put(16500, 8000){\line(1,0){5500}}
\put(12000, 8000){\vector(1,0){200}}
\put(20500, 8000){\vector(1,0){200}}
\put(15500, 8000){\line(0,1){3500}}
\put(16500, 8000){\line(0,1){3500}}
\put(15500, 9900){\vector(0,1){200}}
\put(16500, 9900){\vector(0,-1){200}}
\put(17500, 9900){$\sigma$}
\qbezier(15500,11500)(11000,13500)(16000,14500)
\qbezier(16500,11500)(21000,13500)(16000,14500)
\put(16000,14500){\vector(1,0){200}}
\put(14500,11990){\vector(-2,1){200}}
\put(17500,11990){\vector(-2,-1){200}}
\put(15000, 5000){\vector(1,0){2000}}
\put(17500, 5000){$t$}
\put(15500, 3000){(b)}
%
\end{picture}
\caption{(a) Sigma-quark tadpole diagram in the linear sigma model which is
the mechanism for dynamical mass generation in this model. (b) The
quark line diagram of the Z-graph which corresponds to the sigma-quark
tadpole in (a).}
\end{figure}
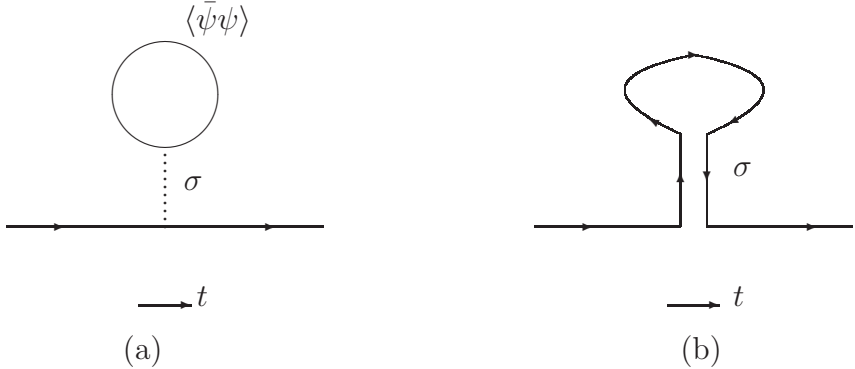

The interpretation we offer for the hyper-fine splitting and
the dynamical quark mass is reminiscent of the chiral quark model of
Manohar and Georgi~\cite{mg84} on which the phenomenological studies
of baryon masses~\cite{gr95,gr96,gpp96} and baryon structure~\cite{cl98}
are based. Arguing that the chiral symmetry breaking
scale $\Lambda_{\chi SB}$ is higher than the confinement scale $\Lambda_{QCD}$,  
they propose the relevant dynamical degrees of freedom to be the 
fundamental quarks,
gluons and the Goldstone bosons in an effective theory at intermediate
scales between $\Lambda_{\chi SB}$ and $\Lambda_{QCD}$. What we observe in
VQCD seems to suggest that the scale for the structure of baryons 
falls just in this range so that the coupling to Goldstone bosons and
dynamical mass generation are evident when QCD and VQCD are compared.

There are other suggestions for the flavor-spin structure of the
quark-quark interaction. These are induced by instantons~\cite{sr89,th76}. 
It is known that the instantons give rise to chiral symmetry breaking 
and generate dynamical quark mass associated with 
$\langle \bar{\Psi}\Psi\rangle$~\cite{ss98,dp84}.
The point-to-point hadronic correlation functions in the instanton liquid model
~\cite{shu93} have been verified by lattice QCD
calculation~\cite{cgh93} and the role of instantons is revealed through
cooling~\cite{cgh94}. Although its direct connection to the cloud degree
of freedom vis-a-vis VQCD is less transparent, the instanton picture,
being the root of chiral symmetry breaking, is expected to reproduce the 
consequences of the chiral quark model.


\section{Symmetry Breaking}

  It is well known that the chiral symmetry ${\rm SU(N_F)_L \times SU(N_F)_R
\times U_V(1)}$ 
of QCD is spontaneously broken to the diagonal ${\rm SU_V(N_f)\times U_V(1)}$.
VQCD, as we have learned in this study, has a different symmetry breaking 
pattern. It starts out with the ${\rm U(2 N_F)}$ symmetry (see Sec. 
\ref{con_sym}) with vector and axial symmetries in the particle-antiparticle
space. Given our lattice simulation, which shows that the pseudoscalar mesons
corresponding to the interpolation fields 
$\bar{u}\gamma_5 v$ and $\bar{v}\gamma_5 u$ become massless at the zero
quark mass limit, the pion decay constant $f_{\pi}$ is non-zero (it may
actually diverge as $1/m_{\pi}$), and the condensate $\langle \bar{u}u\rangle$
and $\langle \bar{v}v\rangle$ do not vanish. We take this as the evidence
for spontaneous breaking of the axial symmetry in Eqs. (\ref{axi1}) and
(\ref{axi2}). This then leads to a ${\rm U_q(N_F) \times U_{\bar{q}}(N_F)}$
symmetry, which is the vector symmetry for the quarks and antiquarks separately.
Again by virtue of the lattice simulation, we find that $SU(6)$ relation
holds quite well for the $g_A^0/g_A^3$ (or $F_A/D_A$), $g_S^3/g_S^0$
(or $D_S/F_S$), and $\mu_n/\mu_p$ ratios. Furthermore, the nucleon and $\Delta$,
and $\rho$ and $\pi$ are nearly degenerate. All these indicate that, although
the $SU(6)$ breaking color-magnetic coupling is present in the VQCD action 
(Eqs. (\ref{S_u_Pauli}), (\ref{S_v_Pauli}), and (\ref{pauli})),
its effects are small. As a result, VQCD has an approximate higher symmetry,
i.e. ${\rm U_q(2 N_F) \times U_{\bar{q}}(2 N_F)}$ where the 2  
represents the spin subgroup SU(2). 
This ${\rm U_q(2 N_F) \times U_{\bar{q}}(2 N_F)}$ is just the non-chiral
${\rm U(6) \times U(6)}$ symmetry of Dashen and Gell-Mann~\cite{dg65} for 
${\rm N_F = 3}$ with quarks and antiquarks in the $({\bf 6, 1})$ and
$({\bf 1, \bar{6}})$ representation respectively. It was proposed as
a `good' symmetry for ``stationary (i.e. bound) and quasi-stationary
(i.e. resonant) states of hadrons at rest''. It is interesting to note
that after stripping off the sea and cloud quarks, we find VQCD 
possesses the same symmetry as in the valence quark model with spin degeneracy.


\section{Conclusion, Analogy to Shell Model and Many Body Theory}

Instead of simulating QCD, we have mutilated it. The valence QCD theory we
have constructed does not honor Lorentz invariance. It also violates unitarity, 
dispersion relation, and cross symmetry. But these are the attributes
shared by the valence quark model which we set out to understand and 
our purpose of this study is to sort out the roles the various 
dynamical quark degrees of freedom play in different observables. 
This is much like the study of the brain~\footnote{We thank T. Cohen for
this analogy.}. One tries to correlate the dysfunction of certain part of the
body with the damage of a specific part of the brain to infer its 
controlling mechanism.   
 
After defining the valence, the cloud, and the sea quarks from the
hadronic tensor in deep inelastic scattering, we have been able to follow these
degrees of freedom to three-point and two-point functions which are
relevant to the quark model at low energies. Upon eliminating the cloud
quarks in the connected insertion with the help of the VQCD action and the 
sea quarks in the disconnected insertion with the quenched VQCD calculation,
we find from the ratios of $g_A^0/g_A^3, g_S^3/g_S^0, \mu_n/\mu_p$ and
the masses of N, $\Delta, \rho$, and $\pi$ that there is an approximate $SU(6)$
symmetry in VQCD which emerges from shaking off the ``dressing'' cloud
and sea quark-antiquark pairs. Its symmetry breaking pattern is distinct
from that of QCD. We summarize the symmetry breaking pattern of QCD and
VQCD in the following chart:
\begin{eqnarray*}
{\rm\Huge\bf QCD: \hspace{2cm} SU(N_F)_L \times SU(N_F)_R \times U_V(1)
\Rightarrow SU_V(N_F) \times U_V(1) } \\
{\rm\Huge\bf VQCD: \hspace{1.8cm} U(2 N_F) \Rightarrow  U_q(N_F) 
\times U_{\bar{q}}(N_F)
\Rightarrow \approx U_q(2 N_F) \times U_{\bar{q}}(2 N_F) }
\end{eqnarray*}
We should point out that the $U_q(2 N_F) \times U_{\bar{q}}(2 N_F)$ symmetry
due to the spin degeneracy is only approximately true. We see that the 
ratio $g_A^0/g_A^3$ in Fig. 21 is not exactly 3/5 and the neutron electric
form factor $G_E^n(q^2)$ in Fig. 26 is not exactly zero. The degeneracies
between $\Delta$ and N and between $\rho$ and $\pi$ are not perfect either.
These indicate that there are still some color-magnetic
coupling induced spin effects. Nevertheless, it is a fairly good
approximate symmetry. What we have demonstrated in this study is that 
QCD has this approximate symmetry in its valence approximation \`{a} la
VQCD action. To our best knowledge, this is the connection between QCD and
the valence quark model.

The relation of the valence quark model and QCD actually is analogous to
that between the shell model of nuclei and the many body theory. It is 
perhaps instructive to point out the parallel developments in the history
of nuclear physics and hadron physics as far as the fermion dynamical
degrees of freedom are concerned. We recall
that the raison d'\^{e}te of the shell model consists of the pattern of
energy levels, the spin and parity quantum numbers of nuclei, and the
Schmidt lines for the magnetic moments of nuclei. Similar reasons,
e.g. the mass pattern of baryons and mesons, SU(3) flavor symmetry, and
the magnetic moments of proton and neutron lent their support for
the existence of the quark model. Later experiments and theoretical
developments in many body theory pointed out the inadequacies of the
shell model and ideas like collectivity of the giant
resonances~\cite{bb59}, pairing through the induced phonon-exchange 
interaction~\cite{kb66}, and core polarization for the
magnetic moments or the Arima-Horie effect~\cite{ah54} are introduced. 
These involve the particle-hole degrees of freedom in the disconnected
insertion which are the core
polarization effects beyond the shell model. With the advent of 
QCD as the fundamental theory of quarks and gluons,
similar ideas are introduced. For example, the resolution of the
U(1) anomaly in terms of the topological susceptibility in
the large $N_c$ analysis by Witten and Veneziano \cite{wv79} is the
schematic model~\cite{bb59} approach to generating the $\eta'$
mass by the collective coupling between quark loops.
The concept of
quark and gluon condensates are certainly related to pairing in the
many body theory. Lack of appreciation for vacuum polarization due to the
sea quarks for flavor-singlet observables in the quark model has led to
the ``proton spin crisis''~\cite{dll95,cheng96} and the $\pi N \sigma$ 
term puzzle~\cite{che76,gls91,dll96}. The importance of Z-graphs for
density dependence was pointed out for the effective nucleon-nucleon
interaction 
~\cite{bk68}, and the higher density effects in the relativistic mean-field
theory are shown largely to be due to the Z-graph with sigma meson exchanges
~\cite{bwb87}. The importance of the cloud quarks in hadrons through the
Z-graphs are just beginning to be unraveled. The violation of the
Gottfried sum rule leading to $\bar{u}(x) \neq \bar{v}(x)$ is shown to be
due to the cloud antiquarks~\cite{ld94}. Furthermore, we have learned in the 
present study that the hyper-fine splitting in baryons and the dynamical 
quark mass are related to the cloud degree of freedom, which are probably 
the most surprising results of VQCD.
 
The valence quark model, as we come to realize it today, is just like the
shell model in nuclear physics. The $U(6) \times U(6)$ symmetry which comes
with the valence quark model as the defining characteristic is not as good a
symmetry as one tends to believe. Even with the supplement of $SU(6)$
breaking one-gluon exchange, it does not capture the richness of
the cloud degree of freedom in various form factors and matrix elements in
the connected insertions. Moreover, the lack of sea degrees of freedom in the
disconnected insertions is responsible for its overestimate of the 
flavor-singlet $g_A^0$ by a factor of $\sim 3$ as well as its underestimate of 
the $\pi N \sigma$ term by a factor $\sim$ 3 -- 4. What it lacks appears to be
the spontaneous breaking chiral symmetry of QCD. This is exemplified in 
hadron spectroscopy where we find that the hyper-fine splitting between N
and $\Delta$ and the dynamical quark mass are related to the cloud quark
in the Z-graphs. 

One lesson we learned in this study is that the valence quark model is not 
necessarily a bad
place to start building an effective theory of hadrons, provided one
knows how to incorporate chiral symmetry and restores the
cloud and sea degrees of freedom. Working in the intermediate
scale between chiral symmetry breaking and confinement
which is appropriate for studying hadron structure and spectroscopy, one may 
start with the chiral quark model~\cite{mg84}. Integrating out the short-range 
part of
the quark field is shown to lead to a very successful effective chiral theory 
of mesons~\cite{li95}. One may extend this to the baryon sector with the
quark coupling to the gluons and mesons~\cite{mg84}. In this way, the
cloud degrees of freedom will show up in the form factors and matrix elements
via meson dominance and meson exchange currents. It can also give rise to
the hyper-fine splitting and the dynamical quark mass. The meson loops on the
quark lines, on the other hand, are responsible for the sea degrees of freedom.
We will study this in the future. 

This work is partially supported by U.S. DOE grant No. DE-FG05-84ER40154. 
The authors would like to thank T. Barnes, C. Bernard, S. Brodsky, G. E. Brown,
T. P. Cheng, T. Cohen, M. Golterman, L. Ya. Glozman, X. Ji, J.-F. Laga\"{e},  
C. S. Lam, B. A. Li, R. McKeown, J. Negele, M. Rho, D. O. Riska, C. Roberts,
J. Rosner, E. Shuryak, A. Thomas, W. Weise, U.-J. Wiese, and A. Williams 
for very useful discussions.

\end{document}